\documentclass[11pt,aps,superscriptaddress,prb]{revtex4}
\usepackage{epsfig}
\usepackage{amsmath}
\usepackage{amssymb}
\usepackage{bbold}

\usepackage{xcolor}
\usepackage{verbatim}  
\usepackage{mathrsfs} 
\usepackage{marvosym} 
\usepackage{leftidx}  
\usepackage{pifont}  
\DeclareMathOperator{\arctanh}{arctanh}

\newcommand{\Ef}{\epsilon_{\textsc{f}}}
\newcommand{\kf}{k_{\textsc{f}}}
\newcommand{\Efso}{E_{\textsc{f}}}   

\newcommand{\as}{a_s}
\newcommand{\ar}{a_r}

\newcommand{\bose}{n_{\textsc{b}}}
\newcommand{\fermi}{n_{\textsc{f}}}

\newcommand{\mub}{\mu_{\textsc{b}}}
\newcommand{\mb}{m_{\textsc{b}}}

\newcommand{\psidagu}{\psi^{\dagger}_{\uparrow}}
\newcommand{\psidagd}{\psi^{\dagger}_{\downarrow}}
\newcommand{\hpsidagu}{\hat\psi^{\dagger}_{\uparrow}}
\newcommand{\hpsidagd}{\hat\psi^{\dagger}_{\downarrow}}
\newcommand{\psiu}{\psi_{\uparrow}}
\newcommand{\psid}{\psi_{\downarrow}}
\newcommand{\hpsiu}{\hat\psi_{\uparrow}}
\newcommand{\hpsid}{\hat\psi_{\downarrow}}
\newcommand{\psidag}{\psi^{\dagger}}
\newcommand{\hpsidag}{\hat\psi^{\dagger}}
\newcommand{\hpsi}{\hat\psi}
\newcommand{\Psidag}{\bar{\Psi}}

\newcommand{\Ek}{E_{\mathbf{k}}}

\newcommand{\xik}{\xi_{\mathbf{k}}}
\newcommand{\gammak}{\gamma_{\mathbf{k}}}

\newcommand{\gammahk}{\gamma_{h\mathbf{k}}}

\newcommand{\tgamma}{\tilde{\gamma}}
\newcommand{\tgammap}{\tilde{\gamma}_{\mathbf{p}}}

\newcommand{\epsk}{\epsilon_{\mathbf{k}}}

\newcommand{\kmq}{{\mathbf{k}-\mathbf{q}}}

\newcommand{\kqp}{_{\mathbf{k}+\mathbf{q}/2}}
\newcommand{\kqm}{_{\mathbf{k}-\mathbf{q}/2}}
\newcommand{\ks}{{\sf k}}
\newcommand{\qs}{{\sf q}}
\newcommand{\q}{{\sf q}}
\newcommand{\p}{{\sf p}}

\newcommand{\vR}{v_{\textsc{r}}}
\newcommand{\vd}{v_{\textsc{d}}}
\newcommand{\vD}{v_{\textsc{d}}}
\newcommand{\vf}{v_{\textsc{f}}}

\newcommand{\tTc}{\tilde{T}_{c}}
\newcommand{\tmu}{\tilde\mu}
\newcommand{\bTc}{\bar{T}_{c}}
\newcommand{\bmu}{\bar\mu}

\newcommand{\tvr}{\tilde{v}_{\textsc{r}}}
\newcommand{\tvd}{\tilde{v}_{\textsc{d}}}

\newcommand{\be}{\begin{equation}}
\newcommand{\ee}{\end{equation}}
\newcommand{\bea}{\begin{eqnarray}}
\newcommand{\eea}{\end{eqnarray}}
\newcommand{\kv}{{\bf{k}}}
\newcommand{\qv}{{\bf{q}}}

\newcommand{\Tr}{\textrm{Tr}}

\newcommand{\bra}{\langle}
\newcommand{\brabra}{\langle\!\langle}
\newcommand{\ket}{\rangle}
\newcommand{\ketket}{\rangle\!\rangle}
\newcommand{\spinup}{\!\uparrow \ }	
\newcommand{\spindown}{\!\downarrow \ }	

\normalsize
\begin{document}


\title{Critical temperature in the BCS-BEC crossover with spin-orbit coupling}
\author{Luca Dell'Anna}
\affiliation{Dipartimento di Fisica e Astronomia ``G. Galilei'', University of Padova, via F. Marzolo 8, 35151 Padova, Italy}
\author{Stefano Grava}
\affiliation{ICFO-Institut de Ciencies Fotoniques, The Barcelona Institute of Science and Technology, 08860 Castelldefels, Barcelona, Spain}

\date{\small\today}


\begin{abstract}

We review the study of the superfluid phase transition in a system of fermions whose interaction can be tuned continuously along the crossover from Bardeen-Cooper-Schrieffer (BCS) superconducting phase to a Bose-Einstein condensate (BEC), also in the presence of a spin-orbit coupling. Below a critical temperature the system is characterized by an order parameter. Generally a mean field approximation cannot reproduce the correct behavior of the critical temperature $T_{c}$ over the whole crossover. 
We analyze the crucial role of quantum fluctuations beyond the mean-field approach useful to find $T_{c}$ along the crossover in the presence of a spin-orbit coupling, within a path integral approach. A formal and detailed derivation for the set of equations useful to derive  $T_c$ is performed in the presence of Rashba, Dresselhaus and Zeeman couplings. In particular in the case of only Rashba coupling, for which the spin-orbit effects are more relevant, the two-body bound state exists for any value of the interaction, namely in the full crossover. As a result the effective masses of the emerging bosonic excitations are finite also in the BCS regime. 

\end{abstract}


\maketitle

\section{Introduction}
Experimental developments in confining, cooling and controlling the strength of the interaction of alkali atoms brought a lot of attention to the physics of the crossover between two fundamental and paradigmatic many-body systems, the Bose-Einstein condensation (BEC) and the Bardeen-Cooper-Schrieffer (BCS) superconductivity.

A system of weakly attracting fermions can be treated in the context of the well-known BCS theory \cite{BCS1}, originally formulated to describe superconductivity in some materials where an effective attractive interaction between electrons can arise from electron-phonon interaction.
This theory predicts a phase transition and the formation of the so-called Cooper pairs \cite{Cooper} with the appearance of a gap in the single particle spectrum. 
The fermions composing these weakly bounded pairs are spatially separated since their correlation length is larger than the mean inter-particle distance; for this reason the pairs cannot be considered as bosons. 
By increasing the strength of the attraction among the fermions, the system could be seen as made of more localized Cooper pairs, or almost bosonic molecules, which may undergo into a true Bose-Einstein condensation. Since there is no evidence of a  broken symmetry by continuously varying the width of the couples, this process realizes a crossover between a BCS and a BEC state. 
The idea of this crossover preceded by far its experimental realization \citep{Eagles} and became a subject of active investigations by several authors. Those studies were first motivated by the purpose of understanding superconductivity in metals at very low electron densities, since in this diluted regime Cooper pairs are smaller than the mean inter-particle distance and, therefore, can be treated like bosons. 
In a particularly relevant pioneering work \citep{NSR} a system of interacting fermions has been investigated to obtain the evolution of the critical temperature of the system along the crossover by changing the strength of the interaction between the BCS and the BEC limits. 
The discovery of the high-temperature superconductors \cite{BernorzMuller}, 
where the coupling between electrons is supposed to be stronger than that in conventional superconductors, 
renewed the interest in the BCS-BEC crossover with the aim to explore the regime of stronger interaction. 
Few years later, an important work \citep{Articolo_tesi1} provided for the first time a path integral formulation of the problem, 
which we will review here, describing the crossover at finite temperature with tunable strength of the coupling between pairs.
The scientific interest in this field was first purely theoretical because the interaction strength in real solid state materials 
cannot be tuned easily. The situation changed drastically after the first realizations of magnetically confined ultracold alkali gases.
A typical alkali gas obtained in laboratory is made by $10^{5}$ to $10^{9}$ atoms, with a density $n$ inside the bulk of a realistic trap in the range from $10^{12}$ to $10^{15}$ $cm^{-3}$. An important property is their diluteness: these densities are such that the mean inter-particle distance is larger than the length scale associated to the atom-atom potential and implying that only two-body physics is relevant. 
Below those densities, the thermalization of the gas after a perturbation becomes too slow, while at higher densities three body collisions become relevant. 
At low temperatures, namely at low kinetic energies, the scattering properties of the atoms can be characterized by a single parameter, the scattering length. 
These gases can be furthermore cooled at such temperatures to enter in a quantum degenerate regime characterized by a thermal de Broglie  wavelength of the same order of the interatomic distance,  i.e. 
   $\lambda_{th}^2=\frac{2\pi\hbar^{2}}{m k_B T}\sim n^{-2/3}$.
The degenerate temperatures, for the densities mentioned before, are smaller than a $\mu K$.  
These are the conditions achieved in the first experimental realization of a Bose-Einstein condensation \citep{firstBEC}.
The importance of such experimental setups does not rely only on these extreme conditions, but also on the possibility of fine tuning the parameters which describe the gas. 
In particular the mechanism of the so-called Feshbach resonance allows one to control the strength of the interaction between two atoms by  simply a magnetic field, for example, making it sufficiently strong to support a new bound state. This phenomenon was first predicted and then  observed experimentally \citep{Inouye,Courteille} in bosonic systems.
Soon after these techniques have been extended to create and manipulate ultracold gases for fermionic systems \citep{JinDeMarco,Feshbach_Fermi,feshbach40K,Chin_feshback_review} employing atomic alkali gases with two different components, a mixture of atoms in two different spin or pseudospin states, where the quantum degeneracy is reached when lowering the temperature the energy of the system ceases to depend on the temperature. 
In this system by Feshbach resonance the interaction among the atoms can be made sufficiently strong to allow the formation of a two-body bound state, called Feshbach molecule, with a lifetime compatible to the time needed to cool the system down and to observe a Bose-Einstein condensation. Later the experimental observation of fermionic pairs was extended to the whole region of the crossover \citep{BCSBECCrossover1,BCSBECCrossover2}, studying also the thermodynamical properties \citep{BCSBECCrossover3,BCSBECCrossover4}. 
A renewed interest in the study of the BCS-BEC crossover arose in 2011 after the experimental realization, in the context of ultracold gases, of a synthetic spin-orbit (SO) interaction \citep{SObose,dalibard,SOfermi,SOfermi2}. In contrast to the spin-orbit interaction in solid state systems, in atomic gases this coupling is produced by means of laser beams and therefore is perfectly tunable. 
The realization of this artificial spin-orbit coupling is a bright example of the versatility and controllability of the ultracold techniques, showing once more that ultracold gases are the perfect experimental playgrounds for testing physical models
hardly realizable by means of other solid state setups. 
These experimental achievements stimulated intense theoretical effort to understand the spin-orbit effects
with the so-called Rashba \cite{rashba} and Dresselhaus \cite{dress} terms in the evolution from BCS to BEC superfluidity, 
considering many SO configurations, in two and three dimensions, at zero and finite temperature, exploring eventually novel phases of matter or topological properties.  
\cite{PhysRevB.83.094515,shenoy2,gong,PhysRevA.84.063618,PhysRevLett.107.195305,iskin,yi,Articolo_tesi2,iskin2,zhou,jiang,sa,chen,zhou2,yangwan,iskin3,sa2,he,luca2,Zhang2,Zhang1,a1,a2,a3,a4,a5,a6,a7,a8,a9,b1,b2,b3,b4,b5,b6,b7,b8,b9,c1,c2,c3,c4,c5} 

The aim of this paper is to review some basic concepts on the BCS-BEC crossover,  
reproduce many results beyond the mean-field level, and extend the study in the presence of a spin-orbit interaction, in a detailed and comprensive way, by means of a path integral approach, with a particular emphasis to the derivation of the critical temperature $T_c$. 
We will show how $T_c$ can be strongly enhanced by the SO couplings in the BCS side, already at the mean-field level, while its increase is softened by the effects of the quantum fluctuations in the intermediate region of the crossover and in the strong coupling regime.

\subsection{Basic concepts in ultracold atomic physics}

To understand the structure of an atom it is necessary to consider the forces among its constituents. For alkali atoms (Li, Na, K, Rb\dots), 
the relevant force is due to Coulomb interaction that acts between the off-shell electron and an the effective charge $Ze$ of the nucleus. This central force is known to lead to a Bohr energy spectrum 
 \begin{equation}
   E_{n}=-\frac{m_r c^{2}}{2}\frac{Z^2\alpha^{2} }{n^{2}} 
 \end{equation}
with $n=1,2,...$, where $ \alpha=\frac{e^{2}}{4\pi\epsilon_{0}\hbar c}\approx\frac{1}{137}$ is the so-called fine structure constant and 
   $m_r\approx m_{e}$ the reduced mass, approximately given by the mass of the electron. 
This spectrum is degenerate in the quantum numbers $L$ and $L_z$, associated to the angular momentum operators $\hat{L}^2$ and $\hat{L}_{z}$. 

The Coulomb interaction, due to its symmetry, preserves the angular momentum and the spin separately. However, in principle,  they are coupled via a spin-orbit term, 
$   \hat{H}_{so}=\frac{\alpha_{so}}{\hbar^{2}}\hat{L}\cdot\hat{S}$, 
therefore, only the total angular momentum $\hat{J}=\hat{L}+\hat{S}$ is conserved. This additional term causes the known correction to the spectrum, called \emph{fine structure}, shifting the degeneracy in the energy spectrum by 
  $ \Delta E_{J}=\frac{\alpha_{so}}{2}\left[J(J+1)-L(L+1)-S(S+1)\right]$. 
However, for ultracold alkali gases, at very low temperatures, 
only the fundamental state is occupied, so that the off-shell electron is in the \emph{s} orbital ($L=0$). In this state $J=S$ and the spin-orbit fine structure can be neglected, $\Delta E_{J}=0$.
However, also the coupling between the spins of the electrons $\hat S$ and of the nucleus $\hat I$ can remove the degeneracies and it is generally approximated by a dipole-dipole interaction
 \begin{equation}
 \label{H_hf}
   \hat{H}_{hf}=\frac{\alpha_{h}}{\hbar^{2}}\hat{I}\cdot\hat{S}
 \end{equation}
This term, again, does not lead to the conservation of both $\hat{I}$ and $\hat{S}$ separately but only of the total spin of the atom $\hat{F}=\hat{I}+\hat{S}$, since ${L}=0$.
Then it is possible to label the states of an alkali atom by the quantum numbers $F$ and $F_{z}$, related to the operators $\hat{F}^2$ and $\hat{F}_{z}$. The effect of this contribution is a small correction to the Bohr spectrum, usually called \emph{hyperfine structure}
 \begin{equation}
   \Delta E_{F}=\frac{\alpha_{h}}{2}\left[F(F+1)-I(I+1)-S(S+1)\right].
 \end{equation}
Since $S=\frac{1}{2}$ and because of the properties of the sums of angular momentums, the quantum number $F$ is allowed only to have the values $F=I\pm\frac{1}{2}$, so the hyperfine correction is given by
 \begin{equation}
   \Delta E_{F=I\pm\frac{1}{2}}=\pm \frac{\alpha_{h}}{2}\left(I+\frac{1}{2}\mp \frac{1}{2}\right)
 \end{equation}

In the presence of an homogeneous magnetic field, this coupling between the latter and the electronic and nuclear spins gives rise to a shift of the energy levels of the atoms, called the Zeeman effect.
An alkali atom at very low energy and in a magnetic field can be described, therefore, by the following Hamiltonian 
 \begin{equation}
   \hat{H}=\hat{H}_{hf}-\gamma_{N}B\hat{I}_{z}+\gamma_E B\hat{S}_{z}
   \label{HhfZee}
 \end{equation}
where the Zeeman couplings are inversely proportional to the mass of the particles, therefore $\gamma_{N}\ll\gamma_E$, since the nucleus is heavier than the electrons.  
As long as the effect of the magnetic field is small, $F$ and $F_z$ can be considered almost good quantum numbers, while 
in general the eigenstates are labeled by $I_z$ e $S_z$, associated to $\hat{I}_{z}$ and $\hat{S}_{z}$. 
Diagonalizing the Hamiltonian in Eq~\eqref{HhfZee} 
we can find the corrections to the Bohr spectrum. For example, in the case of an hydrogen atom $I=\frac{1}{2}$ and we get
 \begin{equation}
   \Delta E_{\pm}=-\frac{\alpha_{h}}{4}
   \pm\frac{\alpha_{h}}{2}\sqrt{1+\left(\frac{\hbar(\gamma_{N}+\gamma_E)B}{\alpha_{h}}\right)^{2}}.
 \end{equation}
The dependence on the magnetic field is quadratic for low intensities and linear for high field. An important experimental feature is that varying the magnetic field it is possible to vary the energy levels of the single atom, 
a property that can be exploited to induce a confining potential for atoms, realizing an atomic trap.

There are two main kinds of atomic traps available, the magnetic traps and the optical dipole traps, which are briefly discussed below.

\paragraph*{Magnetic traps.} 
This technique is based on the concrete possibility of creating position dependent magnetic fields
which, as a consequence, make every hyperfine energies also position dependent. In these conditions an alkali atom 
has a kinetic energy plus an effective potential, that, properly taylored, can be seen as a confining potential for the gas, 
 \begin{equation}
   E_{\text{atom}}(\mathbf{r})=\frac{\hbar^{2}p^{2}}{2m}+V(B(\mathbf{r}))
 \end{equation}
with a confining force $\mathbf{F}(\mathbf{r})=-\nabla V(\mathbf{r})$.
Trapping a gas in a magnetic field is the key to achieve degeneracy temperatures; in these conditions the gas is isolated from the external world getting rid of the main source of heating which is due to the walls of the container.
Typically the magnetic trap is modeled to obtain an harmonic potential, however there are several experimental possibilities  depending only on the ability to create non homogeneous magnetic fields. 
In particular these potentials highly depend on the hyperfine states of the atom and this characteristic can be exploited to select particular states in the trap. Another interesting possibility is to create highly anisotropic potentials by strong confinement in some dimensions, realizing one- or two-dimensional systems. This technique is compatible with the cooling methods of atomic gases but not with the mechanism of the magnetic Feshbach resonance. For the latter reason an alternative trapping method is required.

\paragraph*{Optical traps.}
The optical traps are based upon the interaction between an atom and light of a laser beams. 
The key features of this method are the great stability of the trapped gas and the possibility of making a lot of different geometries. Under proper conditions, called \emph{far-detuning}, the confinement is independent from the internal state of the atom. 
When an atom is put in a laser beam, the electric field $\mathbf{E}(\mathbf{r})=\mathbf{E}_0(\mathbf{r})e^{i\omega t}$ induces an electric dipole moment $\mathbf{p}(\mathbf{r})=\alpha_\omega\mathbf{E}_0(\mathbf{r})e^{i\omega t}$, 
where $\alpha_\omega$ is the polarizability. 
The effective potential given by the iteration between the dipole moment and the electric field is given by a temporal mean value 
 \begin{equation}
   V(\mathbf{r})=
   -\frac{1}{2}\,\overline{\mathbf{p}(\mathbf{r})\cdot\mathbf{E}(\mathbf{r})}
    \end{equation}
which is simply directly proportional to the intensity $\textrm{I}(\mathbf{r})$ of the beam. 
It is then possible to produce an artificial potential, as before, considering a position dependent focalization of the laser so that there is a force acting on the single atom given by
$   \mathbf{F}(\mathbf{r})=-\nabla V(\mathbf{r})\propto\nabla I(\mathbf{r}) $ .
As mentioned, this technique makes possible to easily create different geometries. Indeed, employing two counter-propagating laser in more dimensions for example, \emph{lattice potentials} can be realized, particularly interesting since they can simulate solid state systems in a fully controllable environment.  For a more complete review on this confining technique one can refer to Refs.~\citep{opticaltrap,UCFermi}.

The trapped gas has to be cooled down in order to reach the degeneracy temperatures. This can be done in two steps, by laser cooling and evaporative cooling. \citep{BECtrap}

\paragraph*{Laser or Doppler cooling.} 
The term laser cooling refers to several techniques employed to cool atomic gases, taking advantage of the interaction between light and matter. The most relevant one takes advantage also from the Doppler effect.   
This technique is applied to slow down the mean velocity of the atoms realizing a situation in which an atom absorbs more radiation in a direction and less in the opposite one. This can be done slightly detuning the frequency of the laser below the resonance of the exited state of the atom. Because of the Doppler effect, the atom will see a frequency closer to the resonant one and so it will have an higher absorption probability. In the opposite direction the effect will be reversed. 
The atom, which is now in an exited state, can spontaneously emit a photon. The overall result of the absorption and emission process is to reduce the momentum of the atom, therefore its speed. If the absorption and emission are repeated many times, the average speed, and therefore the kinetic energy of the atoms, will be reduced, and so the temperature. 
By this technique one can reach low temperatures of the order of some $\mu K$, still not enough to get a degenerate gas.

\paragraph*{Evaporative cooling.}

The degenerate temperatures can be reached by the so-called {forced evaporative cooling} method.
A fundamental feature of every trap discussed before is the possibility of 
varying its depth, defined as the energy difference between the bottom and the scattering threshold of the potential. 
Lowering the depth of the trap it is possible to let the most energetic atoms escape; this correspond to extracting energy from the gas at the cost of loosing a considerable quantity of matter. After the thermalization of the gas, the temperature is lowered and this procedure can be repeated.  
The lower limit in temperature now is related to the rate between inelastic collisions, that fix the lifetime of the sample, and the elastic collisions, that assure thermalization and fix the time needed for the cooling process. Loading more atoms in a trap lowers the limit of the lowest temperature achievable but also makes the gas unstable. With the correct compromise this scheme allows to go well below the degenerate temperatures. For alkali atoms the limits in temperatures achievable by evaporative cooling are at the order of $T\sim$pK \citep{evacooling}. 
Once a degenerate gas is obtained, letting the gas expand in free flight by turning off the trap, from the absorption spectrum one 
can access to the temperature of the gas or to the density profile in position space that can be compared with theoretical results.

\subsection{Artificial spin-orbit interaction in ultracold gases}
\label{spin_orbit_term}

Laser techniques can be employed also to build a tunable \emph{synthetic} spin-orbit coupling between two pseudo-spin states of an atom. 
In a previous section it was shown that spin-orbit 
coupling (fine structure) for atoms in the fundamental state ($n=0$, $L=0$) vanishes. 
In order to induce it in laboratory it is necessary to select a gas with two components. In the first realization of this system \citep{SObose} two states 
($|F,F_z\ket$) of {$^{87}$}{Rb}{} were chosen with pseudo-spins $|\uparrow\ket=|1,0\ket$ and $|\downarrow\ket=|1,-1\ket$. 
In order to understand schematically how it works, let us suppose that the atom is in an initial state $|\downarrow\ket$, then, two lasers can be tuned in such a way that one induces a transition to an intermediate exited state and the other induces a stimulated emission from the latter state to $|\uparrow\ket$. 
The inverse process is equally allowed. Due to Doppler effect, this coupling will be necessary momentum-sensitive and, as shown in Refs.~\citep{SObose,dalibard}, the effective generated term is equivalent to a spin-orbit coupling of a spinfull particle moving in a static electric field, 
let us say, along the $z$ axis. An effective additional term in the Hamiltonian can read
 \begin{equation}
   H_{R}=-\mathbf{\mu}_{s}\cdot\textbf{B}_{SO}=v_R(\sigma_{x}k_{y}-\sigma_{y}k_{x})
   \label{HR}
 \end{equation} 
where $B_{SO}$ is the magnetic field seen by the moving particle. This is called a Rashba coupling, in analogy to the same coupling occurring in crystals. 
With the same technique other interactions can be made in laboratory such as a Dresselhaus term 
 \begin{equation}
H_{D}=v_D(\sigma_{x}k_{y}+\sigma_{y}k_{x})
\label{HD}
 \end{equation} 
or a Weyl term $H_{W}=\lambda\,\mathbf{\sigma}\cdot\mathbf{k}$. 
The two lasers are detuned by a frequency $2\delta$ from the Raman resonance 
and couple the states $|\uparrow\ket$ and $|\downarrow\ket$ with the Raman strength $2\Omega$. This generates additional terms, $\Omega\sigma_z$ and $\delta\sigma_y$. We will neglect the detunig term and will consider the Raman coupling as an effective artificial Zeeman term
\begin{equation}
H_{Z}=h\,\sigma_z.
\label{HZ}
 \end{equation} 
By this procedure one can study 
the effects of a spin-orbit interaction in 
ultracold gases, which has the great advantage of being fully under control, allowing 
for fine tuning 
experiments where the couplings can be easily manipulated, not possible otherwise in solid state systems.

\section{Two-body scattering problem}
In the previous section we said that only two-body physics is relevant in a dilute ultracold gas, for that reason we will briefly review the two body problem in a generic central potential.  
We are going to consider only scattering properties after two-body collisions neglecting the three-body ones 
because of the low probability of finding three particles within the range of interaction. Generally interatomic potentials are not known analytically and their approximations are not so easy to use in theoretical and numerical evaluations.  
The density of the gas can fix the interparticle distances and the range of the potential. 
This is what typically happens in the experiments with ultracold dilute gases. 
In this situation an approximate potential, or \emph{pseudo-potential}, can be employed, 
which should exhibit some general properties of the full scattering problem.
A detailed analysis of the scattering problems, specifically for ultracold gases, can be found in Refs.~\citep{Landau,Taylor,Ultracold}.
Scattering theory provides the tools useful to characterize the interaction between particles in terms of few parameters that can be used to create suitable approximate potentials. 
The Schr\"odinger equation is employed to determine the wavefunctions and the scattering amplitudes, then we require that a simple pseudo-potential should reproduce the same results in a low energy limit, namely in the ultracold limit. 
We will assume that the two-body interaction is described by a short-range and spherically symmetric potential.

It is known that Schr\"odinger equation for the wavefunction of two colliding particles can be decomposed in one equation that describes the center-of-mass motion, a free particle equation with mass $M=m_1+m_2$, and another that describes the relative motion. 
With the definition $m=\frac{m_{1}m_{2}}{m_1+m_2}$ for the reduced mass, in the relative frame the Sch\"odinger equation reads
\begin{equation} 
 \left[-\frac{\hbar^{2}\nabla^{2}}{m}+V(\mathbf{r})\right]\psi(\mathbf{r})=
 \left[\hat{H}_{0}+V(\mathbf{r})\right]\psi(\mathbf{r})=
 E\psi(\mathbf{r})\,.
 \label{twobody}
 \end{equation}  
Let us suppose that the potential is short-ranged, so that there exists a characteristic length $r^{*}$ such that the potential can be neglected outside this radius, 
 \begin{equation}
   V(\mathbf{r})\approx0, \ \ \  \forall \ |\mathbf{r}|>r^{*}
 \end{equation} 
In this case Eq.~\eqref{twobody} becomes a free Schr\"odinger equation whose solution is given by 
the composition of an incoming plane wave with momentum $\mathbf{p}$ and a scattering state with momentum $\mathbf{p}'$. It is relevant to look for \emph{elastic collisions}, that can be studied fixing the scattering energy to $E=2\epsilon_{\mathbf{p}}=\frac{p^{2}}{m}$, equal to the incoming wave energy. The solution is given by
 \begin{equation}
 \left[2\epsilon_{\mathbf{p}}-\hat{H}_{0}\right]|\psi_{\mathbf{p}}\rangle=
 \hat{V}|\psi_{\mathbf{p}}\rangle.
 \end{equation} 
The homogeneous solution ($\hat{V}=0$) is simply $|\psi_{\mathbf{p}}\rangle=|\mathbf{p}\rangle$, therefore, it is possible to write an implicit solution, known as the Lippmann-Schwinger equation
 \begin{equation}
|\psi_{\mathbf{p}}\rangle=|\mathbf{p}\rangle+
 \frac{1}{2\epsilon_{\mathbf{p}}-\hat{H}_{0}+i\epsilon}\hat{V}|\psi_{\mathbf{p}}\rangle
 \end{equation} 
with $\epsilon$ an infinitesimally small positive real number. In the coordinate representation, it reads
  \begin{eqnarray}
\nonumber \psi_{\mathbf{p}}(\mathbf{r})&=&
\langle\mathbf{r}|\mathbf{p}\rangle
+\langle\mathbf{r}|\frac{1}{2\epsilon_{\mathbf{p}}-\hat{H}_{0}+i\epsilon}\hat{V}|\psi_{\mathbf{p}}\rangle\\
  &=&\frac{e^{\frac{i}{\hbar}\mathbf{r}\cdot\mathbf{p}}}{(2\pi\hbar)^{3/2}}
-\frac{m}{4\pi\hbar^{2}}\int d^{3}r'
\frac{e^{\frac{i}{\hbar}p|\mathbf{r}-\mathbf{r}'|}}{|\mathbf{r}-\mathbf{r}'|}
\langle\mathbf{r}'|\hat{V}|\psi_{\mathbf{p}}\rangle
\label{eq.19}
 \end{eqnarray}
 At long distances, (see Appendix \ref{app.A} for details) Eq.~(\ref{eq.19}) can be written as 
  \begin{equation}
  \label{eq.20}
\psi_{\mathbf{p}}(\mathbf{r})=\frac{1}{(2\pi\hbar)^{3/2}}
\left\{
e^{\frac{i}{\hbar}\mathbf{r}\cdot\mathbf{p}}
+f(\mathbf{q},\mathbf{p})\frac{e^{\frac{i}{\hbar}pr}}{r}
\right\}
+O(r^{-2})
 \end{equation} 
 after defining $\mathbf{q}\equiv p\,\hat{\mathbf{r}}=\frac{p}{r}\mathbf{r}$, and the scattering amplitude 
 \begin{equation}
 \label{eq.f}
   f(\mathbf{q},\mathbf{p})=
-2\pi^{2}\hbar m\,
\bra{\mathbf{q}}|\hat{V}|\psi_{\mathbf{p}}\rangle\,.
 \end{equation}
The scattering solution at large distances is then composed by an incoming plane wave and an outgoing spherical wave, weighted by the scattering amplitude. 
In terms of the latter it is also possible to define the differential cross section of the process, at large distances, 
 \begin{equation}
   \frac{d\sigma}{d\Omega}=|f(\mathbf{q},\mathbf{p})|^{2}\,.
 \end{equation}
Due to the spherically symmetric potential, the scattering amplitude depends only on the modulus $p$ and the angle $\theta=\arccos(\frac{\mathbf{q}\cdot\mathbf{p}}{p^2})=\arccos(\hat{\mathbf{r}}\cdot\hat{\mathbf{p}})$. It is convenient, therefore, to perform a spherical wave expansion
  \begin{equation}
  \label{f_expand}
   f(\mathbf{q},\mathbf{p})=f(p,\theta)=\sum_{\ell=0}^{\infty}(2\ell+1)f_{\ell}(p)P_{\ell}(\cos\theta)
 \end{equation} 
with $P_{\ell}$ the Legendre polynomials. In order to preserve the normalization of the wavefunction, one has to impose the following condition for the coefficients $f_\ell(p)$ (see Appendix \ref{app.B})
 \begin{equation}
  \label{f_cond}
|1+2i p f_\ell (p)|=1\;\Rightarrow\; 1+2ip f_\ell(p)\equiv e^{2i\delta_\ell(p)}
 \end{equation} 
which defines the phase shift $\delta_\ell(p)$ and where we rescaled, here and in what follows, $p\rightarrow \hbar p$ for simplicity, to get rid of $\hbar$. 
This relation imposes that when the relative momentum vanishes, also the phase shift should be zero, $\delta_{\ell}(0)=0$. The scattering amplitude, at fixed angular momentum, can be rewritten as
 \begin{equation}
   f_{\ell}(p)=\frac{1}{p \cot[\delta_{\ell}(p)] -ip}
   \label{amplitude_delta}
 \end{equation} 
In the radial Schr\"odinger equation (see Appendix \ref{app.C}),  for $\ell\neq0$, besides the potential $V(r)$ there is also the so-called centrifugal barrier, that depends on the angular momentum at which the scattering happens and on the relative distance. In ultracold atoms, at sufficiently low temperatures, the atoms may not have enough energy to overcome this centrifugal barrier, therefore the only relevant contribution to the scattering amplitude will come from s-wave scattering ($\ell=0$). 
To clarify this point one can estimate the angular momentum as the product of the range of the interaction $r^*$ times the momentum given by the inverse of the thermal de Broglie length
 $  \ell\approx \frac{r^*}{\Lambda_{th}}$. 
For ultracold gases the temperatures are typically $T\sim nK$ and, since the thermal wavelength $\Lambda_{th}\sim T^{-1/2}$, it is allowed to consider $\ell\approx 0$, namely, the scattering amplitude is dominated by the s-wave contribution
 \begin{equation}
   f(\mathbf{q},\mathbf{p})\approx f_{0}(p)=\frac{1}{p \cot[\delta_{0}(p)]-ip}\,
 \end{equation}
It is expected that scattering processes modified by the presence of $V(r)$ for $\ell\neq0$ are greatly suppressed at sufficiently low energy. Moreover, also the $\ell=0$ contribution is affected by the low energy hypothesis. As shown in Ref.~\citep{Taylor}, it is possible to define the \emph{scattering length} at fixed $\ell$ considering the low energy limit
 \begin{equation}
   f_{\ell}(p)\underset{p\rightarrow0}{\sim}-a_{\ell}\,p^{2\ell}
 \end{equation}
which comes from the low energy limit of the phase shift 
 \begin{equation}
 \label{delta_psmall}
   \delta_{\ell}(p)\underset{p\rightarrow0}{\sim}p^{2\ell+1}
 \end{equation}
 This can be derived by solving the Schr\"odinger equation as shown in Appendix \ref{app.C}.  
 It is remarkable that ultracold gases {can be cooled down to the point where only one partial wave in the two-body problem becomes  dominant}, the s-wave contribution. 
At low energies ($p\sim1/\Lambda$) it is possible to expand the phase $\delta_{0}(p)=-\as p +O(p^{2})$, so defining the scattering length as
 \begin{equation}
\label{def_as}
   \as\equiv -\lim_{p\rightarrow0}\frac{\delta_{0}(p)}{p}
 \end{equation}
Actually only $a_s=a_{\ell=0}$ has the dimension of a length, contrary to the other terms $a_\ell$ with $\ell>0$.  
 Considering the relation in Eq.~(\ref{amplitude_delta}) and the following expansion
 \begin{equation}
 \label{cot_delta}
   p \,\cot(\delta_{0}(p))\approx-\frac{1}{\as}+\frac{r_{\text{eff}}}{2}p^2
 \end{equation}
the scattering amplitude is then given, at low energy, by the expression
 \begin{equation}
f_{0}(p)\approx
-\frac{\as}{1-\as\frac{r_{\text{eff}}}{2}p^2+i\as p}
\label{scatt_amplitude1}
 \end{equation}
 which shows explicitly that for ultracold gases, under the hypothesis of short-range central potential, at low energy, 
 the two-body scattering depends only on two parameters: $\as$, the s-wave scattering length, and $r_{\text{eff}}$, an effective range proportional to $r^*$. We will not treat $r_{\text{eff}}$, the effective range of the potential, more deeply here, because, for what follows, only the first order expansion of the phase shift is needed.  
 Finally also the cross section admits a low energy limit. Writing $\sigma(p)=\sum_\ell \sigma_\ell(p)$, we have 
 $  \sigma_{\ell\neq0}(p)=\frac{8\pi}{p^2}(2\ell+1)\sin^{2}(\delta_{\ell}(p))
   \underset{p\rightarrow0}{\sim}
   8\pi(2\ell+1)p^{4\ell}$ and
 $  \sigma_{\ell=0}(p)
   \underset{p\rightarrow0}{\sim}
   8\pi\as^{2}$. In particular, from Eqs.~(\ref{cross_nospin}) and (\ref{cross_spin}), one gets 
    $\sigma(p)\underset{p\rightarrow0}{\sim}8\pi\as^{2}$ for spinless bosons, 
    $\sigma(p)\underset{p\rightarrow0}{\sim}p^{4}$ for spinless (polarized) fermions, 
    $\sigma_{S}(p)\underset{p\rightarrow0}{\sim}8\pi\as^{2}$ for fermions in a singlet state, 
    $\sigma_{T(p)}\underset{p\rightarrow0}{\sim}p^{4}$ for fermions in a triplet state. 
These results show that the interaction between polarized fermions is suppressed, while interaction in the singlet channel is dominant in the low energy limit, consistently with the Pauli exclusion principle.

\subsection{Two-body scattering matrix}
\label{Matrice di scattering a due corpi}

It is possible to mimic a realistic potential using the pseudo-potential with the same scattering properties a low energies, namely the same s-wave scattering length. 
Let us introduce the operator $\hat{T}_{2B}$, called $T$ scattering matrix,
 \begin{equation}
   \hat{V}|\psi_{\mathbf{p}}\rangle=\hat{T}_{2B}|\mathbf{p}\rangle
 \end{equation}
By this definition it is possible to rewrite the scattering amplitude as a function of $\hat{T}_{2B}$
 \begin{equation}
   f(\mathbf{q},\mathbf{p})=
-{2\pi^2m}
\langle\mathbf{q}|\hat{V}|\psi_{\mathbf{p}}\rangle=
-{2\pi^2m}
\langle\mathbf{q}|\hat{T}_{2B}|\mathbf{p}\rangle\,.
 \end{equation}
In the same way also the Lippmann-Schwinger equation, introduced earlier, can be written as
 \begin{equation}
   \hat{T}_{2B}(z)=\hat{V}+\hat{V}\frac{1}{z-\hat{H}_{0}+i\epsilon}\hat{T}_{2B}(z)
   \label{T2BLS}
 \end{equation}
whose solution, in terms of $\hat{T}_{2B}$, is the following
 \begin{equation}
\hat{T}_{2B}(z) = \hat{V}+\hat{V}\frac{1}{z-\hat{H}+i\epsilon}\hat{V}\,.
 \end{equation}
This solution can be expressed as an expansion in the potential $\hat{V}$
 \begin{equation}
   \hat{T}_{2B}(z)=\hat{V}\sum_{n=0}^{\infty}
   \left[\frac{1}{z-\hat{H}_{0}+i\epsilon}\hat{V}\right]^{n}\,.
 \end{equation}
Given a complete set of eigenstates of $\hat{H}$, inserting a completeness relation, one can write 
$\hat{T}_{2B}(z) = 
\hat{V}
+\sum_{\alpha}\hat{V}\frac{|\psi_{\alpha}\rangle\langle\psi_{\alpha}|}{z-\epsilon_{\alpha}}\hat{V}
+\frac{\Omega}{(2\pi)^{3}}\int d^{3}p\hat{V}\frac{|\psi_{\mathbf{p}}\rangle\langle\psi_{\mathbf{p}}|}{z-2\epsilon_{\mathbf{p}}}\hat{V}$, 
where $\epsilon_{\alpha}<0$ are the energies of the bound states, 
$|\psi_{\mathbf{p}}\rangle$ the scattering states and $\Omega$ is a volume. It is then clear that the two-body scattering matrix $T_{2B}(z)$ has simple poles 
associated to bound states and a branch cut on the real axis caused by the 
continuum of the scattering states. 
Indeed the transition amplitude given by Eq.~(\ref{scatt_amplitude1}) 
can be also rewritten in terms of the scattering energy $E=\frac{\hbar^{2}p^{2}}{m}$ so that 
 \begin{equation}
\label{TmatrixE}
   T_{2B}(E)\approx 
\frac{1}{2\pi^2m}\,
   \frac{\as}{1- i\as\sqrt{\frac{mE}{\hbar^{2}}} -\frac{m\as r_{\text{eff}}}{2\hbar^{2}}E}
 \end{equation}
At low energies ($E\rightarrow 0^{-}$) it is found that, for positive scattering length, $a_s>0$, this quantity has a simple pole at
 \begin{equation}
   E_{b}=-\frac{\hbar^{2}}{m\as^{2}}\,.
   \label{En_bound}
 \end{equation}
This pole indicates the presence of a bound state close to the scattering 
threshold ($E=0$) and gives an additional meaning to the scattering length $\as$.\\
As a final remark, by Fano-Feshbach resonance method one can control by a magnetic field the value of $a_s$. We refer to Refs. \cite{JinDeMarco,Feshbach_Fermi,feshbach40K,Chin_feshback_review} for a detailed description of this technique.

\subsection{Renormalization of the contact potential}
\label{rin_contact}

Let us consider the contact potential in the position representation
 \begin{equation}
   \hat{V}_{\text{eff}}(\mathbf{r})=-g\ \delta^{(3)}(\mathbf{r})
 \end{equation}
The action of this operator on the eigenstates of the momentum operator is
 \begin{equation}
\label{Vp_contact}
   \bra\mathbf{q}|\hat{V}_{\text{eff}}|\mathbf{p}\ket=
   \bra\mathbf{q}|\int d^{3}r\hat{V}_{\text{eff}}|\mathbf{r}\ket\bra\mathbf{r}|\mathbf{p}\ket=
   -g\bra\mathbf{q}|0\ket\bra0|\mathbf{p}\ket=-\frac{g}{(2\pi)^3}
 \end{equation}
We will want this potential to reproduce the scattering amplitude of a realistic potential. Exploiting the Lippmann-Schwinger equation for $T_{2B}$, written in Eq.~\eqref{T2BLS}, we have 
 \begin{equation}
\label{Lippmann}
   \langle\mathbf{q}|\hat{T}_{2B}|\mathbf{p}\rangle=
   \langle\mathbf{q}|\hat{V}_{\text{eff}}|\mathbf{p}\rangle
   +\int d^{3}k\langle\mathbf{q}|\hat{V}_{\text{eff}}\frac{1}{E-\hat{H}_{0}+i\epsilon}|\mathbf{k}\rangle\langle\mathbf{k}|\hat{T}_{2B}|\mathbf{p}\rangle
 \end{equation}
We now impose that, for the contact potential, the zero energy scattering amplitude should give the same result of a realistic potential. From Eq.~(\ref{TmatrixE}) we have
 \begin{equation}
   T_{2B}(E=0)=\frac{a_s}{2\pi^2m}
 \end{equation}
while from Eq.~(\ref{Lippmann}), using Eq.~(\ref{Vp_contact}), we get
 \begin{equation}
   \frac{1}{g}=
   -\frac{m}{4\pi\as}
   +\int \frac{d^{3}k}{(2\pi)^3}\frac{1}{2\epsilon_{\mathbf{k}}}
   \label{gfixing}
 \end{equation}
This relation may seem inconsistent at first sight due to divergence of the last term. However, as we will see in what follows, this result will turn to be very helpful since this divergence cancels out another divergence appearing in the so-called gap equation, for the order parameter, 
resulting as an elegant renormalization procedure for the contact potential.

\subsubsection{With spin-orbit coupling}
As shown in Refs.~\cite{Zhang1,Zhang2}, in the presence of a spin-orbit coupling, 
the equation useful to cure the ultraviolet divergences is formally the same as Eq.~(\ref{gfixing}), (see Eq.~(\ref{rcSO})), with the only difference that $a_s$ is replaced by the scattering length $a_r$ which contains also a 
dependence on the spin-orbit parameters (see Appendix~\ref{Accoppiamento Spin-Orbita} for more details).

\section{Fermi gas with attractive potential}
\label{sec.5}
In the previous introductory sections  
we showed that the extreme regimes of ultracold gases, namely the diluteness and the low temperatures, allow us to consider only the two-body physics and simplify the real interaction replacing it with a contact potential that is able to reproduce the same scattering properties. 
For these reasons this kind of systems can be described by a set of fermions in three dimensions with a BCS-like Hamiltonian
 \begin{equation}
  H=H_{0}+H_{I}=\int{d^{3}r}
   \left\{  
       \sum_{\sigma} \psidag_{\sigma}(\vec{r})\left(-\frac{\nabla^2}{2m} \right)\psi_{\sigma}(\vec{r})  
       -g\psidagu(\vec{r})\psidagd(\vec{r})\psid(\vec{r})\psiu(\vec{r})                           
   \right\}	\label{H}
 \end{equation} 
The interaction coupling is positive ($g>0$) for an attractive potential. 
In the following section we will review the study of this model by a path integral approach, focusing on the evaluation of the critical temperature as a function of the coupling $g$, along the {crossover} from weak coupling (BCS regime) to strong coupling (BEC regime). 
The Hamiltonian in Eq.~\eqref{H} is associated to the grand canonical partition function in the path integral formalism which is
 \begin{equation}
  \mathcal{Z} = \textrm{Tr}\left(  e^{  -\beta(H-\mu N)  }   \right) 
              = \int D\psidag  D\psi\,e^{-S[\psidag,\psi]}
 \end{equation}
where the action is give by
 \begin{equation}
  S[\psidag,\psi] = \int{d^{3}r}\int_{0}^{\beta}{d\tau} 
   \left\{  
       \sum_{\sigma} \psidag_{\sigma}(\vec{r})\left(\partial_{\tau}-\frac{\nabla^2}{2m}-\mu \right)\psi_{\sigma}(\vec{r}) 
       -g\psidagu(\vec{r})\psidagd(\vec{r})\psid(\vec{r})\psiu(\vec{r}) 
   \right\}
 \end{equation}
It is possible to take advantage of the properties of Gaussian integrals (see for instance Ref.~\citep{Ultracold}) to handle the interacting term  
introducing the complex auxiliary field $\Delta(\mathbf{r},\tau)$
 \begin{equation}
  e^{g\int{d^{3}r}\int_{0}^{\beta}{d\tau}\psidagu\psidagd\psid\psiu}
           \propto \int D\Delta^{*} D\Delta \, e^{\int{d^{3}r}\int_{0}^{\beta}{d\tau} 
 \left[-\frac{|\Delta|^{2}}{g} + \Delta^{*}\psid\psiu +
 \Delta\psidagu\psidagd \right]}
	\label{Hub_Stra}
 \end{equation}
The grand canonical partition function can be rewritten, therefore, as it follows
 \begin{equation}
  \mathcal{Z} = \int D\psidag D\psi \int D\Delta^{*} D\Delta\, e^{-S[\psidag,\psi,\Delta]}
 \end{equation}
with the use of a new action, dependent now also on an auxiliary field
 \begin{equation}
  S[\psidag,\psi,\Delta] = \int d^{3}r\int_{0}^{\beta}d\tau
	\left\{ 
	\sum_{\sigma}\psidag_{\sigma}\left(\partial_{\tau}-\frac{\nabla^2}{2m}-
\mu \right)\psi_{\sigma}+
\frac{|\Delta|^{2}}{g} -\Delta^{*}\psid\psiu -\Delta\psidagu\psidagd\right\}
 \end{equation}
The physical meaning of $\Delta$ will be clear from the saddle point equation. 
It is more convenient to introduce the Nambu spinors
 \begin{equation} 
  \Psi=\begin{pmatrix}
                 \psiu\\
                 \psidagd
             \end{pmatrix} \hspace{1cm}
  \Psidag=\left(\psidagu,\psid\right)
             		  \end{equation} 
so that the action of the system can be written as 
 \begin{equation}
\label{S55}
  S[\Psidag,\Psi,\Delta] = \int{d^{3}r}\int_{0}^{\beta}{d\tau}
	\left[ \frac{|\Delta|^{2}}{g} +\Psidag\mathcal{G}^{-1}\Psi \right]
	+\beta\sum_{\mathbf{k}}\xik
 \end{equation} 
with $\mathcal{G}^{-1}$, the inverse of the interacting Green function, 
defined as
 \begin{equation} 
  \mathcal{G}^{-1}
  =         
                                 \begin{pmatrix}
              -\partial_{\tau}+\frac{\nabla^2}{2m}+\mu  &  \Delta \\
               \Delta^{*}                               &  -\partial_{\tau}-\frac{\nabla^2}{2m}-\mu 
                                 \end{pmatrix}		\label{propagatore}
 \end{equation}
Actually in order to put $S[\psidag,\psi,\Delta]$ in a quadratic form in the spinor representation we exploited the anticommutation relations of the Grassmann fields, getting also the last term in Eq.~(\ref{S55}). 
Defining $z\!=\!e^{-\beta\sum_{\mathbf{k}}\xik}$, this term can be put outside 
the functional integral. 
The grand canonical partition function is the following 
 \begin{equation}
  \mathcal{Z} = z\int D\Delta^{*} D\Delta\, 
               e^{ -\int d^{3}r\int_{0}^{\beta}{d\tau}\frac{\Delta^{2}}{g}}
                \int D\bar{\Psi} D\Psi\,
               e^{ -\int d^{3}r \int_{0}^{\beta}{d\tau}\Psidag\mathcal{G}^{-1}\Psi }
 \end{equation}
Performing the standard Gaussian integral over the fermionic 
degrees of freedom one gets 
 \begin{equation}
  \mathcal{Z} 
            =  z\int D\Delta^{*}D\Delta\, 
              e^{ -\int{d^{3}r}\int_{0}^{\beta}{d\tau}\frac{|\Delta|^{2}}{g} + \textrm{Tr}[\ln(\mathcal{G}^{-1})] } \equiv \int D\Delta^* D\Delta \,e^{-S_{e}[\Delta]}
 \label{Z} 
 \end{equation}
which defines the effective action depending only on the auxiliary field
 \begin{equation}
	  \label{S_eff}  
    S_{e}[\Delta] =\int{d^{3}r}\int_{0}^{\beta}{d\tau}
                    \frac{\left|\Delta(\vec{r},\tau)\right|^{2}}{g} - 
\textrm{Tr}[\ln(\mathcal{G}^{-1})]
		+\beta\sum_{\mathbf{k}}\xik\,.
 \end{equation}
 \subsection{Gap equation}
Starting from this action it is possible to write down the saddle point equation and look for an homogeneous solution. With $\Delta(\vec{r},\tau) = \Delta_0$ the effective action is given by
 \begin{equation}  
    S_{e}[\Delta_0]=
	\frac{\beta{\cal{V}}}{g}|\Delta_0|^{2}
	-\textrm{Tr}[\ln(\mathcal{G}_0^{-1})]
	+\beta\sum_{\mathbf{k}}\xik
 \end{equation}
where ${\cal V}$ is the volume. 
In the momentum space and in the Matsubara frequencies the matrix $\mathcal{G}^{-1}$ defined as in Eq.~\eqref{propagatore} is 
 \begin{equation}
   \mathcal{G}_0^{-1}(\mathbf{k},i\omega_{n})
   =                               \begin{pmatrix}
              i\omega_{n}-\frac{\mathbf{k}^{2}}{2m}+\mu  &  \Delta_0 \\
               \Delta^{*}_0                                &  i\omega_{n}+\frac{\mathbf{k}^{2}}{2m}-\mu
                                  \end{pmatrix}=
                                  \begin{pmatrix}
              i\omega_{n}-\xik  &  \Delta_0 \\
               \Delta^{*}_0                   &  i\omega_{n}+\xik
                                  \end{pmatrix}
 \end{equation}
which has poles in real frequencies at $\pm \Ek$ where 
 \begin{equation}
  \Ek=\sqrt{\xik^{2}+|\Delta_0|^{2}}
 \end{equation}
The minimum of the spectrum is at $\epsk\!=\!\mu$ with the minimum energy $|\Delta_0|$, characterized by the existence of a \emph{gap} compared to the minimum of the spectrum of a free particle, which is zero.
This implies that there is a minimum energetic cost for the creation of an elementary excitation. 
The physical interpretation is that 
the minimum energy requested for breaking a pair is $2|\Delta_0|$.
The value of this minimum energy can be obtained imposing $\frac{\delta S_e}{\delta \Delta_0}=0$, getting 
 \begin{equation}
\frac{1}{g}=\frac{1}{\beta{\cal V}}\sum_{\mathbf{k},i\omega_{n}}\frac{1}{\omega_{n}^2+\Ek^{2}}
 \end{equation}
Performing the standard sum over the Matsubara frequencies 
and going to the continuum limit $\displaystyle{\frac{1}{\cal V}\sum_{\mathbf{k}}\rightarrow\frac{1}{(2\pi)^3}\int d^{3}k}$, 
one gets the so called gap equation 
 \begin{equation}
\frac{1}{g}=\frac{1}{(2\pi)^3}\int d^{3}k \, \frac{\tanh(\beta \Ek/2)}{2\Ek}
\label{gap_eq_def} 
 \end{equation}
The contact potential is a great simplification in the model, however it gives rise to ultraviolet divergences. Actually Eq.~(\ref{gap_eq_def}) diverges linearly with the ultraviolet cut-off.
\subsubsection{BCS superconductors}
The Hamiltonian and the obtained gap equation are the same as those of the BCS theory for conventional superconductors.
However in the BCS theory the ultraviolet behavior is regularized by the existence of a natural cut-off at the Debye frequency $\omega_{D}$ of the underlying lattice. 
Typically, in classical superconductors $\hbar\omega_{D}\ll\Ef=\mu$ and a finite gap can exists only for particles near the Fermi energy. Exploiting the approximation of constant density of states (DOS), $\frac{1}{(2\pi)^3}\int d^{3}k\approx\nu_0\int d\xi$, where $\nu_0$ is the DOS at the Fermi energy, the gap equation becomes
 \begin{equation}
 \label{geBCS}
\frac{1}{g}
\approx
\nu_0\int\limits_{0}^{\hbar\omega_{D}}d\xi \,
      \frac{  \tanh\left(   \frac{(\xi^{2}+|\Delta|^{2})^{1/2}}{2T_{c}}  \right)   }
           {(\xi^{2}+|\Delta|^{2})^{1/2}}\\
\end{equation}
Solving the gap equation Eq.~(\ref{geBCS}) one gets, at $T=0$, 
		\begin{equation}
			|\Delta_{0}|=2\hbar\omega_{D}e^{-\frac{1}{\nu_0 g}}\,.
		\end{equation}
Supposing that at a critical temperature, $T=T_c$, the thermal fluctuations spoil the superconductivity, $\Delta_0(T_c)=0$, one gets
		\begin{equation}
			T_{c}=2\frac{e^\gamma}{\pi}\hbar\omega_{D}\,e^{-\frac{1}{\nu_0 g}}\,,
		\end{equation}
with $\gamma\approx 0.5772$, the Euler-Mascheroni number. Finally, for $T<T_c$ the gap is 
		\begin{equation}
			|\Delta_0(T)|\propto \sqrt{T_c(T_{c}-T)}\,.
		\end{equation}
\subsection{Number equation}
The second equation that we should considered describes the mean number of particles, usually called \emph{number equation}.
Calculations with the full effective action \eqref{S_eff} are difficult, so one has to resort to an approximated scheme expanding the action close to the \emph{classical} solution of the gap equation. Considering $\Delta(x)\!=\!{\Delta}_0+\delta\Delta(x)$, the action can be expanded in the fluctuations of the auxiliary field at the desired order. The zeroth order is equivalent to the mean field approximation. We will then include the fluctuations at the Gaussian level so that 
 \begin{multline}
	S_{e}[\Delta]=S_{e}[{\Delta}_0]+
                        \underbrace{\int dx\left.\frac{\delta S_{e}[\Delta]}{\delta\Delta(x)}\right|_{{\Delta}_0}}_\text{= 0}\delta\Delta(x)+\\
                        +\frac{1}{2}\int dx_{1}dx_{2}\left.\frac{\delta^{2} S_{e}[\Delta]}{\delta\Delta(x_{1})\delta\Delta(x_{2})}\right|_{{\Delta}_0}\delta\Delta(x_{1})\delta\Delta(x_{2})+
                        o(\Delta^{3})
                        \label{expansion}
 \end{multline}

\subsubsection{Mean field}
At the mean field level the action is proportional to the grand canonical potential 
 \begin{equation}
	\Omega[{\Delta}_0,]=\beta S_{e}[\Delta_0]
 \end{equation}
from which is possible to derive the equation for the mean number of particles
 \begin{equation}
	N=-\frac{\partial \Omega}{\partial \mu}=
\sum_{\mathbf{k}}\left[1
+\frac{1}{\beta}\sum_{i\omega_{n}}
\textrm{Tr}\left(\mathcal{G}_0\,\frac{\partial\mathcal{G}_0^{-1}}{\partial \mu}\right)\right]
=\sum_{\mathbf{k}}\left[1
-\frac{1}{\beta}\sum_{i\omega_{n}}\left(\frac{2\xik}{\omega_{n}^{2}+\Ek^{2}}\right)\right]
 \end{equation}
Summing over the Matsubara frequencies one gets 
 \begin{equation}
	N=\sum_{\mathbf{k}}
	\left[
		1-\frac{\xik}{\Ek}\tanh\left(\frac{\Ek}{2T}\right)
	\right]
 \end{equation}
Let us consider the cases at $T=0$ and $T=T_c$. 
\paragraph*{At T=0.}
In order to study the behaviors of the gap $\Delta_0$ and of the chemical potential $\mu$ as functions of the coupling, at $T=0$, at the mean field level, the equations to solve are
 \begin{eqnarray}
		\frac{1}{g}&=&\frac{1}{\cal V}\sum\limits_{\mathbf{k}}
				\frac{1}{2\Ek} \\
		n&=&\frac{1}{\cal V}\sum\limits_{\mathbf{k}}
				\left( 1-\frac{\xik}{\Ek} \right)
 \end{eqnarray}
 where $n=N/{\cal V}$, the particle density. 
\paragraph*{At ${T=T_{c}}$.}
 On the other hand if we want to calculate the critical temperature $T_c$, putting by definition $\Delta_0(T_c)=0$, at the mean field level the equation to solve in terms of $T_c$ are 
 \begin{eqnarray}
		\frac{1}{g}&=&\frac{1}{\cal V}\sum\limits_{\mathbf{k}}
				\frac{\tanh(\frac{\xik}{2T_{c}})}{2\xik} \\
		n&=&\frac{1}{\cal V}\sum\limits_{\mathbf{k}}
				\left[ 1-\tanh\left(\frac{\xik}{2T_{c}}\right) \right]\,.
 \end{eqnarray}
\subsubsection{Inclusion of the Gaussian fluctuations}
\label{Inclusione delle fluttuazioni Gaussiane free}
Let us now include quantum fluctuations beyond the mean field approach at the Gaussian level. The fermionic propagator appearing in Eq.~(\ref{S_eff}), in momentum space, can be written as
 \begin{equation}
	\mathcal{G}^{-1}
=		\begin{pmatrix}
				i\omega_{n}-\xik & \Delta\\
				\Delta^{*} & i\omega_{n}+\xik\\
		\end{pmatrix}
		=\begin{pmatrix}
				i\omega_{n}-\xik & 0\\
				0 & i\omega_{n}+\xik\\
		\end{pmatrix}+
		\begin{pmatrix}
				0 & \Delta\\
				\Delta^{*} & 0\\
		\end{pmatrix}\equiv
	\hat{G}^{-1}+\hat\Delta
 \end{equation}
Expanding the logarithm for small $\Delta$ one gets
 \begin{equation}
	\ln(\mathcal{G}^{-1})=
	\ln(\hat{G}^{-1}(\mathbb{1}+\hat{G}\hat\Delta))
	=\ln(\hat{G}^{-1})+\hat{G}\hat\Delta-\frac{1}{2}\hat{G}\hat\Delta\hat{G}\hat\Delta+O(\Delta^{3})
 \end{equation}
The linear term in $\hat\Delta$ has a vanishing trace, therefore, up to second order the action is given by
 \begin{equation}   
    S_{e}[\Delta] =S_{e}^{(0)}+\int \!d^{3}r\!\int_{0}^{\beta}\!d\tau
                    \frac{|\Delta(\vec{r},\tau)|^{2}}{g} +\frac{1}{2}\textrm{Tr}\left(\hat{G}\hat\Delta\hat{G}\hat\Delta\right) + o(\Delta^{3})
 \end{equation}
Making explicit the quadratic term in $\Delta$ we have
 \begin{equation*}
	\hat{G}\hat\Delta\hat{G}\hat\Delta=
	\begin{pmatrix}
		0 & {G}_{11}\Delta\\
		{G}_{22}\Delta^{*} & 0
	\end{pmatrix}
	\begin{pmatrix}
		0 & {G}_{11}\Delta\\
		{G}_{22}\Delta^{*} & 0
	\end{pmatrix}=
	\begin{pmatrix}
		{G}_{11}\Delta{G}_{22}\Delta^{*} & 0\\
		0 & {G}_{22}\Delta^{*}{G}_{11}\Delta
	\end{pmatrix}
 \end{equation*}
All the calculation can be done in momentum space 
 \begin{equation}
	\frac{1}{2}\textrm{Tr}(\hat{G}\hat\Delta\hat{G}\hat\Delta)=
	\textrm{Tr}({G}_{22}\Delta^{*}{G}_{11}\Delta)
	=\frac{1}{\beta{\cal V}}\sum_{\qs}\Delta^{*}(\qs)\left[
	\sum_{\ks}{G}_{11}(\ks){G}_{22}(\ks-\qs)
					\right]\Delta(\qs)
 \end{equation}
where $\ks=(\mathbf{k},i\omega_n)$ and $\qs=(\mathbf{q},i\nu_m)$ are the four-momenta and $\hat{G}(\ks)$ is given by
 \begin{equation}
	\hat{G}(\ks)=\hat{G}(\mathbf{k},i\omega_n)=
			\begin{pmatrix}
			G_{11} & 0\\
			0 & G_{22}
		\end{pmatrix}	=
		\begin{pmatrix}
			\frac{1}{i\omega_{n}-\xik} & 0\\
			0 & \frac{1}{i\omega_{n}+\xik}
		\end{pmatrix}	
 \end{equation}
 whose components fulfill the property $G_{11}(\ks)=-G_{22}(-\ks)$  
that can be exploited to rewrite the quadratic term. Calling for simplicity $G_{11}(\ks)=G(\ks)$, we have
 \begin{equation}
	\frac{1}{2}\textrm{Tr}(\hat{G}\hat\Delta\hat{G}\hat\Delta)=
	\sum_{\qs}\Delta^{*}(\qs)\left[
		-\frac{1}{\beta{\cal V}}\sum_{k}{G}(\ks){G}(\qs-\ks)
			      \right]\Delta(\qs)\equiv 
	\sum_{\qs}\Delta^{*}(\qs)\chi(\qs)\Delta(\qs)	
 \end{equation}
 where we defined the function $\chi(\qs)$ which, after performing the summation over the Matsubara frequencies $i\omega_{n}$, reads
 \begin{equation}
	\chi(\qs)=
	\frac{1}{\beta{\cal V}}
	\sum_{\mathbf{k},i\omega_{n}}
	\frac{1}{i\omega_{n}-\xik}\cdot
	\frac{1}{i\omega_{n}-i\nu_m+\xi_{\mathbf{q-k}}}=
	\frac{1}{\cal V}\sum_{\mathbf{k}}
			\frac{1-\fermi(\xik)-\fermi(\xi_{\mathbf{q-k}})}{i\nu_m-\xik-\xi_{\mathbf{q-k}}}
 \end{equation}
 where $\fermi$ is the Fermi distribution. 
The action at second order in $\Delta$, corresponding to the inclusion of the Gaussian fluctuation around mean field solution, is 
 \begin{equation}
	S_{e}[\Delta]=S_{e}^{(0)}+\sum_{\q}\Delta^{*}(\q)\Gamma^{-1}(\q)\Delta(\q)	+ o(\Delta^{3})
	\label{actionS}
 \end{equation}
where we introduced the function
 \begin{equation}
 \label{Gamma_q}
	\Gamma^{-1}(\q)=\frac{1}{g}+\frac{1}{\cal V}
		\sum_{\mathbf{k}}
			\frac{1-\fermi(\xik)-\fermi(\xi_{\mathbf{k-q}})}{i\nu_m-\xik-\xi_{\mathbf{k-q}}}
 \end{equation}
The partition function at the Gaussian level is, therefore,
 \begin{equation}
	\mathcal{Z}_{G}=\mathcal{Z}_{0}\int D\Delta^{*} D\Delta \,e^{-\sum_{\q}\Delta^{*}(\q)\Gamma^{-1}(\q)\Delta(\q)}=\mathcal{Z}_{0} \det(\Gamma)
	\label{partition_func_free}
 \end{equation}
with the definition $\mathcal{Z}_{0}=e^{-S_{e}^{(0)}}$, which can be written as 
 \begin{equation}
	\mathcal{Z}_{0} \det(\Gamma)=
	\mathcal{Z}_{0} \ e^{-\ln[\det(\Gamma^{-1})]}=
	\mathcal{Z}_{0} \ e^{-\textrm{Tr}[\ln(\Gamma^{-1})]}=
	e^{-S_{e}^{(0)}-\textrm{Tr}[\ln(\Gamma^{-1})]}
 \end{equation}
the grand canonical potential is shown to be
 \begin{equation}
	\Omega_{G}=-\frac{1}{\beta}\ln(\mathcal{Z}_{G})=\frac{1}{\beta}S_{e}^{(0)}+\frac{1}{\beta}\textrm{Tr}[\ln(\Gamma^{-1})]
 \end{equation}
The number equation, at the Gaussian level, is given by 
 \begin{equation}
 \label{ngauss}
	n=-\frac{1}{\cal V}\frac{\partial \Omega_{G}}{\partial\mu}=n^{(0)}+n^{(2)}
 \end{equation}
 where 
 \begin{equation}
	n^{(2)}=-\frac{1}{\beta{\cal V}}\frac{\partial}{\partial\mu}\textrm{Tr}[\ln(\Gamma^{-1})]=
	-\frac{1}{\beta{\cal V}}\frac{\partial}{\partial\mu}\sum_{\mathbf{q},i\nu_{m}}\ln(\Gamma^{-1}(\mathbf{q},i\nu_m))
 \end{equation}
The first term in Eq.~(\ref{ngauss}) is simply the density of particles at the mean field level while the second one is due to the quantum corrections.
In general if $\ln\left(\Gamma^{-1}(\mathbf{q},z)\right)$, has poles in the complex plane at $z\!=\!z_{j}$ ($j=0,1,2\dots$) and a \emph{branch cut} on the real axis, the sum over Matsubara frequencies can be rewritten as
 \begin{multline}
	\frac{1}{\beta}\sum_{i\nu_{n}}\ln(\Gamma^{-1}(\mathbf{q},i\nu_{n}))=
	\sum_{j}\bose(z_{j})\textrm{Res}\left[\ln(\Gamma^{-1}(\mathbf{q},z_{j}))\right]-\\
	-\frac{1}{2\pi i}\int_{-\infty}^{+\infty}\hspace{-0.2cm}d\omega \ \bose(\omega)\left[\ln(\Gamma^{-1}(\mathbf{q},\omega+i\epsilon))-\ln(\Gamma^{-1}(\mathbf{q},\omega-i\epsilon))\right]\,.
 \end{multline}
In our case the integrand has poles only on the cut, so that this formula reduces to
 \begin{equation}
	\frac{1}{\beta}\sum_{i\nu_{n}}\ln(\Gamma^{-1}(\mathbf{q},i\nu_n))=
	-\frac{1}{2\pi i}\int_{-\infty}^{+\infty}\hspace{-0.2cm}d\omega \ \bose(\omega)\left[\ln(\Gamma^{-1}(\mathbf{q},\omega+i\epsilon))-\ln(\Gamma^{-1}(\mathbf{q},\omega-i\epsilon))\right]
 \end{equation}
 where $\bose$ is the Bose distribution. 
Since $\Gamma(\mathbf{q},\omega)$ has to be real on the real axis, it can be written as a modulus times a phase. Defining a phase shift $\delta(\q)$ as in Ref.~\cite{NSR}, we can write
 \begin{equation}
	\Gamma(\mathbf{q},\omega\pm i\epsilon)^{-1}=\left|\Gamma(\mathbf{q},\omega\pm i\epsilon)^{-1}\right|e^{\pm i\delta(\mathbf{q},\omega+i\epsilon)}
 \end{equation}
therefore
 \begin{equation}
	\frac{1}{\beta}\sum_{i\nu_{n}}\ln(\Gamma^{-1}(\mathbf{q},i\nu_{n}))=\\
	=-\frac{1}{\pi}\int_{-\infty}^{+\infty}d\omega\ \bose(\omega)\delta(\mathbf{q},\omega)
 \end{equation}
At the Gaussian level, therefore, the density of particles can written in the following form 
 \begin{equation}
 \label{n_gauss_correction}
	n=n^{(0)}+\frac{1}{\pi{\cal V}}\sum_{\mathbf{q}}\int_{-\infty}^{+\infty}d\omega\ \bose(\omega)\frac{\partial\delta(\mathbf{q},\omega)}{\partial\mu}\,.
 \end{equation}
The number and the gap equations allow us to find the critical temperature and the chemical potential as functions of the interaction strength. 
In the following section we will solve those equations, first at the mean field level and then with Gaussian fluctuations. 
However, before we proceed with the calculation, due to the presence of a contact potential, we notice that the right-hand-side of Eq.~(\ref{gap_eq_def}) has an ultraviolet divergence that has to be cured.

\section{BCS-BEC crossover}
In this section we will derive the critical temperature $T_{c}$ and the chemical potential at $T_{c}$ as functions of the interaction coupling, discussing the limits of strong and weak attraction, namely the BEC and the BCS limits, respectively, solving the \textit{gap equation} and the \textit{number equation} introduced previously. 
In Sec.~\ref{rin_contact} we showed that an effective contact potential, $V(\mathbf{r})=-g\delta(\mathbf{r})$, can reproduce the same scattering properties of a realistic potential, in the low energy limit, if $g$ and the scattering length $\as$ satisfy the relation in Eq.~(\ref{gfixing}), reported here for convenience
 \begin{equation}
	\frac{m}{4\pi \as}=-\frac{1}{g}+\frac{1}{{\cal V}}
	\sum_{\mathbf{k}}\frac{1}{2\epsilon_{\mathbf{k}}}	   
	 \label{cut-off}
 \end{equation}
Let us substitute the coupling $g$ from Eq.~\eqref{cut-off} in the gap equation Eq.~\eqref{gap_eq_def}, linking in this way the toy model discussed until now with a realistic ultracold gas, through the experimentally accessible scattering length $\as$. In this way the gap equation reads
 \begin{equation}
	-\frac{m}{4\pi \as}=\frac{1}{{\cal V}}\sum_{\mathbf{k}}
	\left(
		\frac{\tanh(\xik/2T_{c})}{2\xik} - \frac{1}{2\epsilon_{\mathbf{k}}}
	\right)			
	\label{gap_eq1}
 \end{equation}
The divergence in the gap equation is exactly canceled out by the divergence in Eq.~\eqref{cut-off}. 
The limits of strong ($g\rightarrow\infty$) and weak coupling ($g\rightarrow0$) have to be reviewed in terms of $\as$ which is now the tuning parameter along the crossover. For this purpose it is useful to introduce an ultraviolet cut-off $\Lambda$ so that Eq.~\eqref{cut-off} becomes
 \begin{equation}
	\frac{m}{4\pi \as}=-\frac{1}{g}+\frac{1}{{\cal V}}
	\sum_{|\mathbf{k}|<\Lambda}\frac{1}{2\epsilon_{\mathbf{k}}}
 \end{equation}
In the limit $g\rightarrow 0$, namely in the BCS limit, the cut-off $\Lambda$ does not play any role and eventually can be sent to infinity, therefore the scattering length is negative and $\frac{1}{\as}\rightarrow-\infty$.\\
In the limit $g\rightarrow\infty$, namely in the BEC limit, we get, instead, 
 \begin{equation*}
	\frac{m}{4\pi \as}=
	\frac{1}{{\cal V}}\sum_{|k|<\Lambda}\frac{1}{2\epsilon_{\mathbf{k}}}=
	\frac{1}{2\pi^{2}}\int_{0}^{\Lambda}dk\frac{k^{2}}{2\epsilon_{\mathbf{k}}}=
	\frac{m}{2\pi^{2}}\int_{0}^{\Lambda}dk=
	\frac{m}{2\pi^{2}}\Lambda
 \end{equation*}
Sending the cut-off to infinity the limit of strong interaction can be captured for $\frac{1}{\as}\rightarrow+\infty$.\\
Since the scattering length comes from the low energy limit of the interaction, it is an intrinsic parameter and not an adjustable property of the potential. However, as already mentioned, by the mechanism of the Feshbach resonance, the scattering length can be modified on a wide range of values, therefore it is the tunable parameter along the crossover.

\subsection{Mean field theory}
\noindent
At the mean field level one has to simultaneously solve the gap and the number equations found previously,
\begin{eqnarray}
\label{gap_eq}
-\frac{m}{4\pi \as}&=&\frac{1}{{\cal V}}\sum_{\mathbf{k}}\left(  \frac{\tanh(\xik/2T_{c})}{2\xik} - \frac{1}{2\epsilon_{\mathbf{k}}} \right)\\	
n&=&\frac{1}{{\cal V}}\sum_{\mathbf{k}}\left[ 1-\tanh\left(  \frac{\xik}{2T_{c}}  \right)  \right] 	
\label{numero}
\end{eqnarray}
We will discuss in what follows the weak (BCS) and the strong (BEC) interacting limits.
\subsubsection{Weak coupling}
\noindent
Let us consider the number equation Eq.~\eqref{numero}, which, in the continuum limit, reminding that $\epsilon_{\mathbf{k}}=\frac{k^{2}}{2m}$ and introducing the density of state, becomes
\begin{equation}
n=\frac{(2m)^{\frac{3}{2}}}{2\pi^{2}}\int_{0}^{\infty}\frac{\sqrt{\epsilon}}
						       {e^{\frac{\epsilon-\mu}{T_{c}}}+1}d\epsilon
\end{equation}
For $T_c$ very small one can resort to the well-known Sommerfield expansion, getting for the density of particle at low temperature
\begin{equation}
n=(2m)^{\frac{3}{2}}\left[
			\frac{1}{3\pi^{2}}\mu^{\frac{3}{2}}+
			\frac{1}{24}\frac{T_{c}^{2}}{\mu^{1/2}}+
			o(T_{c}^{4})
			\right]
\end{equation}
At fixed density $n=\frac{(2m\Ef)^{3/2}}{3\pi^{2}}$, which is the typical experimental situation, this equation of state implies 
$\Ef^{3/2}=\mu^{\frac{3}{2}}+\frac{\pi^{2}}{8}\frac{T_{c}^{2}}{\mu^{1/2}}+o(T_{c}^{4})$, and solving for $\mu$ we find an explicit relationship between the chemical potential and the temperature
\begin{equation}
\mu=\Ef-\frac{\pi^{2}}{12}\frac{T_{c}^{2}}{\Ef}+o(T_{c}^{4})
\end{equation}
The chemical potential is, therefore, almost constant, $\mu\simeq \Ef$. Thermal corrections occur only at second order. Exploiting this result, we can solve the gap equation Eq.~\eqref{gap_eq}, as done in Ref.~\citep{Ultracold}. 
With constant chemical potential, Eq.~\eqref{gap_eq} 
fixes the behavior of the critical temperature $T_c$ as a function of the scattering length. In the continuum limit, after introducing the density of state and after a change of variables, 
$x=\epsilon/\mu$, Eq.~\eqref{gap_eq} becomes
\begin{equation}
	-\frac{\pi}{\kf\as}=
\int_{0}^{\infty} dx\sqrt{x}\left(  \frac{\tanh(\mu(x-1)/2T_c)}{x-1} - \frac{1}{x} \right)
 \end{equation}
At $\mu=\Ef$ and supposing $T_c\ll \Ef$ one gets $\frac{\pi}{\kf\as}=-2\ln(\frac{8e^\gamma\Ef}{e^2\pi T_{c}})$, 
with $\gamma$ the Eulero-Mascheroni number. We remember that $\as$ has a small negative value in the weak coupling limit. Solving in terms of $T_c$ we obtain
\begin{equation}
T_{c}=\frac{8e^{\gamma}}{\pi e^2}\,\Ef \,e^{-\frac{\pi}{2\kf|\as|}}
\label{Tc_BCSlimit}
\end{equation}
a critical temperature exponentially small in the weak coupling limit, in agreement with the BCS theory and consistently with the hypothesis $T_c\ll \Ef$. 
\subsubsection{Strong coupling}
In the strong coupling limit, we will solve first the gap equation Eq.~\eqref{gap_eq} to obtain the chemical potential 
$\mu$ which is expected to be very large and negative, as in the case of free fermions at high temperatures, and that $|\mu(T_{c})|\gg T_{c}$. 
Taking advantage of this ansatz, whose validity will be proved a posteriori, we have
\begin{equation}
		-\frac{m}{4\pi \as}=\frac{m^{3/2}T_{c}^{1/2}}{2\pi^{2}}\int_{0}^{\infty} dx\sqrt{x}\left(  \frac{\tanh(x-\tmu)}{x-\tmu} - \frac{1}{x} \right)
\end{equation} 
where we defined $\tmu=\frac{\mu(T_{c})}{2T_{c}}$.
For $|\tmu|\gg1$, we can approximate $\frac{\tanh(x-\tmu)}{x-\tmu}\approx\frac{1}{x-\tmu}$, so that, rescaling $x\rightarrow -\tmu x$, we get
\begin{equation}
-\frac{m}{4\pi \as}\approx
\frac{m^{3/2}T_c^{1/2}}{2\pi^{2}}\sqrt{-\tmu}\int_{0}^{\infty}\!dx\frac{\sqrt{x}}{x(x+1)}=\frac{m^{3/2}}{2\pi}\sqrt{-\mu}
\end{equation} 
The relation between the chemical potential and the scattering length, in the strong coupling limit, is, therefore, given by
 \begin{equation}
	\mu\approx  -\frac{1}{2m\as^{2}}= \frac{E_b}{2}
	\label{chem_pot_strong}	
 \end{equation}
 where we used Eq.~\eqref{En_bound} (here $\hbar=1$), reminding, from Sec.~\ref{Matrice di scattering a due corpi}, that for positive scattering length the two-body problem supports the existence of a bound state with energy $E_{b}$.
We will use Eq.~\eqref{chem_pot_strong} in the number equation Eq.~\eqref{numero} to derive the critical temperature. After introducing the density of state and after a rescaling, in the limit $\frac{1}{a_s}\rightarrow +\infty$, we have
  \begin{eqnarray}
	\nonumber
	n=\frac{(2m)^{\frac{3}{2}}}{2\pi^{2}}\int_0^\infty\frac{\sqrt{\epsilon}}{e^{\frac{\epsilon-\mu}{T_{c}}}+1}d\epsilon
						 \approx
	\frac{(2mT_{c})^{\frac{3}{2}}}{2\pi^{2}}e^{\frac{\mu}{T_c}}\int_{0}^{\infty}\sqrt{x}e^{-x}dx 
	=\frac{(mT_{c})^{\frac{3}{2}}}{\sqrt{2}\pi^{\frac{3}{2}}}\,e^{-\frac{|E_{b}|}{2T_{c}}}
 \end{eqnarray}
At fixed density of particles along the crossover, $n=\frac{(2m\Ef)^{3/2}}{3\pi^{2}}$, we have $\left(\frac{\Ef}{|E_{b}|}\right)^{\frac{3}{2}}\propto\left(\frac{T_{c}}{|E_{b}|}\right)^{\frac{3}{2}}e^{-\frac{|E_{b}|}{2T_{c}}}$, therefore in the strong coupling limit the leading term for the critical temperature at the mean field level is given by
 \begin{equation}
	{T_{c}}
	\approx \frac{|E_{b}|
	}{3\ln \left( |E_{b}|/\Ef\right)   }, 
	\label{Tdissoc}
 \end{equation}
namely, $T_c$ grows to infinity in the deep strong coupling regime. 
Actually this is the temperature at which Cooper pairs break down due to thermal fluctuations, since it was derived imposing $\Delta(T_{c})=0$, namely in a system of free fermions. It is then naturally expected that when the interaction becomes strong, this approximation fails. As we have seen, at strong coupling, $T_{c}/\Ef\gg 1$,  the gas is out of the degenerate regime. 
Actually, at high temperatures the system is described by a mixture of free particles and pairs, both following almost a Maxwell-Boltzmann distribution
therefore the equilibrium can be imposed by he condition $\mu_{p}=2\mu_{f}$, namely when the chemical potential of the pairs matches twice the  chemical potential of the unpaired fermions. In this picture $T_c$ can be seen as the dissociation temperature for the pairs. 
\subsubsection{Along the crossover}
\begin{figure}
\centering
\includegraphics[width=0.45\textwidth]{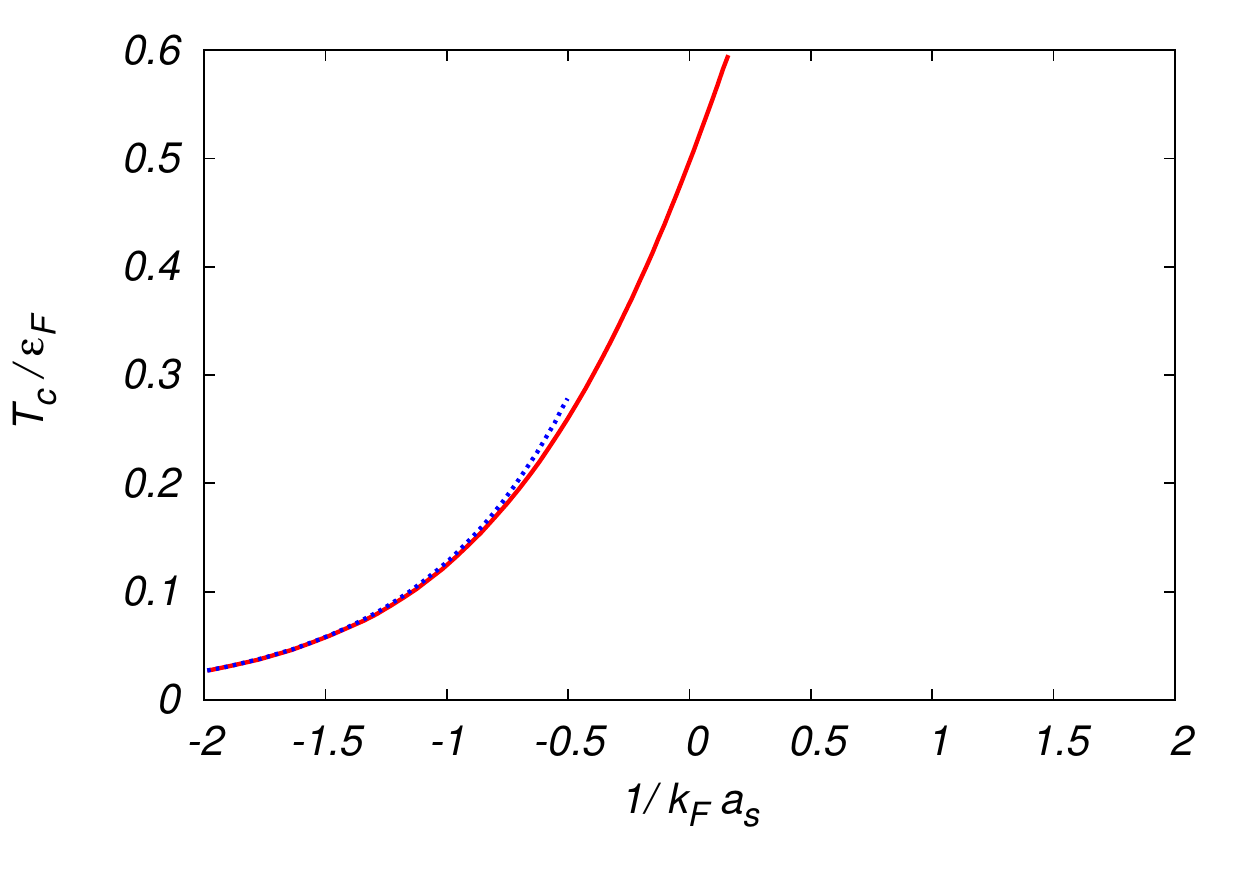}
\includegraphics[width=0.45\textwidth]{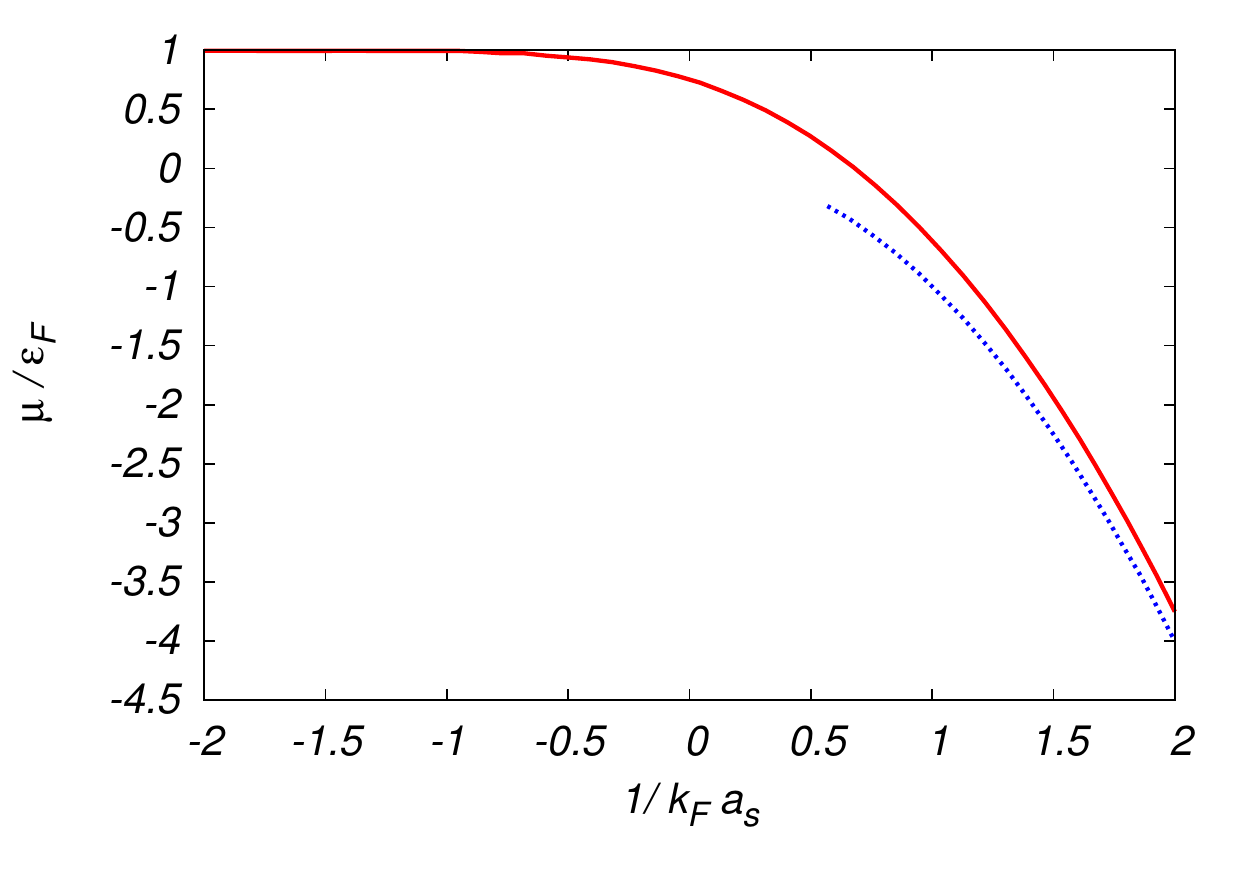}
\caption{Critical temperature $T_c$ and chemical potential $\mu$ at $T_c$ as functions of the inverse of the scattering length $a_s$. The dashed lines are the analytical results from Eq.~\eqref{Tc_BCSlimit} and Eq.~\eqref{chem_pot_strong}.}
\label{fig.meanfield}
\end{figure}

Let us consider Eq.~\eqref{gap_eq} and Eq.~\eqref{numero} in the continuum limit. Defining $y=\frac{1}{\kf\as}$ and $\tilde{\mu}=\frac{\mu}{2T_{c}}$, working at fixed density, we can write
\begin{eqnarray}
\label{eq.gap}
y\left(\frac{\Ef}{T_{c}}\right)^{\frac{1}{2}}=-\frac{\sqrt{2}}{\pi}\int_{0}^{\infty}\sqrt{x}\left(\frac{\tanh(x-\tilde{\mu})}{x-\tilde{\mu}} - \frac{1}{x}\right)dx
\,\equiv \, {I}_g(\tilde{\mu})	\\
\left(\frac{\Ef}{T_{c}}\right)^{\frac{3}{2}}=\frac{3\sqrt{2}}{2}\int_{0}^{\infty}\left(1-\tanh(x-\tilde{\mu}) \right)dx
\,\equiv \, {I}_{0}(\tilde{\mu})
\label{eq.num}
\end{eqnarray}
These integrals can be calculated numerically for different values of $\tilde{\mu}$, building the lists of values for $\left(\frac{\mu}{\Ef},y\right)$ and $\left(\frac{T_{c}}{\Ef},y\right)$. The graphs reported in Fig. \ref{fig.meanfield} are obtained in this way, confirming the analytical results at mean field level.

\subsection{Beyond mean field: Gaussian fluctuations}
As seen previously, the second order expansion of the action is given by Eq.~\eqref{actionS}, which is
 \begin{equation}
	S_{e}[\Delta]=S_{e}^{(0)}+\sum_{\q}\Gamma^{-1}(\q)|\Delta(\q)|^2   + o(\Delta^{3})
 \end{equation}
with $\Gamma^{-1}(\q)$ given by Eq.~\eqref{Gamma_q}. After regularizing the contact interaction through Eq.~\eqref{cut-off}, 
and after a shift of the momenta $\mathbf{k}\rightarrow \mathbf{k}+\mathbf{q}/2$ in order, for simplicity, to get rid of the angular dependence in the denominator on the 
angle defined by $\mathbf{k}\cdot\mathbf{q}=kq\cos\theta$,  we can write
 \begin{equation}
	\Gamma^{-1}(\q)=-\left\{
		\frac{m}{4\pi \as}+
		\frac{1}{\cal V}\sum_{\mathbf{k}}
		\left[
		\frac{1-\fermi(\xi_{\mathbf{k}+\mathbf{q}/2})-\fermi(\xi_{\mathbf{k}-\mathbf{q}/2})}
		{\xi_{\mathbf{k}+\mathbf{q}/2}+\xi_{\mathbf{k}-\mathbf{q}/2} - i\nu_{n}}   -   \frac{1}{2\epsilon_{\mathbf{k}}}
		\right]
			\right\}
 \end{equation}
The sum is invariant under $\mathbf{k}\rightarrow-\mathbf{k}$, therefore the numerator $[1-\fermi(\xi_{\mathbf{k}+\mathbf{q}/2})-\fermi(\xi_{\mathbf{k}-\mathbf{q}/2})]$ can be written as $[1-2\fermi(\xi_{\mathbf{k}+\mathbf{q}/2})]=\tanh\left( \frac{\xi_{\mathbf{k}+\mathbf{q}/2}}{2T_{c}} \right)$, getting
simply
 \begin{equation}
	\Gamma^{-1}(\q)=-\left\{\frac{m}{4\pi \as}+\frac{1}{\cal V}\sum_{\mathbf{k}}
		\left[\frac{ \tanh\left( \frac{\xi_{\mathbf{k}+\mathbf{q}/2}}{2T_{c}} \right)}{\xi_{\mathbf{k}+\mathbf{q}/2}+\xi_{\mathbf{k}-\mathbf{q}/2} - i\nu_{n}}
		-\frac{1}{2\epsilon_{\mathbf{k}}}\right]\right\}
 \end{equation}
In the continuum limit and writing explicitly the spectra $\xi_{\mathbf{k}\pm \mathbf{q}/2}=\frac{|\mathbf{k}\pm \mathbf{q}/2|^2}{2m}$ we have
 \begin{equation}
	\Gamma^{-1}(\q)=-\left\{
		\frac{m}{4\pi \as}+
		\frac{m}{(2\pi)^{2}}\int_{-1}^{1}d\cos\theta\int dkk^{2}
		\left[
		\frac{ \tanh\left( \frac{k^{2}+q^{2}/4-2m\mu+kq \cos\theta}{4mT_{c}} \right)  }    
		     {k^{2}+q^{2}/4-2m\mu-im\nu_{n}}
		-\frac{1}{k^{2}}
		\right]
			\right\}
 \end{equation}
The dependence on $\cos\theta$ is only in the hyperbolic tangent and we can integrate over it, using $\int_{-1}^{1}dx \tanh\left(a+bx\right)=\frac{1}{b}[\ln(\cosh(a+b))-\ln(\cosh(a-b))]$, 
 \begin{equation}
	\int_{-1}^{1}dx\,\tanh\left(\frac{k^2+\frac{q^2}{4}+kq \,x-2m\mu}{4mT_{c}}\right)
	=\frac{4mT_{c}}{kq}\big[A(k,q)-A(k,-q)\big]
 \end{equation}
where
 \begin{equation}
A(k,q)=\ln\left[\cosh\left(\frac{(k+q/2)^2}{4m T_c}-\frac{\mu}{2T_{c}}\right)\right]
 \end{equation}
We have, at the end, the following expression
 \begin{equation}
	\Gamma^{-1}(\q)=-\left\{
		\frac{m}{4\pi \as}+
		\frac{m}{(2\pi)^{2}}\int_{0}^{+\infty} dkk^{2}
		\left[
		\frac {4mT_{c}}{kq}\frac{A(k,q)-A(k,-q)}    
		     {k^{2}+q^{2}/4-2m\mu-im\nu_{n}}
		-\frac{2}{k^{2}}
		\right]
			\right\}\,.
		    \label{eqpercalcnumerico}
 \end{equation} 
 Let us consider the weak and the strong coupling limits.
\subsubsection{Weak coupling}
In the weak coupling limit the chemical potential is finite and the critical temperature and the scattering length go to zero, $\as\rightarrow0^{-}$. 
Without entering into the details we have
\begin{equation}
\label{Gammawc}
\Gamma^{-1}(\q)=-\frac{m}{4\pi} \left(
		\frac{1}{\as}+f(\q,\mu)
	\right)
\end{equation}
where $f(\q,\mu)$ some complex function of the four-momentum $\q$ (momentum and frequency) and chemical potential $\mu$. In the weak coupling limit, approximately
$\Gamma^{-1}(\q)$ does not depend on critical temperature since, for small $T_c$ we have $A(k,q)\sim\frac{1}{2T_c}\left|\frac{(k+q/2)^2}{2m}-\mu\right|$. 
Calculating explicitly the second order correction to the density of particles, we have
 \begin{equation}
 \label{n2wc}
	n^{(2)} = \frac{1}{\beta\cal V}\frac{\partial}{\partial\mu}
	\sum_{\q}\ln\left(\Gamma^{-1}(\q)\right)
	=\frac{1}{\beta\cal V}\sum_{\q}\frac{1}{\frac{1}{\as}+f(\q,\mu)}
	\frac{\partial f(\q,\mu)}{\partial\mu}\underset{\as\rightarrow0^{-}}{\longrightarrow}
	\frac{\as}{\beta\cal V }\sum_{\q}\frac{\partial f(\q,\mu)}{\partial\mu}\propto \as
 \end{equation}
In this limit the correction to the density of particles is proportional to the scattering length and therefore is negligible.
As a result 
the chemical potential will be almost constant, $\mu\approx\Ef$, and since also the gap equation is unchanged, in the limit $\frac{1}{k_F\as}\rightarrow-\infty$, the critical temperature is expected to be almost the same as that obtained at the mean field level.
In other words, the Gaussian fluctuations are not so effective at weak coupling. This expectation will be confirmed by numerical calculations.

\subsubsection{Strong coupling}
\label{Gauss_free_Accopiamento forte}
As we will see, in the strong coupling limit the effective partition function is equivalent to that of a system of free bosons whose critical temperature is that of the BEC. In order to prove this result we first need to manipulate $\Gamma^{-1}$. As we have seen, in the BEC  limit, $\beta\mu\rightarrow-\infty$ and $1/\kf\as\rightarrow\infty$. 
For a large and negative chemical potential
 \begin{equation}
	A(k,q)=\ln\left[\cosh\left(\frac{ \frac{(k+q/2)^2}{2m}-\mu}{2T_{c}}\right)\right]
	\underset{\mu\rightarrow-\infty}{\longrightarrow}
	\frac{ \frac{(k+q/2)^2}{2m}-\mu}{2T_{c}}
 \end{equation}
therefore
 \begin{equation}
	A(k,q)-A(k,-q)\underset{\mu\rightarrow-\infty}{\longrightarrow}
	\frac{ \frac{(k+q/2)^2}{2m}-\mu}{2T_{c}}-\frac{ \frac{(k-q/2)^2}{2m}-\mu}{2T_{c}}=
	\frac{ k q }{2mT_{c}}
 \end{equation}
Inserting this result in Eq.~\eqref{eqpercalcnumerico}, after integration, we get
 \begin{equation}
	\Gamma^{-1}(\q)\approx-\frac{m}{4\pi} 
		\left(  \frac{1}{\as}-\sqrt{\frac{q^{2}}{4}-2m\mu-im\nu_{n}}    \right)
 \end{equation}
Taking advantage of the definition of the bound state $E_{b}=-\frac{1}{m\as^{2}}$, and defining $\mub=2\mu-E_{b}$, we can write
 \begin{equation}
	\Gamma^{-1}(\q)\approx
		\frac{1}{R(\q)}\left[{i\nu_{n}-\left(\frac{q^{2}}{4m}-\mub\right)}\right]
 \end{equation}
 where
$R(\q)=\frac{4\pi}{m^{3/2}}\left(\sqrt{-E_{b}}+\sqrt{\frac{q^{2}}{4m}-2\mu-i\nu_{n}}\right)$, which is simply $R(\q)\approx \frac{8\pi}{m^{2}\as}$ for small fluctuations in the deep BEC limit.
The quantity $\mub$ can be seen as the chemical potential for free bosons, which makes sense only if it is negative. The partition function  takes the form
 \begin{equation}
	\mathcal{Z}_{G}=\mathcal{Z}_{0}\int D\Delta^{*} D\Delta \,e^{-\sum_{\q}\Gamma^{-1}(\q)|\Delta(\q)|^2}
	\approx\mathcal{Z}_{0}\int D\Delta^{*} D\Delta\,
	e^{-\sum_{\q} \left[i\nu_{n}-\big(\frac{q^{2}}{4m}-\mub\big)\right] \frac{|\Delta(\q)|^2}{R(\q)}}
 \end{equation}
Rescaling the fields $\phi(\q)=\frac{\Delta(\q)}{\sqrt{R(\q)}}$ and defining $\mb=2m$ the partition function at the Gaussian level is equivalent to that for free bosons 
with mass $\mb$ (equal to the mass of two fermions)
 \begin{equation}
	\mathcal{Z}_{G} \propto \int D\phi^{*} D\phi\,
	e^{-\sum_{\q}\phi^{*}(\q) \left[i\nu_{n}-\left(\frac{q^{2}}{2\mb}-\mub\right)\right] \phi(\q)}
 \end{equation}
Now we can take advantage of the well-known expression for the critical temperature for the Bose-Einsten condensation which is given by
 \begin{equation}
	T_{c}=\frac{2\pi}{\mb}\left(\frac{\bose}{2\zeta(3/2)}\right)^{\frac{2}{3}}
	\label{Tc_BEC}
 \end{equation}
If in the deep BEC regime all fermions form pairs, therefore $\bose=n/2$. At fixed fermionic density $n=\frac{(2m\Ef)^{2/3}}{3\pi^{2}}$, we have
 \begin{equation}
	T_{c}=\frac{2\pi}{\mb}\left( \frac{\bose}{2\zeta(3/2)}\right)^{\frac{2}{3}}=
	\frac{\pi}{m}\left( \frac{(2m\Ef)^{3/2}}{6\pi^{2}\zeta(3/2)} \right)^{\frac{2}{3}}=\left(\frac{\sqrt{2}}{3\sqrt{\pi}\,\zeta(3/2)}\right)^{\frac{2}{3}}\Ef\,
	\,\approx0.218\,\Ef
 \end{equation}
According to the BEC theory, at the critical point, the effective chemical potential approaches zero from negative values, therefore 
 \begin{equation}
 \label{mu_Eb}
	\mub(T_{c})=2\mu(T_{c})-E_{b}\underset{\frac{1}{\kf\as}\rightarrow\infty}{\longrightarrow}0
 \end{equation}
which is exactly the solution of the gap equation in the strong coupling limit, Eq.~\eqref{chem_pot_strong}.

 \subsubsection{Full crossover}
As shown in Sec.~\ref{Inclusione delle fluttuazioni Gaussiane free}, the set of equations to solve at second order is
 \begin{eqnarray}
 \label{Gaussian_gap}
-\frac{m}{4\pi \as}&=&\frac{1}{\cal V}\sum_{\mathbf{k}}\left(  \frac{\tanh(\xi_{\mathbf{k}}/2T_{c})}{2\xi_{\mathbf{k}}} - \frac{1}{2\epsilon_{\mathbf{k}}} \right)\\
n=n^{(0)}+n^{(2)}&=&
\frac{1}{\cal V}\sum_{\mathbf{k}}\left[ 1-\tanh\left(  \frac{\xi_{\mathbf{k}}}{2T_{c}}  \right)  \right]+
\frac{1}{\pi\cal V}\sum_{\mathbf{q}}\int_{-\infty}^{+\infty}d\omega\ \bose(\omega)\frac{\partial\delta(\mathbf{q},\omega)}{\partial\mu}
	\label{Gaussian_number}
 \end{eqnarray}
with the phase shift $\delta(\mathbf{q},\omega)=-\textrm{Arg}\left(\Gamma^{-1}(\mathbf{q},\omega+i\epsilon)\right)$.
We need, therefore, to find the second term of the number equation, $n^{(2)}$, possibly resorting also to some approximations.

\begin{figure}[h]
\includegraphics[width=0.45\textwidth]{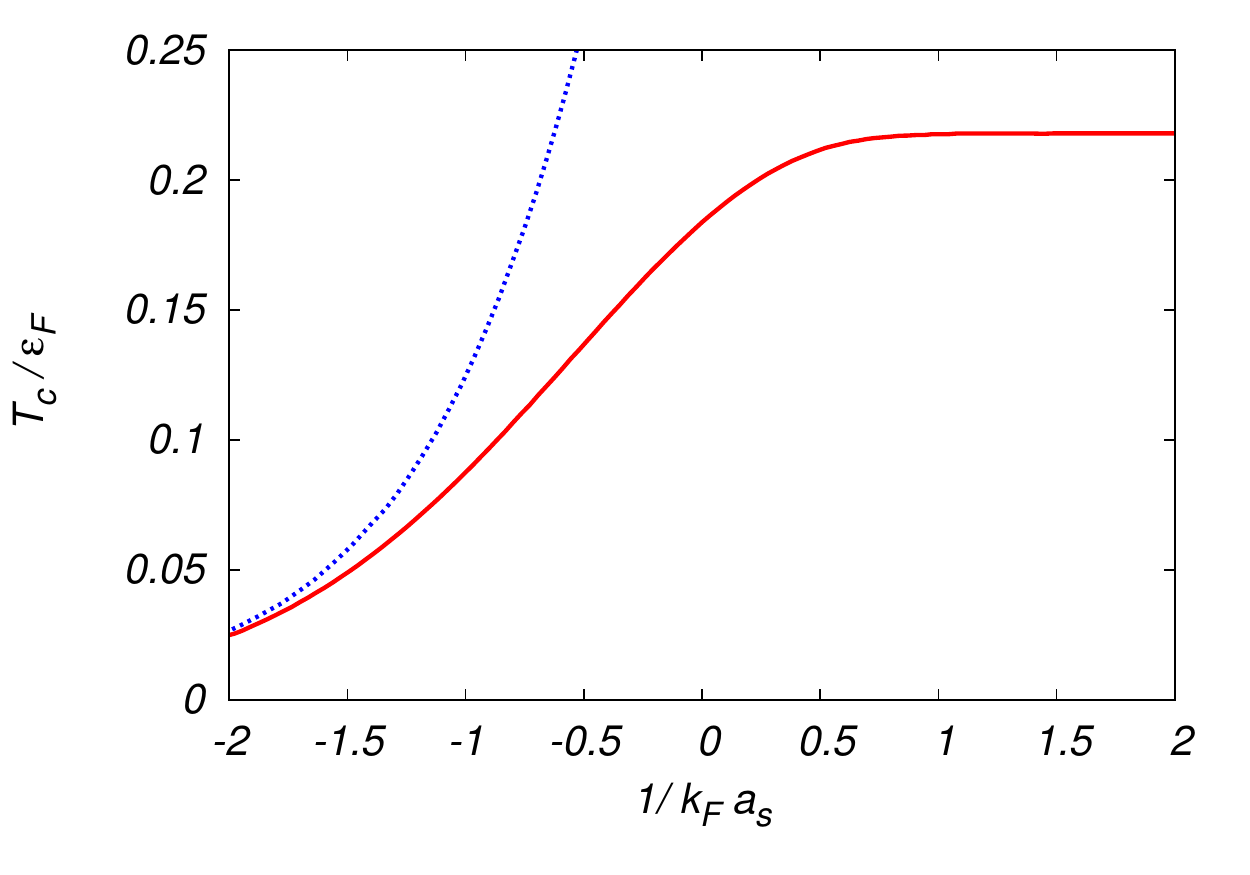}
\includegraphics[width=0.45\textwidth]{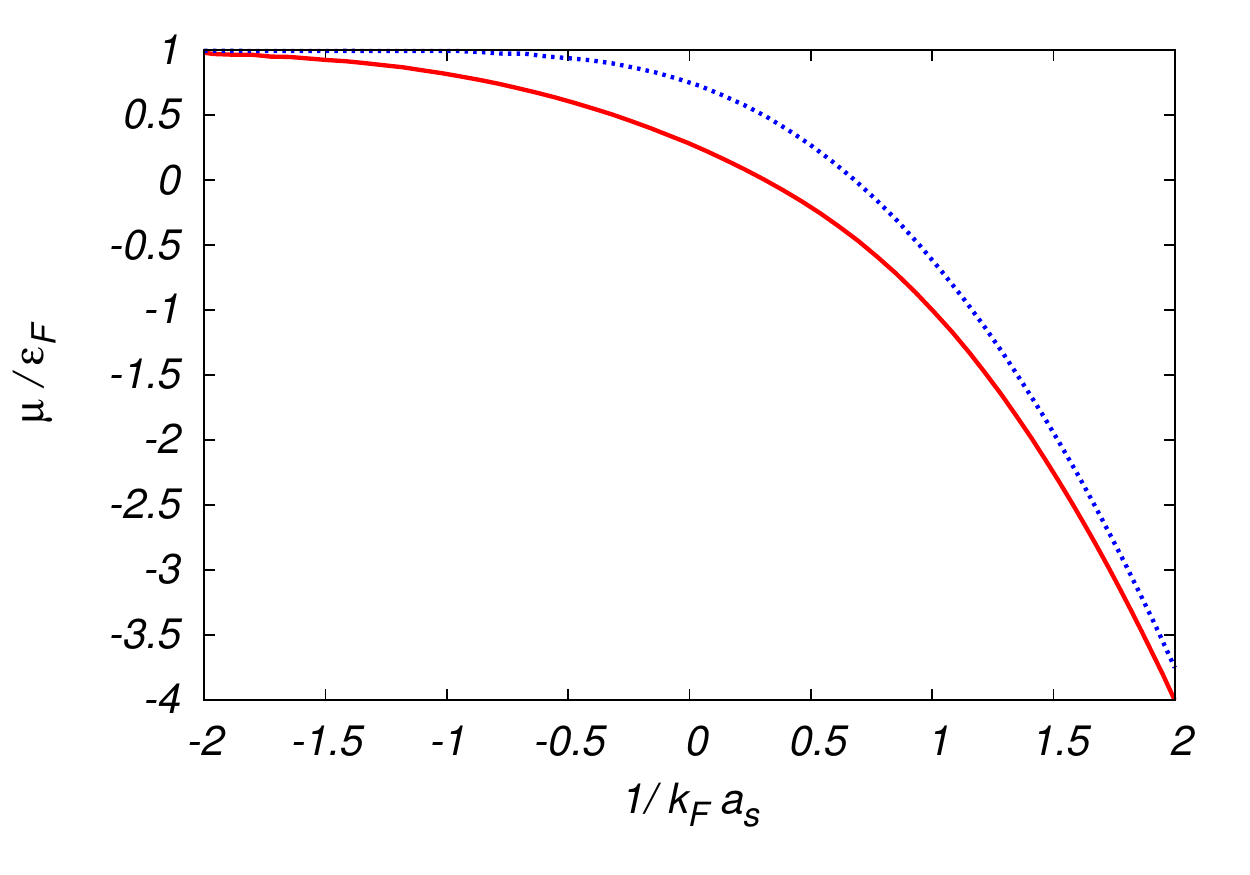}
\caption{Critical temperature $T_c$ and chemical potential $\mu$ at $T_c$ within the bosonic approximation. 
The dashed lines are the mean field results.}
\label{fig.bosonicapprox}
\end{figure}

\paragraph{Bosonic approximation.} 
As we have seen before, the corrections due to quantum fluctuations are more important in the BEC regime while they are negligible in the BCS weak coupling regime. The inverse of the vertex function in the strong coupling limit is
 \begin{equation}
	\Gamma^{-1}(\q)\approx i\nu_{n}-\left(\frac{q^{2}}{2\mb}-\mub\right)
	\label{inv_vert_func}	
 \end{equation}
which describes a system of free bosons, with $\mub=2\mu-E_{b}$ and $\mb=2m$. 
We can use this simple result as a first approximation to include Gaussian fluctuations in order to calculate the critical temperature. 
This approximation consists of simply substituting the second order correction $n^{(2)}$ in the number equation Eq.~\eqref{Gaussian_number} with the density of free bosons whose inverse propagator is given by 
Eq.~\eqref{inv_vert_func}. It is well known that the critical temperature $T_{c}$ of free bosons is given by Eq.~\eqref{Tc_BEC}. Let us suppose that the number of bosonic excitations composed by pairs of fermions is $\bose=2n^{(2)}$, while the rest of fermions remains unpaired. In this approximation, using Eq.~\eqref{Tc_BEC}, we have, therefore,
 \begin{equation}
 \label{n2bosonic}
	n^{(2)} \approx \,2\,\zeta(3/2)\left(\frac{mT_{c}}{\pi}\right)^{3/2} 
 \end{equation} 
 As a result, we have to solve Eq.~\eqref{Gaussian_gap} together with the number equation which, in the bosonic approximation, reads
 \begin{equation}
	n=\frac{1}{\cal V}\sum_{\mathbf{k}}\left[ 1-\tanh\left(  \frac{\xi_{\mathbf{k}}}{2T_{c}}  \right)  \right]+
	2\zeta(3/2)\left(\frac{mT_{c}}{\pi}\right)^{3/2}.
	\label{Gaussian_number_bosons}
 \end{equation}
After defining the following quantities
\begin{equation}
y=\frac{1}{k_F\as},\;\;\;\;\;
\bmu=\frac{\mu}{\Ef},\;\;\;\;\;
\bTc=\frac{T_c}{\Ef}
\end{equation}
in the continuum limit, Eq.~\eqref{Gaussian_gap} and Eq.~\eqref{Gaussian_number_bosons}, at fixed density $n=\frac{\kf^{3}}{3\pi^{2}}$ can be written as
 \begin{eqnarray}
	&& y= -\frac{2}{\pi}\int_{0}^{\infty}\!dx \ x^{2}
	\left(  \frac{\tanh\left(\frac{x^{2}-\bmu}{2\bTc}\right)}{x^{2}-\bmu}-\frac{1}{x^{2}} \right)\\
	&& \frac{2}{3}= \int_{0}^{\infty} dx \ x^{2}
	\left[1-\tanh\left(  \frac{x^{2} -\bmu}{2\bTc}  \right)  \right]+2\,\zeta(3/2)\sqrt{\frac{\pi}{2}}\,\bTc^{3/2}
 \end{eqnarray}
This set of equations can be easily solved numerically, finding the lists of values $(\bTc, \bmu, y)$ which fulfill the two equations.
Some results are reported in Fig~\ref{fig.bosonicapprox}.

\begin{figure}[h]
\centering
\includegraphics[width=0.45\textwidth]{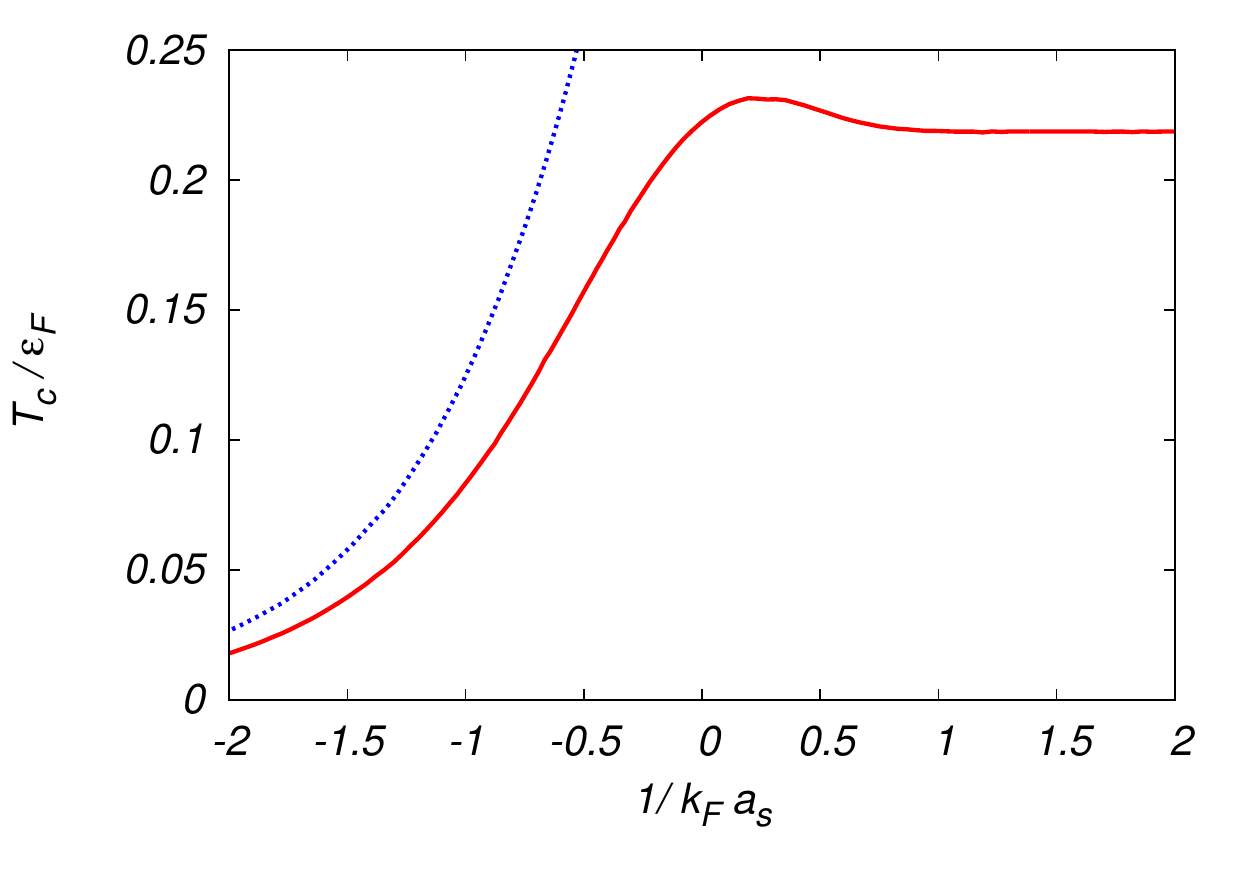}
\includegraphics[width=0.45\textwidth]{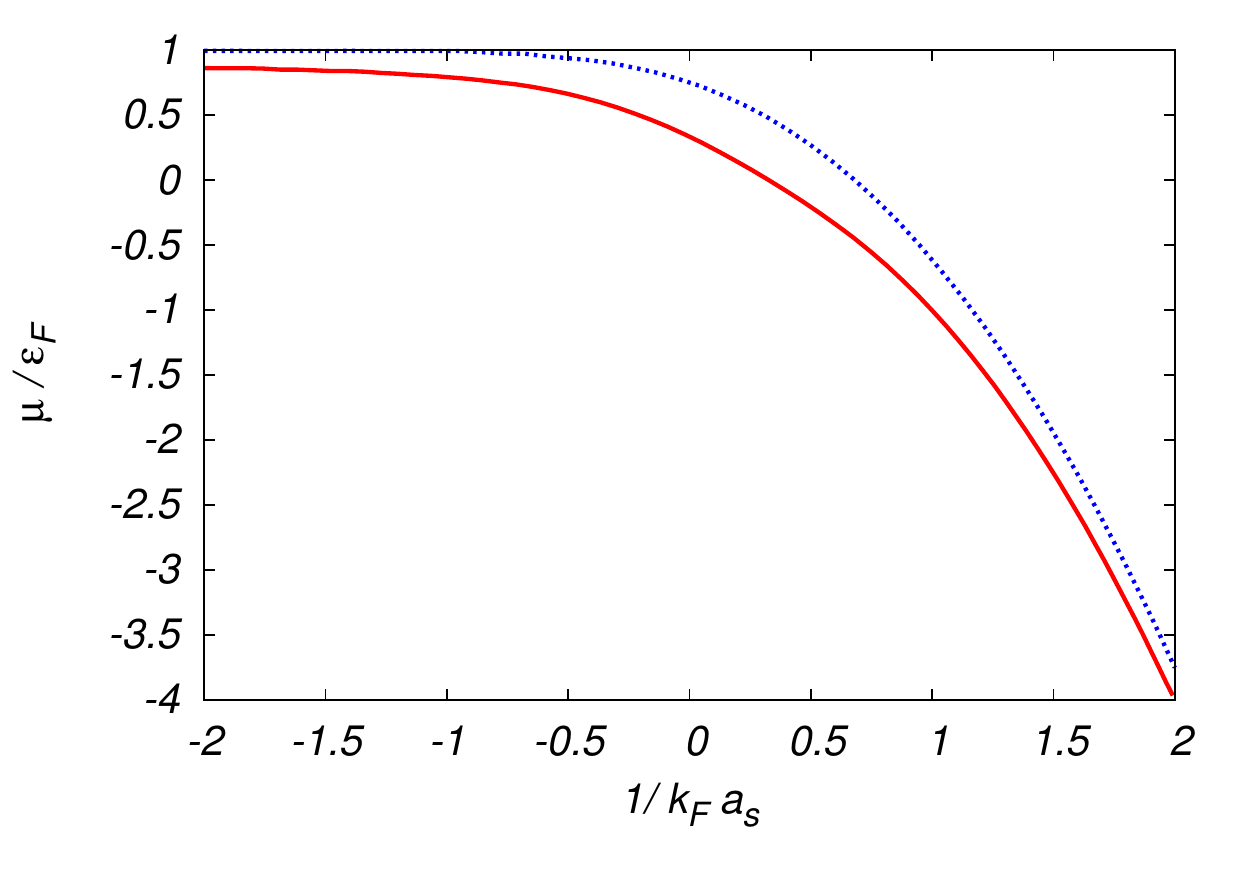}
\caption{Critical temperature $T_c$ and chemical potential $\mu$ at $T_c$ at the Gaussian level. 
The dashed lines are the mean field results.}
\label{Gauss_mu_Tc}
\end{figure}

\paragraph{Exact Gaussian fluctuations.}
Let us work further on the function $\Gamma^{-1}(\mathbf{q},\omega+i\epsilon)$, as obtained in Eq.~\eqref{eqpercalcnumerico}.
With the rescalings $\tilde{\mu}=\frac{\mu}{2T_{c}}$, $p=\frac{k}{\sqrt{4mT_{c}}}$ and $Q=\frac{q}{\sqrt{4mT_{c}}}$, the function $A(k,q)$ becomes 
\begin{equation}
A(k,q)=\ln\left[\cosh\left(\frac{ \frac{(k+q/2)^2}{2m}-\mu}{2T_{c}}\right)\right]=\ln\left[\cosh\left((p+Q/2)^2-\tilde{\mu}\right)\right]=\tilde A(p,Q)
\end{equation}
Making the same substitutions in the vertex function, after a Wick rotation in the frequencies ($\nu_n\rightarrow \omega +i\epsilon$, with $\epsilon>0$ an infinitesimal quantity), we can define $\Gamma^{-1}(\mathbf{q},\omega+i\epsilon)=\widetilde{\Gamma}^{-1}(\mathbf{Q},\nu+i\epsilon)$,  with the rescaled real frequency $\nu=\frac{\omega}{4T_{c}}$,
\begin{equation}
\widetilde{\Gamma}^{-1}(\mathbf{Q},\nu+i\epsilon)
		=-\left(\frac{m}{4\pi}\right)
			\left\{
		\frac{1}{\as}+
		\frac{\sqrt{4mT_{c}}}{\pi Q}\int_{0}^{\infty} dp
		\left[\frac{p\left[\tilde A(p,Q)-\tilde A(p,-Q)\right]}    
		     {p^{2}+Q^{2}/4-\tilde{\mu}-\nu -i\epsilon}-2Q\right]
			\right\}
\end{equation}
We remind that, 
looking at Eq.~\eqref{n_gauss_correction}, the interesting part of the function $\widetilde{\Gamma}^{-1}({\bf Q},\nu+i\epsilon)$ is actually its argument, therefore, its prefactor is irrelevant. 
Within the range of integration the integrand has a pole in $p_{0}=\sqrt{\tilde{\mu}+\nu-Q^{2}/4}$, if $p_0$ real and positive. Defining the function
\begin{equation}
t(p,Q)=\frac{p\left[\tilde A(p,Q)-\tilde A(p,-Q)\right]-2Q(p^2-p_0^2)}{p+p_{0}}
\end{equation}
we can write 
\begin{equation}
\widetilde{\Gamma}^{-1}(\mathbf{Q},\nu+i\epsilon)=
-C\left[Q \,y\left(\frac{\Ef}{T_{c}}\right)^{1/2}+\frac{\sqrt{2}}{\pi}\int_{0}^{+\infty}
	\frac{t(p,Q)}{p-p_{0}-i\epsilon}
	dp\right]
	\label{Gamma_tilde}
\end{equation}
with the prefactor $C=-\left(\frac{m}{2}\right)^{3/2}  \frac{\sqrt{T_c}}{\pi Q}$. 
The integral in Eq.~\eqref{Gamma_tilde} can be evaluated knowing its Cauchy principal value 
\begin{equation}
\int_{0}^{+\infty}
	\frac{t(p,Q)}{p-p_{0}\pm i\epsilon}
	dp=\mathcal{P}\int_{0}^{+\infty}\frac{t(p,Q)}{p-p_{0}}dp\mp i\pi \int_{0}^{+\infty}\delta(p-p_0)t(p,Q)
\end{equation}
For $(\tilde{\mu}+\nu-Q^{2}/4)>0$, namely for $p_0$ real and positive, $\Gamma^{-1}$ has an imaginary part. We can write the vertex function $\widetilde{\Gamma}^{-1}$ and its argument $\tilde\delta$ as it follows
\begin{equation}
\widetilde{\Gamma}^{-1}(\mathbf{Q},\nu)=-C\left(X+iY\right),\;\;\;\;\;\;\;\;\;\tilde\delta(\mathbf{Q},\nu)=
\arctan\left({\frac{Y}{X}}\right)
\end{equation}
where the real and imaginary parts of $\widetilde{\Gamma}^{-1}$ are given by
\begin{eqnarray}
&&X=Q\,y\left(\frac{\Ef}{T_{c}}\right)^{{1}/{2}}+\frac{\sqrt{2}}{\pi}\mathcal{P}\int_0^\infty\frac{t(p,Q)}{p-p_{0}}dp\\
&&Y=\sqrt{2}\ t(p_{0},Q)=\frac{1}{\sqrt{2}}\left[\tilde A(p_0,Q)-\tilde A(p_0,-Q)\right]
\end{eqnarray}
We obtained an analytic expression for the imaginary part of the vertex function. We can now calculate the contribution to the density of particles due to quantum Gaussian fluctuations from Eq.~\eqref{n_gauss_correction}, which
by rescaling all the parameters, reads
\begin{eqnarray}
n^{(2)}&=&
\frac{1}{\pi\cal V}\sum_{\mathbf{q}}\int_{-\infty}^{+\infty}d\omega\ \bose(\omega)\frac{\partial\delta(\bf{q},\omega)}{\partial\mu}=\\
\nonumber&=&\frac{(4mT_{c})^{3/2}}{\pi^{3}}\int_{0}^{+\infty}dQ\ Q^{2}
	\int_{-\infty}^{+\infty}d\nu\frac{1}{e^{4\nu}-1}\frac{\partial\tilde{\delta}(\bf{Q},\nu)}{\partial\tilde{\mu}}
\equiv\frac{(2mT_{c})^{3/2}}{3\pi^{2}}\,{I}_{2}\left(\tilde{\mu},y\sqrt{\frac{\Ef}{T_{c}}}\,\right)
\end{eqnarray}
Writing the number of particles in terms of $\Ef$, the final number equation can be rewritten as
\begin{equation}
\left(\frac{\Ef}{T_{c}}\right)^{3/2}=
{I}_{0}(\tilde{\mu})+ {I}_{2}\left(\tilde{\mu},y\sqrt{\frac{\Ef}{T_{c}}}\,\right)
\end{equation}
We have, therefore, to solve this equation together with the gap equation Eq.~\eqref{eq.gap}. The complete set of equations, obtained going beyond the mean field theory, with the inclusion of the full Gaussian fluctuations contained in the function $I_2$, is, therefore,
\begin{eqnarray}
&&y\left(\frac{\Ef}{T_{c}}\right)^{1/2}=
{I}_g(\tilde{\mu})\\
&&\left(\frac{\Ef}{T_{c}}\right)^{3/2}=
{I}_0(\tilde{\mu})+
{I}_2\big(\tilde{\mu},{I}_g(\tilde{\mu})\big)
\label{eq_state_crit}
\end{eqnarray}
This set of equations can be solved numerically with {root-finding} algorithms, finding the values for $(y,{\mu},T_c)$ which satisfy simultaneously the two equations.  The results are shown in Fig.~\ref{Gauss_mu_Tc}. 
In particular Eq.~\eqref{eq_state_crit} is actually an equation of state which put in relation $T_{c}$ and $\mu$. 
As one can see from Fig.~\ref{Gauss_mu_Tc}, the quantum fluctuations are more important in the BEC regime, namely in the strong coupling limit, as already discussed.  
The failure of the mean field theory in correctly reproducing the behavior for the critical temperature along the crossover is due to the fact that, in that scheme, the number equation describes a set of non interacting fermions, disregarding, 
therefore, the two-body physics which becomes relevant specially when the attraction becomes strong.
We showed that going towards strong coupling, the partition function of the system actually tends to that of a system of bosons which can be seen as composed by pairs of fermions. 

\section{Fermi gas with spin-orbit interaction, Zeeman term and attractive potential}
The aim of the next sections is to see how the critical temperature along the BCS-BEC crossover, described in the previous section, is modified by the presence of the spin-orbit (SO) coupling discussed in section Sec.~\ref{spin_orbit_term}. 
Let us consider the following Hamiltonian $H=H_{0}+H_{I}$, in the momentum space, made by a single-particle term
 \begin{equation}
	H_{0}=\sum_{\mathbf{k}}\left(\hpsidagu(\mathbf{k}),\hpsidagd(\mathbf{k})\right)
		\left\{ 
	\frac{k^2}{2m}\mathbb{1}+\vR(\sigma_x k_y-\sigma_y k_x)+\vd(\sigma_x k_y+\sigma_y k_x)+h\sigma_z
		 \right\}
		\begin{pmatrix}
			\hpsiu(\mathbf{k}) \\ \hpsid(\mathbf{k})
		\end{pmatrix}
 \end{equation} 
 and an interacting term 
 \begin{equation}
	H_{I}=-\frac{g}{\cal V}\sum_{\mathbf{k},\mathbf{k}',\mathbf{q}}\hpsidagu(\mathbf{k}+\mathbf{q})\hpsidagd(-\mathbf{k})\hpsid(\mathbf{q}-\mathbf{k}')\hpsiu(-\mathbf{k}')
 \end{equation}
where $\sigma_x=
\begin{pmatrix} 
	0 & 1 \\ 
	1 & 0 
\end{pmatrix}$ 
and $\sigma_y=
\begin{pmatrix} 
	0 & -i \\ 
	i & 0 
\end{pmatrix}$
are Pauli matrices. 
The new ingredient with respect to the previous case is the spin-orbit interaction, contained in the two terms of the single-particle Hamiltonian, respectively the Rashba term, $\vR(\sigma_x k_y-\sigma_y k_x)$, Eq.~\eqref{HR}, and the Dresselhaus term, $\vd(\sigma_x k_y+\sigma_y k_x)$, Eq.~\eqref{HD}. We included also an effective artificial Zeeman term $h\sigma_z$, Eq.~\eqref{HZ}, which can be tuned by a Raman laser \cite{SObose}.
This Hamiltonian is associated, in the path-integral formalism, by the Grassmann variables $\psi$ and $\psidag$, to the partition function
 \begin{equation}
	\mathcal{Z}=\textrm{Tr}\left( e^{-\beta(H-\mu N)}\right)=\int D\psidag D\psi\,e^{-S[\psidag,\psi]}
 \end{equation}
with the action
 \begin{multline}
	S[\psidag,\psi]= \sum_{\ks,s}\psidag_{s}(\ks)
		\left[ i\omega_{n}+\frac{k^2}{2m}-\mu \right]
					\psi_{s}(\ks) \ -\frac{g}{\beta\cal V}\sum_{\ks,\ks',\q}\psidagu(\ks+\q)\psidagd(-\ks)\psid(\q-\ks')\psiu(\ks')\\
	+\sum_{\ks}\left(\psidagu(\ks),\psidagd(\ks)\right)
		\bigg[ \vR(\sigma_x k_y-\sigma_y k_x)+\vd(\sigma_x k_y+\sigma_y k_x)+h \sigma_z \bigg]
		\begin{pmatrix}
			\psiu(\ks) \\ \psid(\ks)
		\end{pmatrix}
 \end{multline}
where $\psi_{s}(\ks)$ and $\psidag_{s}(\ks)$ are Grassmann variables depending on the four-momentum $\ks=(\omega_n,{\bf k})$. 
As already seen, the term with four fermions can be decoupled by means of an Hubbard-Stratonovich transformation, introducing the auxiliary field $\Delta(\q)$
 \begin{equation*}
  e^{\frac{g}{\beta\cal V}\sum\limits_{\ks,\ks',\q}\psidagu(\ks+\q)\psidagd(-\ks)\psid(-\ks')\psiu(\ks'+\q)}
 =\int D\Delta^* D\Delta \ e^{\sum\limits_{\q} 
	\left[  
		-\frac{\beta\cal V}{g}|\Delta|^{2} +
		\sum\limits_{\ks}\left(\Delta^{*}(\q)\psid(-\ks)\psiu(\ks+\q) +
		\psidagu(k+\q)\psidagd(-\ks)\Delta(\q)\right)
	\right]}
 \end{equation*}
 in this way the partition function takes the form 
$ \mathcal{Z}=\int D\psidag D\psi\int D\Delta^* D\Delta\,e^{-S[\psidag,\psi,\Delta]}$
where, after defining 
\begin{equation}
\gamma(\mathbf{k})=\vR(k_x+ik_y)+\vd(k_x-ik_y)
\label{gamma_k}
\end{equation}
the new action reads
 \begin{multline}
 \label{SppDD}
	S[\psidag,\psi,\Delta]= \sum_{\ks,s}\psidag_{s}(\ks)
		\left[ i\omega_{n}+\xik^{(s)} \right]
					\psi_{s}(\ks)
	+\sum_{\ks}\psidagd(\ks)\gamma^{*}(\mathbf{k})\psiu(\ks)
	+\sum_{\ks}\psidagu(\ks)\gamma(\mathbf{k})\psid(\ks)\\
	+\sum_{\q}\frac{\beta\cal V}{g}|\Delta(\q)|^2
		-\sum_{\q,\ks}\Delta^{*}(\q)\psid(-\ks)\psiu(\ks+\q)
		-\sum_{\q,\ks}\psidagu(\ks+\q)\psidagd(-\ks)\Delta(\q)
 \end{multline}
 where $\xik^{(\uparrow)}=\xik+h$ and $\xik^{(\downarrow)}=\xik-h$. 
It is now convenient to introduce the following Nambu--Jona-Lasinio multispinors
 \begin{equation} 
  \Psi(\ks)=\begin{pmatrix}
		\psiu(\ks)\\
		\psidagd(-\ks)\\
		\psid(\ks)\\
		\psidagu(-\ks)
             \end{pmatrix}\ \ \ \ \textrm{and} \ \ \ \
  \Psidag(\ks)=\left(\psidagu(\ks),\psid(-\ks),\psidagd(\ks),\psiu(-\ks)\right)
 \end{equation}
so that Eq.~\eqref{SppDD} can be written as
 \begin{equation}
 \label{SppDD_matrix}
	S[\psidag,\psi,\Delta]=
\frac{\beta\cal V}{g}\sum_{\q}|\Delta(\q)|^2
+\frac{1}{2}\sum_{\ks,\p}\Psidag(\ks)\,\mathcal{G}^{-1}(\ks,\p)\Psi(\p)
+\beta\sum_{\mathbf{k}}\xik
 \end{equation}
 where the sum $\beta\sum_{\mathbf{k}}\xik$ comes from interchanging the fermionic fields and the corresponding equal-time limiting procedure in order to write the action in a matrix form \cite{Ultracold}, namely 
 $\sum\limits_{{\bf k},s}\hpsidag_{s}({\bf k}) \xik \hpsi_{s}({\bf k})= \frac{1}{2}\sum\limits_{{\bf k}}\xik[\hpsidagu({\bf k})\hpsiu({\bf k})-\hpsiu(-{\bf k})\hpsidagu(-{\bf k}) +\hpsidagd({\bf k})\hpsid({\bf k})-\hpsid(-{\bf k})\hpsidagd(-{\bf k})]+\sum\limits_{\mathbf{k}}\xik$. 
In Eq.~\eqref{SppDD_matrix} we also introduced the following inverse Green function 
 \begin{equation}
 \label{G_so}
	\mathcal{G}^{-1}(\ks,\p)=
	\begin{pmatrix}
	(i\omega_{n}+\xik+h)\delta_{\ks,\p} & -\Delta\text{($\p$-$\ks$)} & \gamma(\mathbf{k})\delta_{\ks,\p} & 0\\
	-\Delta^{*}\text{($\ks$-$\p$)} & (i\omega_{n}-\xik+h)\delta_{\ks,\p} & 0 & -\gamma(-\mathbf{k})\delta_{\ks,\p}\\
	\gamma^{*}(\mathbf{k})\delta_{\ks,\p} & 0 & (i\omega_{n}+\xik-h)\delta_{\ks,\p} & \Delta\text{($\p$-$\ks$)}\\
	0 & -\gamma^{*}(-\mathbf{k})\delta_{\ks,\p} & \Delta^{*}\text{($\ks$-$\p$)} & (i\omega_{n}-\xik-h)\delta_{\ks,\p}\\
	\end{pmatrix}
 \end{equation}
 where $\gamma(-\mathbf{k})=-\gamma(\mathbf{k})$. 
Being the action quadratic, we can perform a Gaussian integral in the Grassmann variables getting an effective theory which depends only on the auxiliary complex field
 \begin{equation}
	\mathcal{Z}=\int D\psidag D\psi D\Delta^{*} D\Delta \,e^{-S[\psidag,\psi,\Delta]}
\propto \int D\Delta^{*} D\Delta \,e^{S_{e}[\Delta]}
 \end{equation}
obtaining an effective action depending only on the pairing function 
 \begin{equation}
	S_{e}[\Delta]=
\frac{\beta\cal V}{g}\sum_{\q}|\Delta(\q)|^2
-\frac{1}{2}\ln\left[\det\left(\mathcal{G}^{-1}\right)\right]
+\beta\sum_{\mathbf{k}}\xik\,.					
\label{S_eff2}
 \end{equation}

\subsection{Gap equation}
 
As done in Sec.~\ref{sec.5}, we first search for the homogeneous pairing which minimizes the action, selecting only the contribution at $\q=0$ in momentum space. Calling $\Delta(\q=0)=\Delta_0$ the 
determinant appearing in Eq.~(\ref{S_eff2}) is simply
 \begin{equation}
	\det\left(\mathcal{G}^{-1}\right)=
	\prod_{\mathbf{k},i\omega_{n},j}(i\omega_{n}-E_{j\kv})
 \end{equation}
 where the energies are
 \bea
&&E_{1\kv}=\sqrt{\xi_\kv^2+\gamma_{h\kv}^2+|\Delta_0|^2-2\sqrt{\xi^2_\kv
\gamma_{h\kv}^2+|\Delta_0|^2 h^2}}\\
&&E_{2\kv}=\sqrt{\xi_\kv^2+\gamma_{h\kv}^2+|\Delta_0|^2+2\sqrt{\xi^2_\kv\gamma_{h\kv}^2+|\Delta_0|^2h^2}}
\eea
and where
\be
\label{gammahk}
\gamma_{h\kv}=\sqrt{|\gamma(\kv)|^2+h^2}=\sqrt{(\vR+\vd)^2 k_y^2+(\vR-\vd)^2 k_x^2+h^2}
\ee
with $\gamma(\kv)=\vR(k_y+ik_x)+\vd(k_y-ik_x)$, Eq.~(\ref{gamma_k}), and 
$\xi_{\kv}=k^2/2m-\mu$. 
$\vR$ and $\vd$ are the Rashba and Dresselhaus velocities 
and $h$ is the Zeeman field.
The effective action, therefore, takes the form
 \begin{equation}
	S_{e}[{\Delta_0}]=
	\frac{\beta\cal V}{g}{|\Delta_0|}^{2}
	-\frac{1}{2}\sum_{\mathbf{k},i\omega_{n},j}\ln\left(i\omega_{n}-E_{j\kv}\right)
	+\beta\sum_{\mathbf{k}}\xik	
	\label{Se0}
 \end{equation}
The saddle point equation sets the mean field solution and is given by the condition $\frac{\partial S_{e}}{\partial\Delta_0}=0$
\be
\frac{\Delta_0^*}{g}=-\frac{1}{2\beta \cal V}\sum_{\mathbf{k},i\omega_{n},j}\frac{1}{i\omega_{n}-E_{j\kv}}\frac{\partial E_{j\kv}}{\partial\Delta_0}
\ee
After summing over the Matsubara frequencies the gap equation, in the presence of spin-orbit interaction and Zeeman term, reads
\be
\frac{1}{g}=\frac{1}{4\cal V}\sum_\kv\left[
\frac{\tanh({\beta E_{1\kv}}/{2})}{E_{1\kv}}
\left(1-\frac{h^2}{\sqrt{\xi^2_\kv\gamma_{h\kv}^2+|\Delta_0|^2h^2}}\right)+
\frac{\tanh({\beta E_{2\kv}}/{2})}{E_{2\kv}}\left(1+\frac{h^2}{\sqrt{\xi^2_\kv \gamma_{h\kv}^2+|\Delta_0|^2h^2}}\right)\right].
\ee

\subsection{Number equation}
As seen in the previous case, from the full action we can not calculate the partition function exactly. The approximation scheme that we will  employ is based again on the expansion of the action in the gap field, as in Eq.~(\ref{expansion}). The zeroth order is equivalent to the mean field approximation. We will then include the fluctuations at the Gaussian level. 
\subsubsection{Mean field}
The effective action, at mean field level, is the action in Eq.~\eqref{Se0}, evaluated at the homogeneous solution of the gap equation, corresponding to the saddle point approximation. 
The associated grand canonical potential is simply $\Omega[{\Delta}_0,]=\beta S_{e}[\Delta_0]$. 
The mean number of particles, defined by $N=-\frac{\partial \Omega}{\partial \mu}$ is, therefore, given by
 \begin{equation}
	N=\frac{1}{2\beta}
	\frac{\partial}{\partial \mu}\sum_{\mathbf{k},i\omega_{n},j}\ln\left(i\omega_{n}-E_{j\kv}\right)
-\frac{\partial}{\partial \mu}\sum_{\mathbf{k}}\xik=
\sum_{\mathbf{k}}\left(1-\frac{1}{2\beta}\sum_{i\omega_{n},j}\frac{1}{i\omega_{n}-E_{j\kv}}\frac{\partial E_{j\kv}}{\partial \mu}\right)
 \end{equation}
Summing over the Matsubara frequencies, the number equation, at the mean-field level, reads
\be
N=\sum_{\kv}\left[1-
\frac{\tanh({\beta E_{1\kv}}/{2})}{2E_{1\kv}}
\left(\xi_{\kv}-\frac{\xi_{\kv}\gamma_{h\kv}^2}{\sqrt{\xi^2_\kv\gamma_{h\kv}^2+|\Delta_0|^2h^2}}\right)
-\frac{\tanh({\beta E_{2\kv}}/{2})}{2E_{2\kv}}
\left(\xi_{\kv}+\frac{\xi_{\kv}\gamma_{h\kv}^2}{\sqrt{\xi^2_\kv\gamma_{h\kv}^2+|\Delta_0|^2h^2}}\right)\right]
\ee
Let us consider the cases at $T=0$ and $T=T_c$. 

\paragraph*{At T=0.} 
In order to study the behavior of the gap and of the chemical potential as functions of the coupling, at $T=0$, the equations to solve at mean field level are
 \begin{eqnarray}
\hspace{-0.5cm}		\frac{1}{g}&=&\frac{1}{4\cal V}\sum_\kv\left[
\frac{1}{E_{1\kv}}
\left(1-\frac{h^2}{\sqrt{\xi^2_\kv\gamma_{h\kv}^2+|\Delta_0|^2h^2}}\right)+
\frac{1}{E_{2\kv}}\left(1+\frac{h^2}{\sqrt{\xi^2_\kv \gamma_{h\kv}^2+|\Delta_0|^2h^2}}\right)\right]\\
\hspace{-0.5cm}			n&=&\frac{1}{\cal V}\sum_{\kv}\left[1-
\frac{1}{2E_{1\kv}}
\left(\xi_{\kv}-\frac{\xi_{\kv}\gamma_{h\kv}^2}{\sqrt{\xi^2_\kv\gamma_{h\kv}^2+|\Delta_0|^2h^2}}\right)
-\frac{1}{2E_{2\kv}}
\left(\xi_{\kv}+\frac{\xi_{\kv}\gamma_{h\kv}^2}{\sqrt{\xi^2_\kv\gamma_{h\kv}^2+|\Delta_0|^2h^2}}\right)\right]
 \end{eqnarray}
 where $n=N/{\cal V}$ is the particle density. 
This set of equations, for $h=0$, 
has been solved in Ref.~\cite{Articolo_tesi2}. We will focus, instead, on the finite temperature case, in particular at the critical point.

\paragraph*{At T=$T_{c}$.}
If we want to calculate the critical temperature $T_c$, putting
by definition $\Delta_0(T_c)=0$, at the mean field level the equation to solve in terms of $T_c$ are the following
 \begin{eqnarray}
 \label{gapTc}
	\frac{1}{g}&=&\frac{1}{4\cal V}\sum_{\mathbf{k}}
	\left[
	\frac{ \tanh\left(\frac{\xik-\gamma_{h\kv}}{2T_{c}}\right)  }{\xik-\gamma_{h\kv}}\left(1-\frac{h^2}{\xi_\kv \gamma_{h\kv}}\right)
	+\frac{ \tanh\left(\frac{\xik+\gamma_{h\kv}}{2T_{c}}\right) }{\xik+\gamma_{h\kv}}\left(1+\frac{h^2}{\xi_\kv \gamma_{h\kv}}\right)
	\right]\\
	n&=&\frac{1}{\cal V}\sum_{\mathbf{k}}\left\{1
		-\frac{1}{2}\left[\tanh\left(\frac{\xik-\gamma_{h\kv}}{2T_{c}} \right)
		+\tanh\left(\frac{\xik+\gamma_{h\kv}}{2T_{c}} \right)\right]
			\right\}
	\label{mean_field_SO6}	
 \end{eqnarray}
These equations should be solved in order to obtain the critical temperature and the chemical potential, at $T_{c}$, as functions of the coupling $g$. 
The latter can be also written as 
\be
n=\frac{1}{\cal V}
\sum_{\kv}\left(1-\frac{\sinh(\xi_\kv/T_c)}{\cosh(\xi_\kv/T_c)+\cosh(\gamma_{h\kv}/T_c)}\right).
\ee

\subsubsection{Inclusion of Gaussian fluctuations at ${T=T_{c}}$}

The inadequacy of the mean field approach, also with SO coupling, in the strong coupling regime will be clarified later, 
and its reason is the same as before, based on the fact that the number equation in Eq.~(\ref{mean_field_SO6}) is that of non-interacting fermions.
It is, therefore, necessary to extend the analysis including Gaussian fluctuations around the mean field solution. 
We will get a second order term in the action, 
$	S_{e}[\Delta]=S_{e}^{(0)}+S_{e}^{(2)}$, 
in addition to $S_{e}^{(0)}$, the action of the non-interacting system but in the presence of spin-orbit and Zeeman terms. 
As already seen, the following action has already a quadratic term while the trace has to be expanded, 
 \begin{equation}   
    S_{e}[\Delta] =\int{d^{3}r}\int_{0}^{\beta}{d\tau}
                    \frac{|\Delta(\vec{r},\tau)|^{2}}{g} - \frac{1}{2}\Tr(\ln(\mathcal{G}^{-1}))
			+\beta\sum_{\mathbf{k}}\xik\,.
 \end{equation}
The inverse of the Green function can be written in momentum space and in the Matsubara frequency space as in Eq.~(\ref{G_so}). It can be written 
as 
\be
 \mathcal{G}^{-1}(\ks,\p)=\hat{G}^{-1}(\ks,\p)+\hat\Delta(\ks,\p)
\ee
where
 \begin{eqnarray}
	\hat{G}^{-1}(\ks,\p)&=&\begin{pmatrix}
	(i\omega_{n}+\xik+h) & 0 & \gamma(\kv) & 0\\
	0 & (i\omega_{n}-\xik+h) & 0 & \gamma(\kv)\\
	\gamma^*(\kv) & 0 & (i\omega_{n}+\xik-h) & 0\\
	0 & \gamma^*(\kv) & 0 & (i\omega_{n}-\xik-h)\\
	\end{pmatrix}\delta_{\ks,\p}\\
	\hat\Delta(\ks,\p)&=&\begin{pmatrix}
	0 & -\Delta\text{($\p$-$\ks$)} & 0 & 0\\
	-\Delta^{*}\text{($\ks$-$\p$)} & 0 & 0 & 0\\
	0 & 0 & 0 & \Delta\text{($\p$-$\ks$)}\\
	0 & 0 & \Delta^{*}\text{($\ks$-$\p$)} & 0\\
	\end{pmatrix}
 \end{eqnarray}
We can now perform the series expansion of the logarithmic function
 \begin{multline}
	\ln(\mathcal{G}^{-1})=
	\ln({G}^{-1}+\hat\Delta)=
	\ln\left({G}^{-1}(\mathbb{1}+{G}\hat\Delta)\right)=
	\ln({G}^{-1})+{G}\hat\Delta-\frac{1}{2}{G}\hat\Delta{G}\hat\Delta+O(\Delta^{3})
 \end{multline}
The first term gives 
the action of the non-interacting system in the presence of spin-orbit and Zeeman terms. The second term 
has a vanishing trace, as well as all the terms with odd powers, $\Tr(G\hat\Delta)^{2n+1}=0$. 
The action at second order in $\hat\Delta$ is, then, given by
 \begin{equation}   
 \label{Se_real}
    S_{e}[\Delta] =S_{e}^{(0)}+\int\!d^{3}r\!\int_{0}^{\beta}\!d\tau
                    \frac{|\Delta(\vec{r},\tau)|^{2}}{g} +\frac{1}{4}\Tr\left({G}\hat\Delta{G}\hat\Delta\right) + o(\Delta^{3})
 \end{equation}
The inverse of ${G}^{-1}$, the non-interacting Green function, in momentum space, is given by
 \begin{equation}
 \label{Gmatrix}
	{G}(\ks,\p)={G}(\ks)\,\delta_{\ks,\p}=
	\begin{pmatrix}
	\frac{i\omega_{n}+\xik-h}{(i\omega_{n}+\xik)^{2}-\gammahk^{2}} & 0 & \frac{-\gamma(\kv)}{(i\omega_{n}+\xik)^{2}-\gammahk^{2}} & 0\\
	0 & \frac{i\omega_{n}-\xik-h}{(i\omega_{n}-\xik)^{2}-\gammahk^{2}} & 0 & \frac{-\gamma(\kv)}{(i\omega_{n}-\xik)^{2}-\gammahk^{2}}\\
	\frac{-\gamma^*(\kv)}{(i\omega_{n}+\xik)^{2}-\gammahk^{2}} & 0 & \frac{i\omega_{n}+\xik+h}{(i\omega_{n}+\xik)^{2}-\gammahk^{2}} & 0\\
	0 & \frac{-\gamma^*(\kv)}{(i\omega_{n}-\xik)^{2}-\gammahk^{2}} & 0 & \frac{i\omega_{n}-\xik+h}{(i\omega_{n}-\xik)^{2}-\gammahk^{2}}\\
	\end{pmatrix}\delta_{\ks,\p}
 \end{equation}
Making the matrix product $G\hat\Delta G\hat\Delta$ and taking the trace we get
 \begin{equation}
 \label{TrGDGD}
\frac{1}{4}\Tr({G}\hat\Delta{G}\hat\Delta)=
\frac{1}{2}\Tr\big(
{G}_{22}\Delta^{*}{G}_{11}\Delta
-{G}_{24}\Delta^{*}{G}_{31}\Delta
+{G}_{44}\Delta^{*}{G}_{33}\Delta
-{G}_{42}\Delta^{*}{G}_{13}\Delta\big).
 \end{equation}
 where $G_{ij}$ are the matrix elements of $G$. In momentum space Eq.~(\ref{TrGDGD}) reads
   \begin{equation}
	\frac{1}{4}\Tr({G}\hat\Delta{G}\hat\Delta)=\sum_{\q}\Delta^{*}(\q)\,\chi({\q})\,\Delta(\q)
 \end{equation}
where we define the function
\begin{equation}
\label{chi1}
\chi({\q})=
	\frac{1}{2\beta\cal V}
		\sum_{\ks}\big[
		{G}_{11}(\ks){G}_{22}(\ks-\q)+{G}_{33}(\ks){G}_{44}(\ks-\q)
		-{G}_{13}(\ks){G}_{42}(\ks-\q)-{G}_{31}(\ks){G}_{24}(\ks-\q)
	\big]
 \end{equation}
The Fourier transform of the first term in Eq.~(\ref{Se_real}) is $\int\!d^{3}r\!\int_{0}^{\beta}\!d\tau \frac{|\Delta(\vec{r},\tau)|^{2}}{g}=\sum_\q \frac{|\Delta(\q)|^{2}}{g}$, therefore the action, at the second order in $\Delta$, namely at the Gaussian level, can be written as
 \begin{equation}
S_{e}[\Delta]=S_{e}^{(0)}+\sum_{\q}\Delta^{*}(\q) \, \Gamma^{-1}(\mathbf{\q}) \, \Delta(\q)	+ o(\Delta^{3})
 \end{equation}
where the same notation for $\Gamma^{-1}({\q})$, used for the case without SO coupling, has been adopted
\be
\Gamma^{-1}({\q})=\frac{1}{g}+\chi({\q})
\ee
but where now $\chi({\q})$ is more complicated than before. From Eqs.~(\ref{Gmatrix}) and (\ref{chi1}) we get
\be
\chi({\q})=\chi({\bf q},i\nu_n)=\frac{1}{\beta\cal V}\sum_{\kv,\omega_n}\frac{(i\omega_{n}+\xik)(i\omega_{n}+i\nu_m-\xi_\kmq)+h^2-\textrm{Re}[\gamma^*(\kv)\gamma(\kmq)]
}{\left((i\omega_{n}+\xik)^{2}-\gammahk^{2}\right)\left((i\omega_{n}-i\nu_n-\xi_\kmq)^{2}-\gamma_{h\hspace{0.01cm}\kmq}^{2}\right)}
\ee
It is convenient to make the shift $\kv\rightarrow \kv+\frac{\qv}{2}$ and, after performing the sum over the Matsubara frequencies, we obtain
\bea
\nonumber \chi(\qv, i\nu_n)=\frac{1}{4 \cal V}\sum_{\kv}\left\{
\Big(1+C^h_{\kv,\qv}\Big)\frac{1-\fermi\left(\xi_{\kv+\qv/2}-\gamma_{h\hspace{0.01cm}\kv+\qv/2}\right)-\fermi\left(\xi_{\kv-\qv/2}-\gamma_{h\hspace{0.01cm}\kv-\qv/2}\right)}{i\nu_n-\xi_{\kv+\qv/2}-\xi_{\kv-\qv/2}+\gamma_{h\hspace{0.01cm}\kv+\qv/2}+\gamma_{h\hspace{0.01cm}\kv-\qv/2}}\right.\\\nonumber\left.
+\Big(1+C^h_{\kv,\qv}\Big)\frac{1-\fermi\left(\xi_{\kv+\qv/2}+\gamma_{h\hspace{0.01cm}\kv+\qv/2}\right)-\fermi\left(\xi_{\kv-\qv/2}
+\gamma_{h\hspace{0.01cm}\kv-\qv/2}\right)}{i\nu_n-\xi_{\kv+\qv/2}-\xi_{\kv-\qv/2}-\gamma_{h\hspace{0.01cm}\kv+\qv/2}
-\gamma_{h\hspace{0.01cm}\kv-\qv/2}}\right.\\\nonumber
\left.+\Big(1-C^h_{\kv,\qv}\Big)\frac{1-\fermi\left(\xi_{\kv+\qv/2}-\gamma_{h\hspace{0.01cm}\kv+\qv/2}\right)-\fermi\left(\xi_{\kv-\qv/2}
+\gamma_{h\hspace{0.01cm}\kv-\qv/2}\right)}{i\nu_n-\xi_{\kv+\qv/2}-\xi_{\kv-\qv/2}+\gamma_{h\hspace{0.01cm}\kv+\qv/2}
-\gamma_{h\hspace{0.01cm}\kv-\qv/2}}\right.\\\left.
+\Big(1-C^h_{\kv,\qv}\Big)\frac{1-\fermi\left(\xi_{\kv+\qv/2}+\gamma_{h\hspace{0.01cm}\kv+\qv/2}\right)
-\fermi\left(\xi_{\kv-\qv/2}-\gamma_{h\hspace{0.01cm}\kv-\qv/2}\right)}{i\nu_n-\xi_{\kv+\qv/2}
-\xi_{\kv-\qv/2}-\gamma_{h\hspace{0.01cm}\kv+\qv/2}+\gamma_{h\hspace{0.01cm}\kv-\qv/2}}\right\}
\label{chi}
\eea
where 
\be
C^h_{\kv,\qv}=\frac{\textrm{Re}[\gamma^*(\kv+\frac{\qv}{2})\gamma(\kv-\frac{\qv}{2})]-h^2}{\gamma_{h\hspace{0.01cm}\kv+\qv/2}\,\gamma_{h\hspace{0.01cm}\kv-\qv/2}}
=
\frac{(\vR+\vD)^2 (k_y^2-\frac{q_y^2}{4})+(\vR-\vD)^2 (k_x^2-\frac{q_x^2}{4})-h^2}{\gamma_{h\hspace{0.01cm}\kv+\qv/2}\,\gamma_{h\hspace{0.01cm}\kv-\qv/2}}
\ee
with
\be
\gamma_{h\,\kv\pm \qv/2}=\sqrt{(\vR+\vD)^2 \left(k_y\pm{q_y}/{2}\right)^2+(\vR-\vD)^2 \left(k_x\pm{q_x}/{2}\right)^2+h^2}
\ee 
For $\vD=0$, namely with only the Rashba coupling, we have 
\be
C^h_{\kv,\qv}=\frac{|\kv_\perp|^2-|\qv_\perp/2|^2-(h/\vR)^2}{\sqrt{[|\kv_\perp-\qv_\perp/2|^2+(h/\vR)^2][|\kv_\perp+\qv_\perp/2|^2+(h/\vR)^2]}}
\ee
where $\kv_\perp=(k_x,k_y,0)$ are momenta perpendicular to the $z$-direction, the Zeeman direction. \\
For $\vR=\vD=v/2$, namely for equal Rashba and Dresselhaus couplings, we have, instead,
\be
C^h_{\kv,\qv}=\frac{v^2(k_y^2-q_y^2/4)-h^2}{\sqrt{[v^2(k_y-q_y/2)^2+h^2][v^2(k_y+q_y/2)^2+h^2]}}
\ee
which, for $h=0$, namely without the Zeeman field, it reduces to simply $C^0_{\kv,\qv}=1$.\\
Employing the symmetry $\kv\rightarrow -\kv$ under the sum we can simplify 
Eq.~(\ref{chi}) as it follows
\bea
\nonumber \chi(\qv, i\nu_n)=\frac{1}{4\cal V}\sum_{\kv}\left\{
\Big(1+C^h_{\kv,\qv}\Big)\left[
\frac{1-2\fermi\left(\xi_{\kv+\qv/2}-\gamma_{h\hspace{0.01cm}\kv+\qv/2}
\right)}{i\nu_n-\xi_{\kv+\qv/2}-\xi_{\kv-\qv/2}+\gamma_{h\hspace{0.01cm}\kv+\qv/2}+\gamma_{h\hspace{0.01cm}\kv-\qv/2}}\right.
\right.\\\nonumber\left.
+\frac{1-2\fermi\left(\xi_{\kv+\qv/2}+\gamma_{h\hspace{0.01cm}\kv+\qv/2}\right)}
{i\nu_n-\xi_{\kv+\qv/2}-\xi_{\kv-\qv/2}-\gamma_{h\hspace{0.01cm}\kv+\qv/2}
-\gamma_{h\hspace{0.01cm}\kv-\qv/2}}\right]
\\\left.
+\Big(1-C^h_{\kv,\qv}\Big)
\frac{2-2\fermi\left(\xi_{\kv+\qv/2}
-\gamma^h_{h\hspace{0.01cm}\kv+\qv/2}\right)-2\fermi\left(\xi_{\kv-\qv/2}
+\gamma^h_{h\hspace{0.01cm}\kv-\qv/2}\right)}{i\nu_n-\xi_{\kv+\qv/2}-\xi_{\kv-\qv/2}
+\gamma^h_{h\hspace{0.01cm}\kv+\qv/2}-\gamma^h_{h\hspace{0.01cm}\kv-\qv/2}|}
\right\}
\label{chi2}
\eea
or, in terms of the hyperbolic tangent, 
\bea
\nonumber\chi(\qv,i\nu_n)=\frac{1}{4\cal V}\sum_\kv\left\{\Big(1+C^h_{\kv,\qv}\Big)
\left[\frac{\tanh\left(\frac{\xi_{\kv+\qv/2}-\gamma_{h\hspace{0.01cm}\kv+\qv/2}}{2T}\right)}
{i\nu_n-\xi_{\kv+\qv/2}-\xi_{\kv-\qv/2}+\gamma_{h\hspace{0.01cm}\kv+\qv/2}+\gamma_{h\hspace{0.01cm}\kv-\qv/2}}\right.\right.
\\\nonumber\left.+\frac{\tanh\left(\frac{\xi_{\kv+\qv/2}+\gamma_{h\hspace{0.01cm}\kv+\qv/2}}{2T}\right)}
{i\nu_n-\xi_{\kv+\qv/2}-\xi_{\kv-\qv/2}-\gamma_{h\hspace{0.01cm}\kv+\qv/2}-\gamma_{h\hspace{0.01cm}\kv-\qv/2}}\right]
\\\left.
+\Big(1-C^h_{\kv,\qv}\Big)
\frac{\tanh\left(\frac{\xi_{\kv+\qv/2}-\gamma_{h\hspace{0.01cm}\kv+\qv/2}}{2T}\right)+
\tanh\left(\frac{\xi_{\kv-\qv/2}+\gamma_{h\hspace{0.01cm}\kv-\qv/2}}{2T}\right)}
{i\nu_n-\xi_{\kv+\qv/2}-\xi_{\kv-\qv/2}+\gamma_{h\hspace{0.01cm}\kv+\qv/2}-\gamma_{h\hspace{0.01cm}\kv-\qv/2}}
\right\}
\label{chi3}
\eea
We found, therefore, that $\Gamma^{-1}(\q)$ is much more complicated that that obtained in the previous standard case without spin-orbit coupling. However, the partition function has the same form as that shown in Eq~(\ref{partition_func_free}), and, therefore, it is possible to apply the same procedure seen before (in Sec.~\ref{Inclusione delle fluttuazioni Gaussiane free}) to find the grand canonical potential and the density of particles. This is again due to the fact that the analytical continuation of $\Gamma^{-1}(\mathbf{q},z)$ has only a branch-cut of simple poles in the real axis. The density of particles is, then, given by
\be
\label{eqn0n2}
n=n^{(0)}+n^{(2)}
\ee
where $n^{(0)}$ is the mean field contribution as in Eq.~(\ref{mean_field_SO6}) while the second order term is given by
 \begin{equation}
 \label{eqn2}
	n^{(2)}=\frac{1}{\beta\cal V}\frac{\partial}{\partial\mu}\Tr(\ln(\Gamma^{-1}))=
	\frac{1}{\pi\cal V}\sum_{\mathbf{q}}\int_{-\infty}^{+\infty}d\omega\ \bose(\omega)\frac{\partial\delta(\mathbf{q},\omega)}{\partial\mu}
 \end{equation}
where the phase shift $\delta(\mathbf{q},\omega)$ is defined as before
 \begin{equation}
 \label{eqdelta}
	\delta(\mathbf{q},\omega)=\lim_{\epsilon\rightarrow0} \textrm{Arg}\left(\Gamma(\mathbf{q},\omega+ i\epsilon)^{-1}\right)
 \end{equation}
 after a Wick rotation $i\nu_n\rightarrow \omega+i\epsilon$.

\subsection{Critical point}
Taking the static and homogeneous limit of Eq.~(\ref{chi3}), 
after noticing that, at $\qv=0$, 
\be
C^h_{\kv,0}= 1-\frac{2h^2}{\gamma_{h \kv}^2}
\ee 
we have that, at $T=T_c$, $\chi(0,0)$ is simply given by
\bea
\nonumber\chi(0,0)&=&
-\frac{1}{4\cal V}
\sum_{\kv}\left\{\left(1-\frac{h^2}{\gamma_{h\kv}^2}\right)\left[
\frac{\tanh\left(\frac{\xi_\kv-\gamma_{h\kv}}{2T_c}\right)}{\xi_{\kv}-\gamma_{h\kv}}+\frac{\tanh\left(\frac{\xi_\kv+\gamma_{h\kv}}{2T_c}\right)}{\xi_{\kv}+\gamma_{h\kv}}\right]
\right.\\&&
+\left.\frac{h^2}{\gamma_{h\kv}^2}\left[
\frac{\tanh\left(\frac{\xi_\kv-\gamma_{h\kv}}{2T_c}\right)+\tanh\left(\frac{\xi_\kv+\gamma_{h\kv}}{2T_c}\right)}
{\xi_{\kv}}\right]\right\}
\eea
We find that the equation $\Gamma^{-1}(0)=0$, namely 
\be
\label{gapTc2}
\frac{1}{g}=-\chi(0,0)
\ee
is exactly the same equation reported in Eq.~(\ref{gapTc}) after reshuffling the terms.
The set of equations that has to be solved, at the Gaussian level, in order to find $T_c$ and $\mu$ at $T_c$ is given by Eq.~(\ref{gapTc2}), together with Eq.~(\ref{eqn0n2}).

\section{BCS-BEC crossover with spin-orbit coupling}

For the same reason discussed before, the introduction of a contact potential gives rise to divergences that can be regularized by the same  procedure. For simplicity we will focus on the case without Zeeman term, $h=0$. 
As shown in Appendix \ref{Rinormalizzazione nel caso spin-orbita} and in Refs.~\cite{Zhang1,Zhang2}, 
the presence of a SO coupling does not alter the form of the renormalization condition for a contact pseudo-potential. The only caution to be taken is to replace $\as$ with $\ar$, a scattering length which depends both on the interatomic potential and on the SO term
\begin{equation}
\frac{m}{4\pi \ar}=-\frac{1}{g}+\frac{1}{\cal V}\sum_{\mathbf{k}}\frac{1}{2\epsilon_{\mathbf{k}}}  
\label{renorm_sostitution}
\end{equation}
therefore, the arguments used to implement the BCS and BEC limits in terms of the scattering length remains unchanged.

\subsection{Mean field}
At the mean field level the equations to solve along the crossover are  Eqs.~(\ref{gapTc}), (\ref{mean_field_SO6}). Implementing Eq.~(\ref{renorm_sostitution}), for $h=0$, the equations useful to derive the critical temperature $T_c$ and the chemical potential $\mu$ within the mean field theory become
 \begin{eqnarray}
	\label{gap_so6}
	-\frac{m}{4\pi\ar}=&  \frac{1}{4\cal V}\sum_{\mathbf{k}}
	\left\{
	\frac{ \tanh(\frac{\xik-\gammak}{2T_{c}})  }{\xik-\gammak}
	+\frac{ \tanh(\frac{\xik+\gammak}{2T_{c}}) }{\xik+\gammak}
	-\frac{2}{\epsilon_{\mathbf{k}}}
	\right\}\\
	n=\frac{1}{\cal V}\sum_{\mathbf{k}}&\left\{1
		-\frac{1}{2}\tanh\left(\frac{\xik-\gammak}{2T_{c}} \right)
		-\frac{1}{2}\tanh\left(\frac{\xik+\gammak}{2T_{c}} \right)
			\right\}	
	\label{mean_field_SO7}
 \end{eqnarray}
 where $\gammak\equiv\gamma_{0\kv}$, as reported in Eq.~(\ref{gammahk}) with $h=0$.
Also in this case the number equation is the same as that of free fermions, therefore we expect that the mean field level will fail in correctly reproducing the thermodynamic quantities in the strong coupling limit.\\
Let us consider Eq.~(\ref{gap_so6}). Contrary to the case without SO, the mass cannot be eliminated from the equations by simply redefining the momenta, but it can be incorporated in the couplings $\tvd=\frac{\vd}{\vf}=\frac{m\vd}{\kf}$ and similarly for $\vR$.
In the continuum limit, with the natural change of variables and substitutions
	$\vec{p}=\frac{1}{\kf}\vec{k}$, $\tmu=\frac{\mu}{\Ef}$, 
	$\tTc=\frac{T_{c}}{\Ef}$, 
	$\tvr=\frac{\vR}{\vf}$,  
	$\tvd=\frac{\vd}{\vf}$,  
	$\tilde{\gamma}_\mathbf{p}(\tvr,\tvd)=2\gamma_\mathbf{k}(\vR, \vd)$, 
the gap equation can be written as
 \begin{equation}
	-\frac{m}{4\pi\ar}
	=\frac{\kf^{3}}{4(2\pi)^{3}\Ef}\int d^{3}p
	\left\{
	\frac{  \tanh( \frac{p^{2}-\tmu-\tgammap}{2\tTc} ) }
	     {p^{2}-\tmu-\tgammap}
	+\frac{ \tanh(\frac{p^{2}-\tmu+\tgammap}{2\tTc}) }
	      {p^{2}-\tmu+\tgammap}
	-\frac{2}{p^{2}}
	\right\}
	\label{gap_continuum}
 \end{equation}
 while the number equation, Eq.~(\ref{mean_field_SO7}), becomes
  \begin{equation}
  n= \kf^3 \int \frac{d^{3}p}{(2\pi)^3}\left[1
		-\frac{1}{2}\tanh\left(\frac{p^{2}-\tmu-\tgamma_{\mathbf{p}}}{2\tTc}\right)
		-\frac{1}{2}\tanh\left(\frac{p^{2}-\tmu+\tgamma_{\mathbf{p}}}{2\tTc}\right)
			\right].
  \end{equation}

\subsubsection{Weak coupling}
At low temperature we expect that the chemical potential of a system of weakly interacting fermions is almost constant 
 \begin{equation}
	\mu\underset{T\rightarrow0}{\approx}\Efso(\Ef,\vR,\vd)
	\label{ansatz_chem}
 \end{equation}
and therefore, in this regime, $\frac{\mu}{T}\gg1$. Actually, one can shown perturbativly \cite{Articolo_tesi2,luca2} that $\mu\approx \Ef-{m}(\vR+\vd)^2/2$. In this limit, the number equation can be neglected due to the ansatz Eq.~(\ref{ansatz_chem}), therefore   Eq.~(\ref{gap_continuum}) is enough to self-consistently determine the critical temperature $T_c$ as a function of $\ar$. 
As we will see, the critical temperature, in the BCS part of the crossover, exhibits an exponential behavior as in the case without SO, however $T_c$ will be considerably improved by turning on the spin-orbit coupling. We will also see that, in this weak coupling limit $T_c$ is almost unaltered by the inclusion of Gaussian fluctuations.

\subsubsection{Strong coupling}
In the strong coupling limit, supposing that $\mu/T_{c}\rightarrow-\infty$, which can be verified {\it a posteriori},  Eq.~(\ref{gap_continuum}) can be used to determine the chemical potential of the system, in fact the dependence in the equations on the critical temperature disappears making the following approximation
 \begin{equation}
	\tanh\left( \frac{p^{2}-\tmu-\tgammap}{2\tTc} \right)\approx1
 \end{equation}
so that Eq.~(\ref{gap_continuum}), after introducing the dimensionless coupling $y=\frac{1}{\kf\ar}$, becomes simply
  \begin{equation}
	y=-\frac{2}{\pi^{2}}\int_{0}^{\infty}dp_x\int_{0}^{\infty}dp_y\int_{0}^{\infty}dp_z
	\left\{
	\frac{1}{p^{2}-\tmu-\tgammap}
	+\frac{1}{p^{2}-\tmu+\tgammap}
	-\frac{2}{p^{2}}
	\right\}
	\label{mean_field_strong_coup}
 \end{equation}
where
  \begin{equation}
	\tgammap=2\sqrt{(\tvr+\tvd)^{2}\,p_{x}^{2}+(\tvr-\tvd)^{2}\,p_{y}^{2}}
 \end{equation}
We can integrate over $p_z$ getting
 \begin{equation}
	y=-\frac{1}{\pi}\int_{0}^{\infty}dp_x\int_{0}^{\infty}dp_y
	\left\{
	\frac{1}{\sqrt{p_x^{2}+p_y^{2}-\tmu-\tgammap}}
	+\frac{1}{\sqrt{p_x^{2}+p_y^{2}-\tmu+\tgammap}}
	-\frac{2}{\sqrt{p_x^{2}+p_y^{2}}}
	\right\}
	\label{mean_field_strong_coup2}
 \end{equation}
For simplicity, let us consider the case with pure Rashba coupling, so that $\tgammap=2\tvr\sqrt{p_{x}^{2}+p_{y}^{2}}\equiv 2\tvr |p_{\perp}|$. 
The above equation can be written as 
 \begin{equation}
	y=-\frac{1}{2}\int_{0}^{\infty}dp_{\perp}p_{\perp}
	\left\{
	\frac{1}{\sqrt{p_{\perp}^{2}-2\tvr p_{\perp}-\tmu}}+
		\frac{1}{\sqrt{p_{\perp}^{2}+2\tvr p_{\perp}-\tmu}}
	-\frac{2}{p_{\perp}}
	\right\}
 \end{equation}
With the substitutions $p_{\perp}\rightarrow p_{\perp}\pm\tvr$ in the first two integrals we have
\be
\label{eq>}
	y=-\frac{1}{2}\lim_{\Lambda\rightarrow \infty}\left\{\int_{-\tvr}^{\Lambda}dp_{\perp}
	\frac{p_{\perp}+\tvr}{\sqrt{p_{\perp}^{2}-\tvr^2-\tmu}}+
	\int_{\tvr}^{\Lambda}dp_{\perp}\frac{p_{\perp}-\tvr}{\sqrt{p_{\perp}^{2}-\tvr^2-\tmu}}
	-2\int_{0}^{\Lambda}dp_{\perp}
	\right\}
\ee
where $\Lambda$ is the ultraviolet cut-off which can be sent to infinity, $\Lambda\rightarrow \infty$. Eq.~(\ref{eq>}) can be rewritten as
\be
\label{eq>>}
	y=\lim_{\Lambda\rightarrow \infty}\left\{
	\Lambda-\int_{\tvr}^{\Lambda}dp_{\perp}
	\frac{p_{\perp}}{\sqrt{p_{\perp}^{2}-\tvr^2-\tmu}}\right\}-
	\frac{\tvr}{2}\int_{-\tvr}^{\tvr}dp_{\perp}\frac{1}{\sqrt{p_{\perp}^{2}-\tvr^2-\tmu}}
\ee
which gives 
\be
\label{mu_sc_vr}
y=\sqrt{-\tmu}-\tvr\,\arctanh\left(\frac{\tvr}{\sqrt{-\tmu}}\right)\approx \sqrt{-\tmu}-\frac{\tvr^2}{\sqrt{-\tmu}}
\ee
Taking the square, $y^2\approx -\tmu-2\tvr^2$, and using the definitions $\tmu=\mu/\Ef$ and $\tvr=\vR/\vf$, reminding that $y=\frac{1}{\kf\ar}$, we get
\be
\label{muSOsc}
\mu= -\frac{1}{2m\ar^2}-m\vR^2+o(a_r)
\ee
Therefore, in the strong coupling limit (BEC limit), $1/\ar\rightarrow \infty$, we found $\mu\simeq -\frac{1}{2m\ar^{2}}$, namely the spin-orbit interaction can be neglected and we get the same expression for the chemical potential as that obtained previously in Eq.~(\ref{chem_pot_strong}), but with $\ar$ instead of $\as$. 
We expect that also in this case the critical temperature in the strong coupling limit at the mean field level is not accurate therefore we will not discuss it here in details. 

\subsubsection{Full crossover}
Let us consider Eq.~(\ref{gap_so6}), whose form, in the continuum limit and with rescaled dimensionless parameters, 
	$\vec{p}=\frac{1}{\kf}\vec{k}$, $\tmu=\frac{\mu}{\Ef}$, $\tTc=\frac{T_{c}}{\Ef}$, $\tvr=\frac{\vR}{\vf}$,  $\tvd=\frac{\vd}{\vf}$,  
	$\tilde{\gamma}_\mathbf{p}(\tvr,\tvd)=2\gamma_\mathbf{k}(\vR, \vd)$, is reported in Eq.~(\ref{gap_continuum}), which can be rewritten in terms of $y=\frac{1}{\kf\ar}$ as it follows
 \begin{equation}
 \label{y_eq0}
	y=
	-\int \frac{d^{3}p}{(2\pi)^2}
	\left[
	\frac{  \tanh( \frac{p^{2}-\tmu-\tgammap}{2\tTc} ) }
	     {p^{2}-\tmu-\tgammap}
	+\frac{ \tanh(\frac{p^{2}-\tmu+\tgammap}{2\tTc}) }
	      {p^{2}-\tmu+\tgammap}
	-\frac{2}{p^{2}}
	\right]\equiv 
	I_{y}^{\textsc{so}}(\tmu,\tTc,\tvr,\tvd)
 \end{equation}
The number equation, Eq.~(\ref{mean_field_SO7}), in the continuum limit and in terms of the same dimensionless parameters, is 
 \begin{eqnarray}
\nonumber n
	&=&
	\frac{\kf^{3}}{3\pi^2}\left\{3\pi^2\int \frac{d^{3}p}{(2\pi)^3}\left[1
		-\frac{1}{2}\tanh\left(\frac{p^{2}-\tmu-\tgamma_{\mathbf{p}}}{2\tTc}\right)
		-\frac{1}{2}\tanh\left(\frac{p^{2}-\tmu+\tgamma_{\mathbf{p}}}{2\tTc}\right)
			\right]\right\}\\
	&\equiv&\frac{\kf^{3}}{3\pi^{2}}\,I_{n}^{\textsc{so}}(\tmu,\tTc,\tvr,\tvd)
	\label{n_eq0}
 \end{eqnarray}
Reminding that the density of particles is $n=\frac{(2m\Ef)^{3/2}}{3\pi^{2}}=\frac{\kf^{3}}{3\pi^{2}}$, the equations to solve are
 \begin{eqnarray}
 \label{y_eq}
 	y=I_{y}^{\textsc{so}}(\tmu,\tTc,\tvr,\tvd)\\
	1=I_n^{\textsc{so}}(\tmu,\tTc,\tvr,\tvd)
	 \label{n_eq}
 \end{eqnarray}
\begin{figure}[h]
\centering
\includegraphics[width=0.45\textwidth]{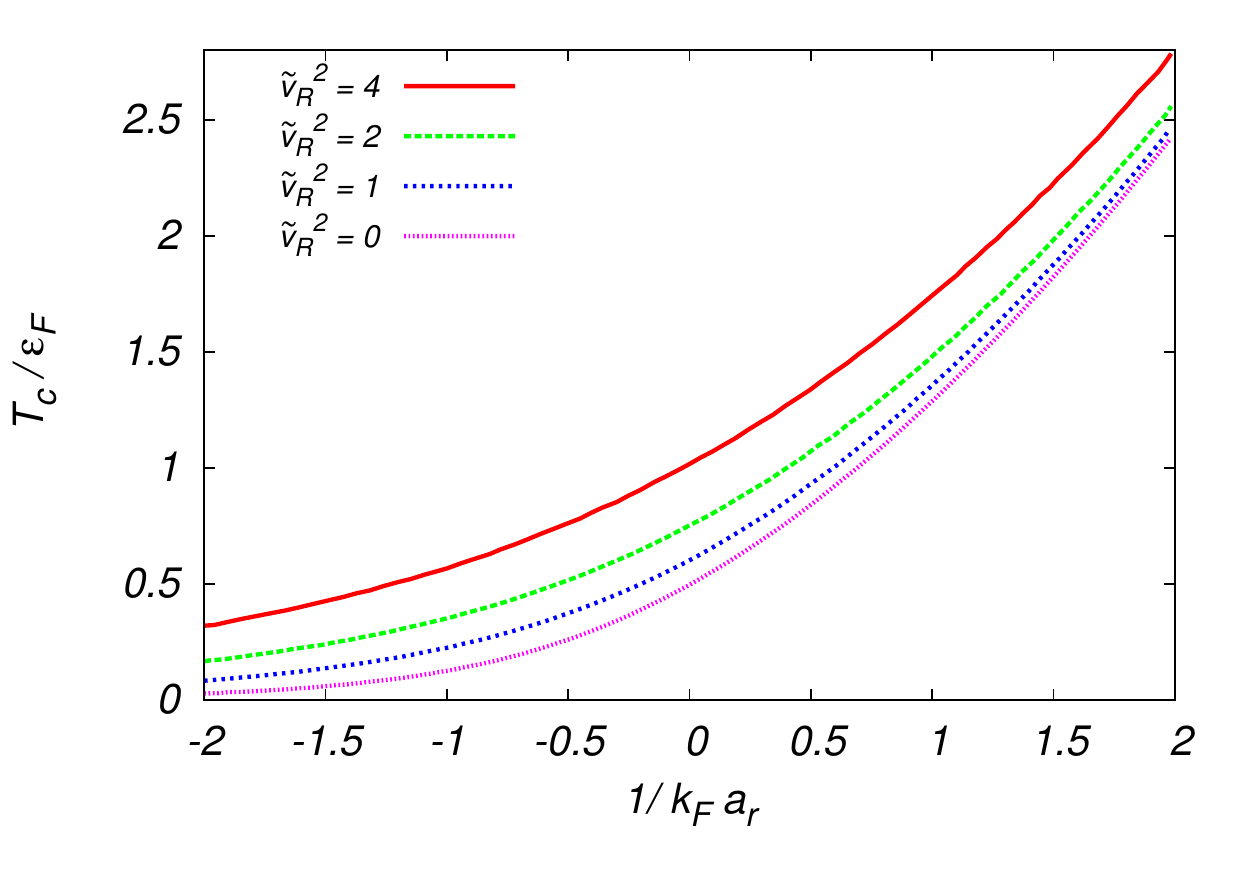}
\includegraphics[width=0.45\textwidth]{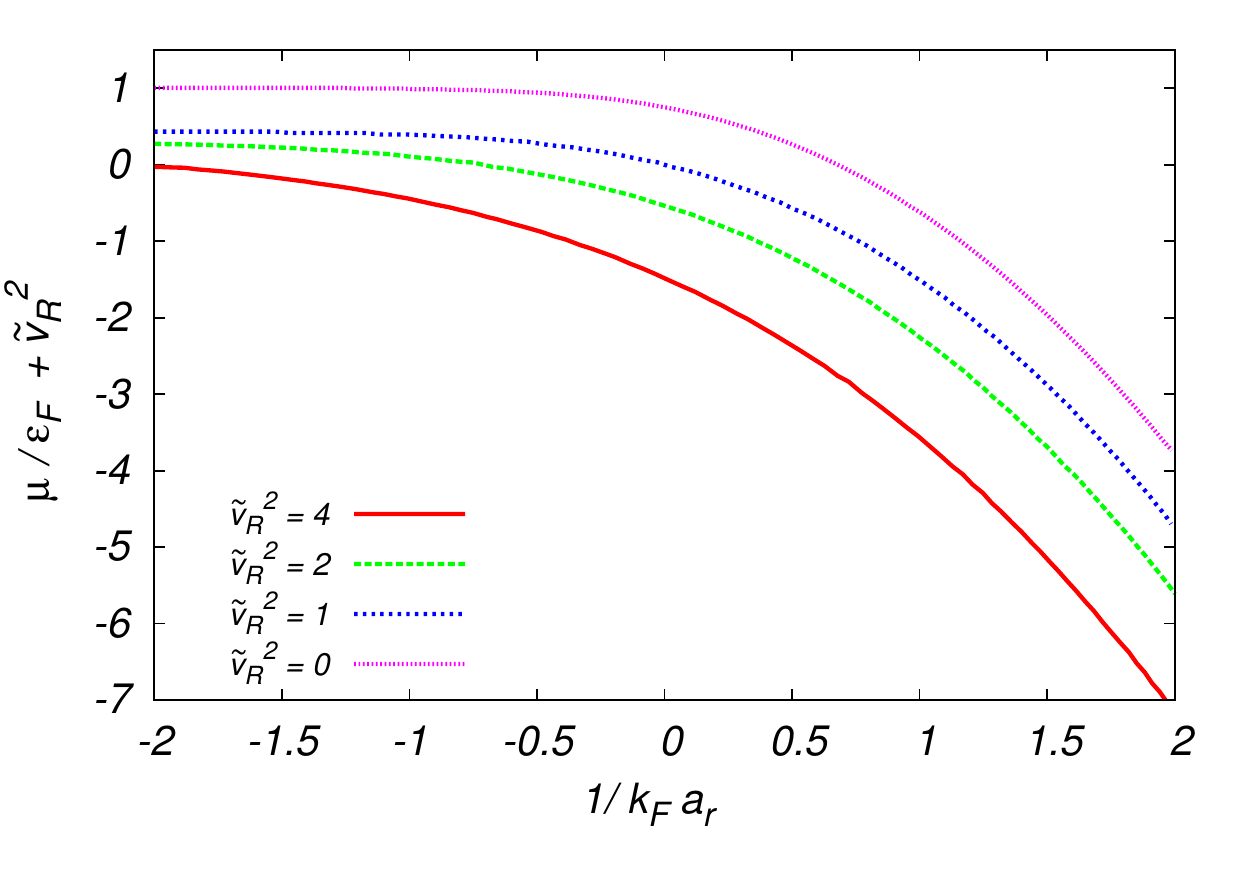}
\caption{Critical temperature $T_c$ and chemical potential $\mu$ as functions of $y$, at the mean field level, for $\tvd=0$ and for different values of $\tvr$.}
\label{fig.meanfieldSO1}
\end{figure}
\begin{figure}[h]
\centering
\includegraphics[width=0.45\textwidth]{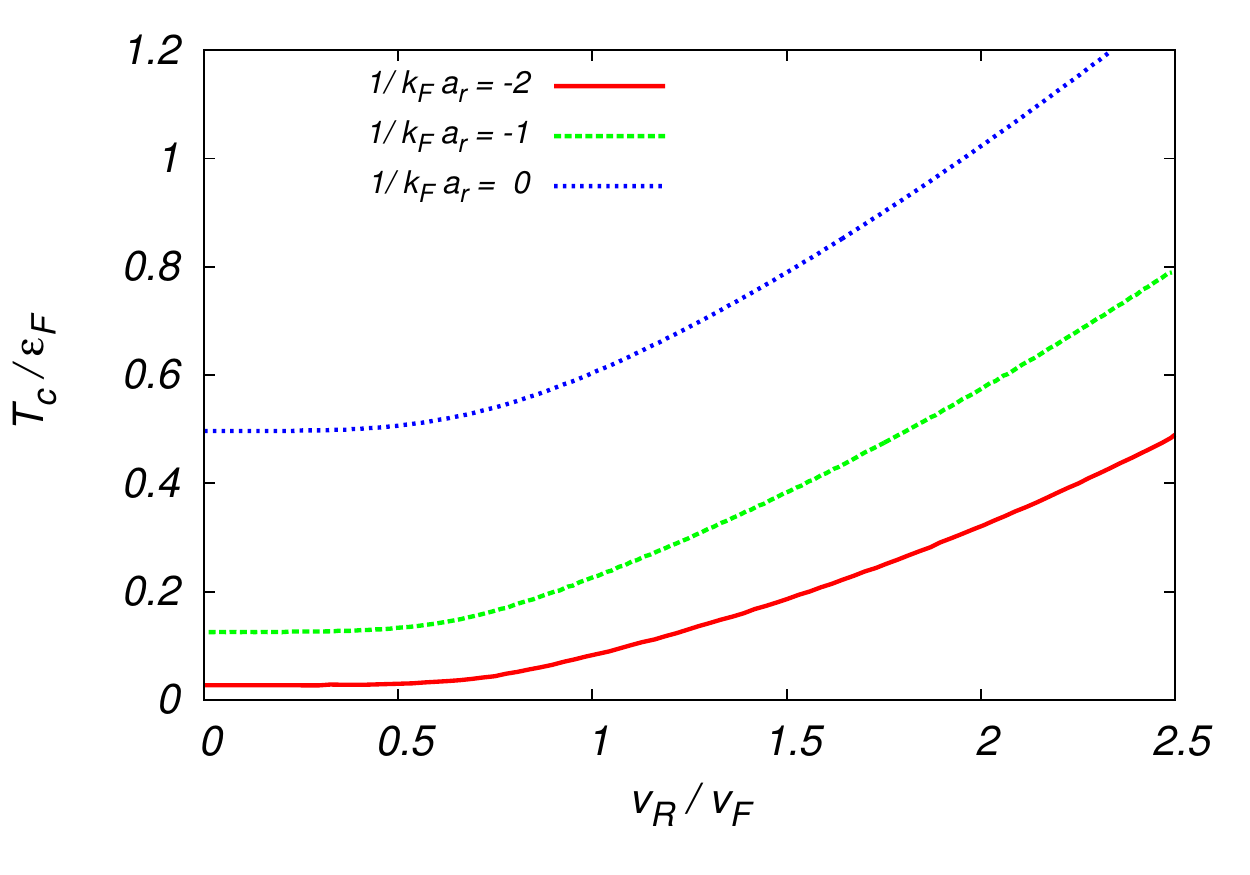}
\caption{Critical temperature $T_c$ as a function of $\vR$, at $\tvd=0$ and for some values of $\frac{1}{\kf \ar}$.}
\label{fig.meanfieldSO2}
\end{figure}
We can search for the solutions of these two equations, Eqs.~(\ref{y_eq}) and (\ref{n_eq}), resorting to a multidimensional root-finding algorithm: 
for any fixed values of the couplings $y$, $\tvr$ and $\tvd$ we look for the critical temperature $T_c$ and the chemical potential $\mu$ that satisfy 
those equations simultaneously. 
In Figs. \ref{fig.meanfieldSO1} and \ref{fig.meanfieldSO2} we report some results for the case with only Rashba coupling.

\paragraph*{Special case: equal Rashba and Dresselhaus couplings.} For $\vR=\vd\equiv v/2$ we have $\gammak=v|k_y|$, or, in terms of the rescaled quantities $p_y=k_y/\kf$ and $\tilde v=v/\vf$, 
\be
\tgammap=2 \tilde v |p_y|
\ee
Both in Eq.~(\ref{y_eq0}) and in Eq.~(\ref{n_eq0}) there are functions like ${F}(p_y^2\pm 2v|p_y|)$ under the integral
\be
\frac{1}{2}\int_{-\infty}^{+\infty}dp_y \left[F\left(p_y^2-2 \tilde v |p_y|-\tmu\right)+F\left(p_y^2+2 \tilde v |p_y|-\tmu\right)\right]
\ee
in a symmetric combination so that we can remove the absolute value symbols 
\be
\frac{1}{2}\int_{-\infty}^{+\infty}dp_y \left[F\left(p_y^2- 2\tilde v p_y-\tmu\right)+F\left(p_y^2+ 2\tilde v p_y-\tmu\right)\right]
\ee
so that after changing the variables $p_y \rightarrow p_y\pm \tilde v$ we get simply
\be
 \int_{-\infty}^{+\infty}dp_y \,F\left(p_y^2- \tilde{v}^2-\tmu\right)
\ee
As a general result, for the case with equal Rashba and Dresselhaus couplings at the mean field level the effect of the spin-orbit interaction is just 
a rigid shift of the chemical potential. The critical temperature $T_c$, is therefore the same as that without SO (see Fig.~\ref{fig.meanfield}) while $\mu$ at finite $v$ is linked to that without SO ($v=0$) in the following way
\be
\frac{\mu(v)}{\Ef}= \frac{\mu(0)}{\Ef}-\frac{v^2}{\vf^2}
\ee
which means $\mu(v)=\mu(0)-mv^2/2$, in agreement with the result obtained by gauge-transforming the fermionic fields \cite{luca2}.

\subsection{With Gaussian fluctuations}
Let us now recap the general results with the inclusion of Gaussian fluctuations reporting the full set of equations, Eqs.~(\ref{gapTc}), (\ref{mean_field_SO6}) with Eqs.~(\ref{chi3})-(\ref{eqdelta}) together with the regularization Eq.~(\ref{renorm_sostitution}), useful to derive the chemical potential $\mu$ and the critical temperature $T_c$ as functions of the effective interaction parameter, the inverse scattering length,
 \be
 \label{gapTcnew}
	\frac{m}{4\pi\ar}=-\frac{1}{4\cal V}\sum_{\mathbf{k}}
	\left[
	\frac{ \tanh\left(\frac{\xik-\gamma_{h\kv}}{2T_{c}}\right)  }{\xik-\gamma_{h\kv}}\left(1-\frac{h^2}{\xi_\kv \gamma_{h\kv}}\right)
	+\frac{ \tanh\left(\frac{\xik+\gamma_{h\kv}}{2T_{c}}\right) }{\xik+\gamma_{h\kv}}\left(1+\frac{h^2}{\xi_\kv \gamma_{h\kv}}\right)
	-\frac{2}{\epsilon_{\mathbf{k}}}
	\right]
\ee
\be
	n=\frac{1}{\cal V}\sum_{\mathbf{k}}\left\{1
		-\frac{1}{2}\left[\tanh\left(\frac{\xik-\gamma_{h\kv}}{2T_{c}} \right)
		+\tanh\left(\frac{\xik+\gamma_{h\kv}}{2T_{c}} \right)\right]\right\}
		+\frac{1}{\cal V}
		\sum_{\mathbf{q}}
		\int_{-\infty}^{+\infty}\frac{d\omega}{\pi}\ \bose(\omega)\frac{\partial\delta(\mathbf{q},\omega)}{\partial\mu}
	\label{mean_field_SOnew}	
 \ee
with
\be
\gamma_{h \kv}=\sqrt{(\vR+\vD)^2 k_y^2+(\vR-\vD)^2 k_x^2+h^2}
\ee
 \bea
 \label{delta_new}
	\delta(\mathbf{q},\omega)&=&\textrm{Arg}\left(\Gamma^{-1}(\mathbf{q},\omega+i0^{+})\right)\\
	\Gamma^{-1}(\qv,i\nu_n)&=&-\frac{m}{4\pi\ar}+\chi(\qv,i\nu_n)+\frac{1}{\cal V}\sum_{\mathbf{k}}\frac{1}{2\epsilon_{\mathbf{k}}}
	\label{gamma_222}
\eea
\bea
\nonumber\chi(\qv,i\nu_n)=\frac{1}{4\cal V}\sum_\kv\left\{\Big(1+C^h_{\kv,\qv}\Big)
\left[\frac{\tanh\left(\frac{\xi_{\kv+\qv/2}-\gamma_{h\hspace{0.01cm}\kv+\qv/2}}{2T_c}\right)}
{i\nu_n-\xi_{\kv+\qv/2}-\xi_{\kv-\qv/2}+\gamma_{h\hspace{0.01cm}\kv+\qv/2}+\gamma_{h\hspace{0.01cm}\kv-\qv/2}}\right.\right.\\
\nonumber\left.+\frac{\tanh\left(\frac{\xi_{\kv+\qv/2}+\gamma_{h\hspace{0.01cm}\kv+\qv/2}}{2T_c}\right)}
{i\nu_n-\xi_{\kv+\qv/2}-\xi_{\kv-\qv/2}-\gamma_{h\hspace{0.01cm}\kv+\qv/2}-\gamma_{h\hspace{0.01cm}\kv-\qv/2}}\right]\\
\left.+\Big(1-C^h_{\kv,\qv}\Big)
\frac{\tanh\left(\frac{\xi_{\kv+\qv/2}-\gamma_{h\hspace{0.01cm}\kv+\qv/2}}{2T_c}\right)+
\tanh\left(\frac{\xi_{\kv-\qv/2}+\gamma_{h\hspace{0.01cm}\kv-\qv/2}}{2T_c}\right)}
{i\nu_n-\xi_{\kv+\qv/2}-\xi_{\kv-\qv/2}+\gamma_{h\hspace{0.01cm}\kv+\qv/2}-\gamma_{h\hspace{0.01cm}\kv-\qv/2}}
\right\}
\label{chinew}
\eea
where
\be
\label{Chkq}
C^h_{\kv,\qv}=\frac{(\vR+\vD)^2 (k_y^2-\frac{q_y^2}{4})+(\vR-\vD)^2 (k_x^2-\frac{q_x^2}{4})-h^2}{\gamma_{h\hspace{0.01cm}\kv+\qv/2}\,\gamma_{h\hspace{0.01cm}\kv-\qv/2}}
\ee
We will consider for simplicity the case with $h=0$ but let us first discuss two special cases. 

\subsubsection{Special case: equal Rashba and Dresselhaus}
Let us consider the case where $h=0$ and $\vR=\vD$, namley when Rashba and Dresselhauf couplings are equal. The equations are
\be
 \label{gapTcERD}
	\frac{m}{4\pi\ar}=-\frac{1}{4\cal V}\sum_{\mathbf{k}}
	\left[
	\frac{ \tanh\left(\frac{\xik-\gamma_{\kv}}{2T_{c}}\right)  }{\xik-\gamma_{\kv}}
	+\frac{ \tanh\left(\frac{\xik+\gamma_{\kv}}{2T_{c}}\right) }{\xik+\gamma_{\kv}}
	-\frac{2}{\epsilon_{\mathbf{k}}}
	\right]
\ee
\be
	n=\frac{1}{\cal V}\sum_{\mathbf{k}}\left\{1
		-\frac{1}{2}\left[\tanh\left(\frac{\xik-\gamma_{\kv}}{2T_{c}} \right)
		+\tanh\left(\frac{\xik+\gamma_{\kv}}{2T_{c}} \right)\right]\right\}
		+\frac{1}{\cal V}
		\sum_{\mathbf{q}}
		\int_{-\infty}^{+\infty}\frac{d\omega}{\pi}\ \bose(\omega)\frac{\partial\delta(\mathbf{q},\omega)}{\partial\mu}
	\label{SOERD}	
 \ee
where $\gamma_{\kv}=\gamma_{0\kv}$ and since $\vR=\vd$,
\be
\gamma_{\kv}=(\vR+\vD) |k_y|\equiv v |k_y|
\ee
while $\delta(\qv,\omega)$ is defined in Eq.~(\ref{delta_new}). Since $C^0_{\kv,\qv}=1$, we have that $\chi(\qv,i\nu_n)$ is simply given by
\be
\chi(\qv,i\nu_n)=\frac{1}{2\cal V}\sum_\kv
\left[\frac{\tanh\left(\frac{\xi_{\kv+\qv}-\gamma_{\kv+\qv}}{2T_c}\right)}
{i\nu_n-\xi_{\kv+\qv}-\xi_{\kv}+\gamma_{\kv+\qv}+\gamma_{\kv}}
+\frac{\tanh\left(\frac{\xi_{\kv+\qv}+\gamma_{\kv+\qv}}{2T_c}\right)}
{i\nu_n-\xi_{\kv+\qv}-\xi_{\kv}-\gamma_{\kv+\qv}-\gamma_{\kv}}
\right]
\label{chiERD}
\ee
As already discussed, at the mean field level, after rescaling $k_y\rightarrow k_y\pm mv$ we obtain Eqs.~(\ref{gap_eq}) and (\ref{numero}), namely the mean-field equations without SO, but where the chemical potential is $\mu+mv^2/2$, so that Eqs.~(\ref{gapTcERD}), (\ref{SOERD}) can be simplified as it follows
 \be
 \label{gap_ERD}
\frac{m}{4\pi \as}=-\frac{1}{2\cal V}\sum_{\mathbf{k}}\left[  \frac{\tanh\left(\frac{(\xi_{\mathbf{k}}-{mv^2}/{2})}{2T_{c}}\right)}{\xi_{\mathbf{k}}-{mv^2}/{2}} - \frac{1}{\epsilon_{\mathbf{k}}} \right]
\ee
\be
n=
\frac{1}{\cal V}\sum_{\mathbf{k}}\left[ 1-\tanh\left(  \frac{\xi_{\mathbf{k}}-{mv^2}/{2}  }{2T_{c}}  \right)  \right]+
\frac{1}{\pi\cal V}\sum_{\mathbf{q}}\int_{-\infty}^{+\infty}d\omega\ \bose(\omega)\frac{\partial\delta(\mathbf{q},\omega)}{\partial\mu}
	\label{number_ERD}
 \ee

\subsubsection{Special case: only Zeeman term}
For completness, let us consider the case without spin-orbit couplings but in the presence of only a Zeeman term. In this case $\gamma_{h\kv}=h$. 
The equations become simply
 \be
	\frac{m}{4\pi\ar}=-\frac{1}{4\cal V}\sum_{\mathbf{k}}
	\left\{
	\frac{1}{\xik} \left[
	\tanh\left(\frac{\xik-h}{2T_{c}}\right)  + \tanh\left(\frac{\xik+h}{2T_{c}}\right) \right]
	-\frac{2}{\epsilon_{\mathbf{k}}}
	\right\}
\ee
\be
	n=\frac{1}{\cal V}\sum_{\mathbf{k}}\left\{1
		-\frac{1}{2}\left[\tanh\left(\frac{\xik-h}{2T_{c}} \right)
		+\tanh\left(\frac{\xik+h}{2T_{c}} \right)\right]\right\}
		+\frac{1}{\cal V}
		\sum_{\mathbf{q}}
		\int_{-\infty}^{+\infty}\frac{d\omega}{\pi}\ \bose(\omega)\frac{\partial\delta(\mathbf{q},\omega)}{\partial\mu}
 \ee
with $\delta(\qv,\omega)$ still defined by Eq.~(\ref{delta_new}) but where $\chi(\qv,i\nu_n)$, since $C^h_{\kv,\qv}=-1$,  is simply given by
\be
\chi(\qv,i\nu_n)=\frac{1}{2\cal V}\sum_\kv
\left[
\frac{\tanh\left(\frac{\xi_{\kv+\qv/2}-h}{2T_c}\right)+
\tanh\left(\frac{\xi_{\kv-\qv/2}+h}{2T_c}\right)}
{i\nu_n-\xi_{\kv+\qv/2}-\xi_{\kv-\qv/2}}
\right]
\ee
These are the equations to solve for finding the critical temperature of a polarized Fermi gas \cite{Parish,Sheehy} which should be the same in the presence of a Rabi coupling. 

\subsubsection{Weak coupling}
In this limit the chemical potential is finite and the critical temperature and the scattering length go to zero, in particular $\as\rightarrow0^{-}$. 
As argued in the case without SO coupling, $\Gamma^{-1}(\q)$ can be expressed as in Eq.~(\ref{Gammawc}). 
As a result, in this limit, the Gaussian correction to the density of particles is proportional to the scattering length and, therefore, is negligible,  as shown in Eq.~(\ref{n2wc}). Since the chemical potential is almost constant and since also the gap equation is unchanged, in the limit $\frac{1}{k_F\as}\rightarrow-\infty$, the critical temperature is expected to be almost the same as in the mean field level, 
therefore, the Gaussian fluctuations are not so effective at weak coupling, as we will verify by numerical calculations.

\subsubsection{Strong coupling}
\label{Gaussian_strong_coupling}

In the strong coupling limit, $\ar\rightarrow 0^+$, namely in the BEC limit, with SO coupling we found that $\mu(T_{c})$ diverges with the inverse scattering length $y=\frac{1}{\kf\ar}$, Eq.~(\ref{muSOsc}), getting the same formal result as that without SO coupling, 
	$\mu\rightarrow -\frac{1}{2m\ar}$. 
Let us now calculate the critical temperature $T_c$ as a function of the $y$. In what follows we will consider a generic spin-orbit coupling but without Zeeman term, $h=0$. 

\begin{figure}[h]
\centering
\includegraphics[width=0.5\textwidth]{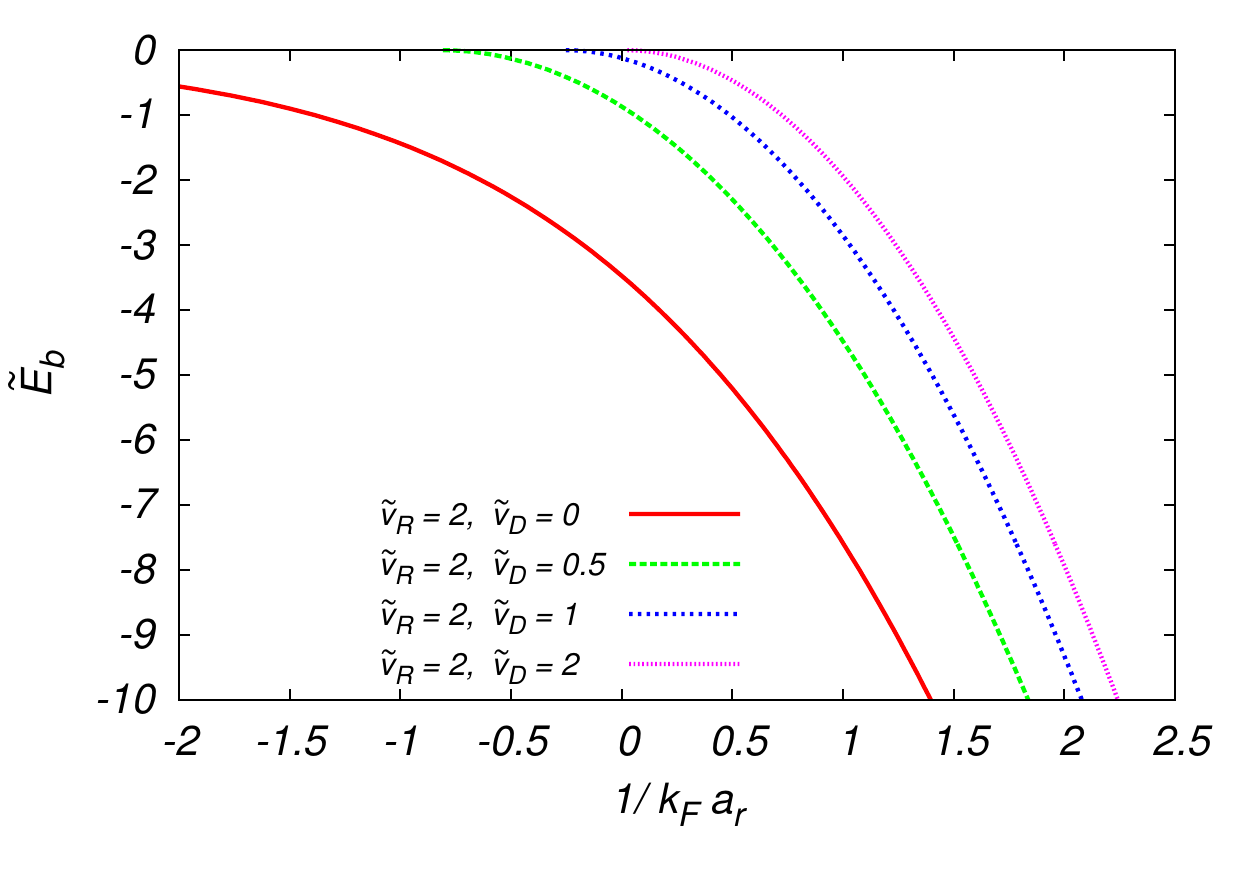}
\caption{ 
Bound state energy $\tilde E_b$ for a mixed spin-orbit coupling, keeping $\vR$ fixed and varying $\vd$. 
For $\vd=\vR$ the binding energy exists only for $y\ge 0$, in the BEC side, and is exactly the same curve one obtains in the absence of spin-orbit coupling. For pure Rashba (or pure Dresselhaus) the binding energy is always present, also in the BCS side while for any mixed SO coupling there is a threshold value of $y$ for the existence of the binding energy.}
\label{Eb_vs_y_so}
\end{figure}

In the case without SO term the inverse of the vertex function, $\Gamma^{-1}(\q)$, in the strong coupling limit, can be seen as the inverse propagator of free bosons, Eq.~(\ref{inv_vert_func}), with an effective mass and a redefined chemical potential, $\mub=2\mu-E_{b}$, which undergo the Bose-Einstein condensation below the critical temperature. 
The calculation in the presence of SO coupling is more difficult to perform due to the absence of spherical symmetry, however we can adopt the method applied in Refs.~\citep{PhysRevA.84.063618,PhysRevLett.107.195305,PhysRevB.83.094515}, doing a second order expansion in the momenta for $\Gamma^{-1}(\q)$ in the BEC limit.
The two-body physics emerges from $\Gamma^{-1}(\q)$, Eq.~(\ref{gamma_222}),  
 in the so-called vacuum limit, discarding the Fermi distribution, $\fermi\rightarrow0$, and putting the chemical potential to zero \citep{PhysRevA.84.063618,PhysRevLett.107.195305,PhysRevB.83.094515}. What remains is the inverse of the two-body scattering matrix, $T_{2B}^{-1}(\mathbf{q},i\nu_n)\equiv \Gamma^{-1}(\mathbf{q},i\nu_n)|_{\fermi,\mu=0}$, which, 
reads
 \begin{multline}
	T_{2B}^{-1}(\mathbf{q},i\nu_n)=
	-\frac{m}{4\pi\ar}+\frac{1}{4\cal V}\sum_{\mathbf{k}}\left[ \frac{2}{\epsilon_{\mathbf{k}}}
	+\frac{\left(1+C^0_{\mathbf{k},\mathbf{q}}\right)}
	{i\nu_n-\epsilon\kqm-\epsilon\kqp-\gamma\kqm-\gamma\kqp }+\right. \\
	\frac{\left(1+C^0_{\mathbf{k},\mathbf{q}}\right)}
	{i\nu_n-\epsilon\kqm-\epsilon\kqp+\gamma\kqm+\gamma\kqp}
	+\left. 
	\frac{2\left(1-C^0_{\mathbf{k},\mathbf{q}}\right)}
	{i\nu_n-\epsilon\kqm-\epsilon\kqp+\gamma\kqp-\gamma\kqm}\right]
	\label{twobody_scattering_matrix}
 \end{multline}
\paragraph*{Bound state.}
From the two-body scattering matrix we can evaluate the scattering energies and the existence of a two-particle bound state. 
We saw in a previous section, Sec.~\ref{Matrice di scattering a due corpi}, 
how $T_{2B}(z)$ has simple poles at the bound states and a branch cut at the scattering states. In order to prove the existence of a bound state in the presence of a spin-orbit term it is enough, therefore, to impose that, at vanishing momentum,
 \begin{equation}
 \label{T1}
	T_{2B}^{-1}\left(\mathbf{q}=0,z\!=\!{\cal E}_b\right)=0
 \end{equation}
From Eq.~(\ref{Chkq}) we have $C^0_{\kv,0}=1$, so that Eq.~(\ref{T1}) implies the following equation
 \begin{equation}
	-\frac{m}{4\pi\ar}+\frac{1}{2\cal V}\sum_{\mathbf{k}}
	\left\{
	\frac{1}{{\cal E}_b-2\epsk-2 \gammak }
	+\frac{1}{{\cal E}_b-2\epsk+2 \gammak }
	+\frac{1}{\epsilon_{\mathbf{k}}}
	\right\}=0
	\label{bound_state_eq}
 \end{equation}
that should be solved in terms of the bound state 
 energy, measured from the threshold energy for free fermions, as a function of the coupling $y$. This can be done by simple numerical techniques. In the continuum limit, introducing the following dimensionless quantities
	${\bf p}=\frac{\bf k}{\kf}$, $\tilde {\cal E}_b=\frac{{\cal E}_b}{\Ef}$, 
	$\tTc=\frac{T_{c}}{\Ef}$, 
	$\tvr=\frac{\vR}{\vf}$,  
	$\tvd=\frac{\vd}{\vf}$  and 
	$\tilde{\gamma}_\mathbf{p}(\tvr,\tvd)=2\gamma_\mathbf{k}(\vR, \vd)$, we get 
\begin{equation}
	y+\frac{2}{\pi^{2}}\int_{0}^{\infty}dp_{x}\int_{0}^{\infty}dp_{y}\int_{0}^{\infty}dp_{z}
	\left\{
	\frac{1}{p^{2}+\tgammap-\tilde {\cal E}_b/2}
	+\frac{1}{p^{2}-\tgammap-\tilde {\cal E}_b/2}
	-\frac{2}{p^{2}}
	\right\}=0
	\label{bound_state_eq_riscalata}
 \end{equation}
which has the same form of Eq.~(\ref{mean_field_strong_coup}). 
The chemical potential and the bound state energy, $E_b=\Ef \tilde E_b$, measured at the threshold energy $m(\vR+\vd)^2$, in terms of dimensionless quantities, 
\be
\label{calE}
\tilde {\cal E}_b=\tilde E_b-2(\tvr+\tvd)^2
\ee
satisfy the same equation, therefore, in the the strong coupling limit, they are connected by the relation $\mu\underset{y\rightarrow\infty}{\longrightarrow}\frac{{\cal E}_b}{2}=\frac{E_b}{2}+\frac{m(\vR+\vd)^2}{2}$. In the deep BEC, neglecting the second term, we have
\be
\mu\underset{y\rightarrow\infty}{\longrightarrow}E_{b}/2
\ee
which is the same result obtained in the absence of SO coupling, Eq.~(\ref{mu_Eb}). 
Taking advantage of Eq.~(\ref{mu_sc_vr}), in the case of only Rashba coupling ($\vd=0$), we can write an analytical solution
\be
\label{y_Eb_vr}
y=\sqrt{\tvr^2-\frac{\tilde E_b}{2}}-{\tvr} \arctanh\left(\tvr\sqrt{\frac{2}{2\tvr^2-\tilde E_b}}\right)
\ee 
Inverting Eq.~(\ref{y_Eb_vr}) we can get $E_b$ as a function of $y$. For equal Rashba and Dresselhaus couplings, $\vR=\vD$,  we have, instead,
\be
y=\sqrt{-\frac{\tilde E_b}{2}}
\ee
which is the same result as that without SO coupling. 
For the more general case one can easily resort to a numerical computation to solve Eq.~(\ref{bound_state_eq_riscalata}), or, after analytically integrating over $p_z$, Eq.~(\ref{mean_field_strong_coup2}) with $\tilde{\cal E}_b/2$ instead of $\tmu$. 
The results are shown in Fig.~\ref{Eb_vs_y_so}. \\
For pure Rashba coupling, Fig.~\ref{Eb_vs_y_so}, a bound state exists also in the BCS part of the crossover, where the scattering length is negative. Actually by looking at Eq.~(\ref{y_Eb_vr}), for any value of $y$, a negative value of $E_b$ exists. 
Moreover, in Fig.~\ref{Eb_vs_y_so} we show that turning on a Dresselhaus term ($\vd\neq 0$), greatly affects the existence of the bound state in the BCS side.
The very important feature is, therefore, that a bound state exists also for negative scattering length, namely in the BCS side. The effect is relevant for a pure Rashba coupling while it is suppressed in the presence of mixed Rashba and Dresselhaus couplings. For equal Rashba and Dresselhaus couplings the bound state exists only in the BEC side, exactly as in the case without SO. 

\paragraph*{Effective masses.}  
At strong coupling two fermions can be treated as a single bosonic particle with a double mass. However in the presence of a spin-orbit interaction we have to reconsider the evaluation of the effective mass.  
This can be done at low momenta requiring that the dispersion relation of free bosons in the strong coupling limit (${1}/{\kf\as}\rightarrow+\infty$) can be written as
 \begin{equation}
	\epsilon_{b}(\mathbf{q})={\cal E}_{b}+\frac{q_{x}^{2}}{2m_x}+\frac{q_{y}^{2}}{2m_{y}}+\frac{q_{z}^{2}}{2m_z}
	\label{bound_state_spectrum}
 \end{equation}
where ${\cal E}_{b}$, $m_x$ and $m_y$ depend on both the strength of the interaction and the spin-orbit couplings. 
The effective masses along $x$ and $y$ directions can be generally different in the presence of mixed Rashba and Dresselhaus terms. 
These masses can be derived imposing that the inverse of the scattering matrix vanishes when measured right at $\epsilon_{b}(\mathbf{q})$ as in Eq.~\eqref{bound_state_spectrum}
 \begin{equation}
	T_{2B}^{-1}\left(\mathbf{q},z=\epsilon_{b}(\mathbf{q})\right)=0
 \end{equation}
At low energy and momentum this condition is imposed to the second order expansion in $\mathbf{q}$
 \be
	T_{2B}^{-1}\left(\mathbf{q},\epsilon_{b}(\mathbf{q})\right) \approx
	T_{2B}^{-1}\left(0,{\cal E}_{b}\right)+
	{\cal F}_x({\cal E}_{b},m_{x})\,q_{x}^{2}+{\cal F}_y({\cal E}_{b},m_{y})q_{y}^{2}+{\cal F}_z({\cal E}_{b},m_z)q_{z}^{2}=0
 \ee
 which implies that
 \be
 \label{conditions}
 		T_{2B}^{-1}\left(0,{\cal E}_{b}\right)=0, \;\;\;
		{\cal F}_x({\cal E}_{b},m_{x})=0, \;\;\;
		{\cal F}_y({\cal E}_{b},m_{y})=0, \;\;\;
		{\cal F}_z({\cal E}_{b},m_z)=0
 \ee
The first equation in Eq.~(\ref{conditions}) is the same encountered before for the bound state energy, Eq.~(\ref{T1}).
By choosing $m_z=2m$ we get immediately that ${\cal F}_z({\cal E}_{b},2m)=0$ since, in this way, 
$T_{2B}^{-1}\left(\mathbf{q},\epsilon_{b}(\mathbf{q})\right)$ 
loses the dependence on $q_z$. Actually the two-body scattering matrix reads
 \bea
\nonumber    \hspace{-0.8cm} T_{2B}^{-1}\left(\mathbf{q},\epsilon_{b}(\mathbf{q})\right)&=&
	-\frac{m}{4\pi\ar}+\frac{1}{4\cal V}\sum_{\mathbf{k}}  \frac{2}{\epsilon_{\mathbf{k}}}   \\
\nonumber	&&+\frac{1}{4\cal V}\sum_{\mathbf{k}}
	\frac{\left(1+C^0_{\mathbf{k},\mathbf{q}}\right)}
	{{\cal E}_{b}+\frac{q_{x}^{2}}{2m_x}+\frac{q_{y}^{2}}{2m_{y}}+\frac{q_{z}^{2}}{4m}-\epsilon\kqm-\epsilon\kqp-\gamma\kqm-\gamma\kqp}\\
\nonumber	&&+\frac{1}{4\cal V}\sum_{\mathbf{k}}
	\frac{\left(1+C^0_{\mathbf{k},\mathbf{q}}\right)}
{{\cal E}_{b}+\frac{q_{x}^{2}}{2m_x}+\frac{q_{y}^{2}}{2m_{y}}+\frac{q_{z}^{2}}{4m}-\epsilon\kqm-\epsilon\kqp+\gamma\kqm+\gamma\kqp}\\
&& +\frac{1}{2\cal V}\sum_{\mathbf{k}}
	\frac{\left(1-C^0_{\mathbf{k},\mathbf{q}}\right)}
	{{\cal E}_{b}+\frac{q_{x}^{2}}{2m_x}+\frac{q_{y}^{2}}{2m_{y}}+\frac{q_{z}^{2}}{4m}-\epsilon\kqm-\epsilon\kqp+\gamma\kqp -\gamma\kqm}
 \eea
Introducing the following definitions 
\be
\label{XY}
X=\frac{2m}{m_{x}}-1,\;\;\;\;\;
Y=\frac{2m}{m_{y}}-1
\ee
the two-body scattering matrix can be written as it follows
 \bea
\nonumber \hspace{-0.5cm} T_{2B}^{-1}\left(\mathbf{q},\epsilon_{b}(\mathbf{q})\right)=
	-\frac{m}{4\pi\ar}+\frac{1}{\cal V}\sum_{\mathbf{k}}\frac{1}{2\epsk}
	+\frac{1}{4\cal V}\sum_{\mathbf{k}}
	\frac{\left(1+C^0_{\mathbf{k},\mathbf{q}}\right)}
	{{\cal E}_{b}+\frac{X}{4m}q_{x}^{2}+\frac{Y}{4m}q_{y}^{2}-2\epsk-\gamma\kqm-\gamma\kqp}\\
\nonumber	+\frac{1}{4\cal V}\sum_{\mathbf{k}}
	\frac{\left(1+C^0_{\mathbf{k},\mathbf{q}}\right)}
	{{\cal E}_{b}+\frac{X}{4m}q_{x}^{2}+\frac{Y}{4m}q_{y}^{2}-2\epsk+\gamma\kqm+\gamma\kqp}\\
	+\frac{1}{2\cal V}\sum_{\mathbf{k}}
	\frac{\left(1-C^0_{\mathbf{k},\mathbf{q}}\right)}
	{{\cal E}_{b}+\frac{X}{4m}q_{x}^{2}+\frac{Y}{4m}q_{y}^{2}-2\epsk+\gamma\kqp-\gamma\kqm}
	\label{twobody_expansion}
 \eea
Expanding for small $\qv$ the following quantities
\bea
&&C^0_{\kv,\qv}\approx 1-\frac{(\vR^{2}-\vd^{2})^{2}}{2\gammak ^{4}}\left(k_{x}q_{y}-k_{y}q_{x}\right)^2
\\
&&\gamma\kqm+\gamma\kqp\approx 2\gammak+\frac{(\vR^{2}-\vd^{2})^{2}}{4\gammak^{3}}\left(k_{x}q_{y}-k_{y}q_{x}\right)^2\\
&&\gamma\kqp-\gamma\kqm \approx \frac{(\vR+\vd)^2}{\gammak}k_x q_x+\frac{(\vR-\vd)^2}{\gammak}k_y q_y
\eea
the two-body scattering matrix, at leading orders, reads
\bea
\nonumber&& T_{2B}^{-1}\left(\mathbf{q},\epsilon_{b}(\mathbf{q})\right) \approx 
-\frac{m}{4\pi\ar}+\frac{1}{2\cal V}\sum_{\mathbf{k}}
	\left[\frac{1}{{\cal E}_b-2\epsk-2 \gammak } +\frac{1}{{\cal E}_b-2\epsk+2 \gammak }
	+\frac{1}{\epsilon_{\mathbf{k}}}\right]\\
\nonumber&&	-\frac{1}{4\cal V}\left[
	\frac{X}{m}\sum_{\mathbf{k}}\frac{({\cal E}_b-2\epsk)^2+4 \gammak^2}{\left(({\cal E}_b-2\epsk)^2-4 \gammak^2\right)^2}	
	+\left(\vR^2-\vd^2\right)^2\sum_{\mathbf{k}}\frac{16\,k_y^2}{\left(({\cal E}_b-2\epsk)^2-4 \gammak^2\right)^2({\cal E}_b-2\epsk)}
	\right] q_x^2\\
\nonumber&&	-\frac{1}{4\cal V}\left[
	\frac{Y}{m}\sum_{\mathbf{k}}\frac{({\cal E}_b-2\epsk)^2+4 \gammak^2}{\left(({\cal E}_b-2\epsk)^2-4 \gammak^2\right)^2}	
	+\left(\vR^2-\vd^2\right)^2\sum_{\mathbf{k}}\frac{16\,k_x^2}{\left(({\cal E}_b-2\epsk)^2-4 \gammak^2\right)^2({\cal E}_b-2\epsk)}
	\right] q_y^2\\
	&&\equiv T_{2B}^{-1}\left(0,{\cal E}_{b}\right)+{\cal F}_x({\cal E}_{b},m_{x})\phantom{\Big |}q_{x}^{2}+{\cal F}_y({\cal E}_{b},m_{y})\phantom{\Big |}q_{y}^{2}
\eea
The first term is the scattering matrix at $\qv=0$, therefore, $T_{2B}^{-1}\left(0,{\cal E}_{b}\right)=0$, by definition of the shifted bound state energy ${\cal E}_b$. Solving ${\cal F}_x({\cal E}_{b},m_{x})=0$ in terms of $X$ and ${\cal F}_y({\cal E}_{b},m_{y})=0$ in terms of $Y$, and from Eq.~(\ref{XY}), we can derive the values of the effective masses
 \bea
	\frac{2m}{m_{x}}=1-m\left(\vR^{2}-\vd^{2}\right)^{2}\frac{{\cal B}_{x}}{{\cal A}}\\
	\frac{2m}{m_{y}}=1-m\left(\vR^{2}-\vd^{2}\right)^{2}\frac{{\cal B}_{y}}{{\cal A}}
 \eea
where 
 \bea
 	\label{A}
	{\cal A}&=&\frac{1}{\cal V}\sum_{\kv}\frac{({\cal E}_b-2\epsk)^2+4 \gammak^2}{\left(({\cal E}_b-2\epsk)^2-4 \gammak^2\right)^2}\\
	\label{Bx}
	{\cal B}_{x}&=&\frac{1}{\cal V}\sum_{\kv}\frac{16\,k_y^2}{\left(({\cal E}_b-2\epsk)^2-4 \gammak^2\right)^2({\cal E}_b-2\epsk)}\\
	{\cal B}_{y}&=&\frac{1}{\cal V}\sum_{\kv}\frac{16\,k_x^2}{\left(({\cal E}_b-2\epsk)^2-4 \gammak^2\right)^2({\cal E}_b-2\epsk)}
	\label{By}
 \eea
In the specific case of equal Rashba and Dresselhaus couplings ($\vR=\pm\vd$), as already seen, the bound state energy is not distinguishable from the one in absence of SO coupling, which means that a bound state is present only for $y\ge 0$, namely only in the BEC side, where the effective masses are simply $2m/m_x=2m/m_y=1$. \\
In the continuum limit, rescaling the variables 
	${\bf p}=\frac{\bf k}{\kf}$, $\tmu=\frac{\mu}{\Ef}$, 
	$\tilde {\cal E}_b=\frac{{\cal E}_b}{\Ef}$, 
	$\tvr=\frac{\vR}{\vf}$, 
	$\tvd=\frac{\vd}{\vf}$, 
	$\tgammap(\tvr,\tvd)=2\gammak(\vR, \vd)$, Eqs.~\eqref{A}, \eqref{Bx} and \eqref{By} in terms of the dimensionless parameters become
 \begin{eqnarray}
 \label{calA}
	{\cal A}	&=&\frac{\kf^{3}}{\Ef^{2}}\int \frac{d\mathbf{p}}{(2\pi)^3}
	\frac{(\tilde{\cal E}_b-2p^2)^2+4 \tgammap^2}{\left((\tilde{\cal E}_b-2p^2)^2-4 \tgammap^2\right)^2}=\frac{\kf^{3}}{\Ef^{2}}\tilde{\cal A}\\
	 \label{calBx}
	{\cal B}_{x}&=&\frac{\kf^{5}}{\Ef^{5}}\int \frac{d\mathbf{p}}{(2\pi)^3}
	\frac{16\,p_y^2}{\left((\tilde{\cal E}_b-2 p^2)^2-4 \tgammap^2\right)^2(\tilde{\cal E}_b-2p^2)}=\frac{\kf^{5}}{\Ef^{5}}\tilde{\cal B}_x\\
	{\cal B}_{y}&=&\frac{\kf^{5}}{\Ef^{5}}\int \frac{d\mathbf{p}}{(2\pi)^3}
	\frac{16\,p_x^2}{\left((\tilde{\cal E}_b-2 p^2)^2-4 \tgammap^2\right)^2(\tilde{\cal E}_b-2p^2)}=\frac{\kf^{5}}{\Ef^{5}}\tilde{\cal B}_y
	 \label{calBy}
 \end{eqnarray}
The effective masses can be written in terms of dimensionless quantities
 \bea
 \label{calmx}
	\frac{2m}{m_{x}}=1-8\left(\tvr^{2}-\tvd^{2}\right)^{2}\frac{\tilde{\cal B}_{x}}{\tilde{\cal A}}\\
	\frac{2m}{m_{y}}=1-8\left(\tvr^{2}-\tvd^{2}\right)^{2}\frac{\tilde{\cal B}_{y}}{\tilde{\cal A}}
	 \label{calmy}
 \eea
For the case with a pure Rashba coupling, for which $m_{x}=m_{y}$, we can perform the integrals $\tilde{\cal A}$, $\tilde{\cal B}_x$ and $\tilde{\cal B}_y$ analytically, obtaining 
\begin{figure}[h]
\centering
\includegraphics[width=0.5\textwidth]{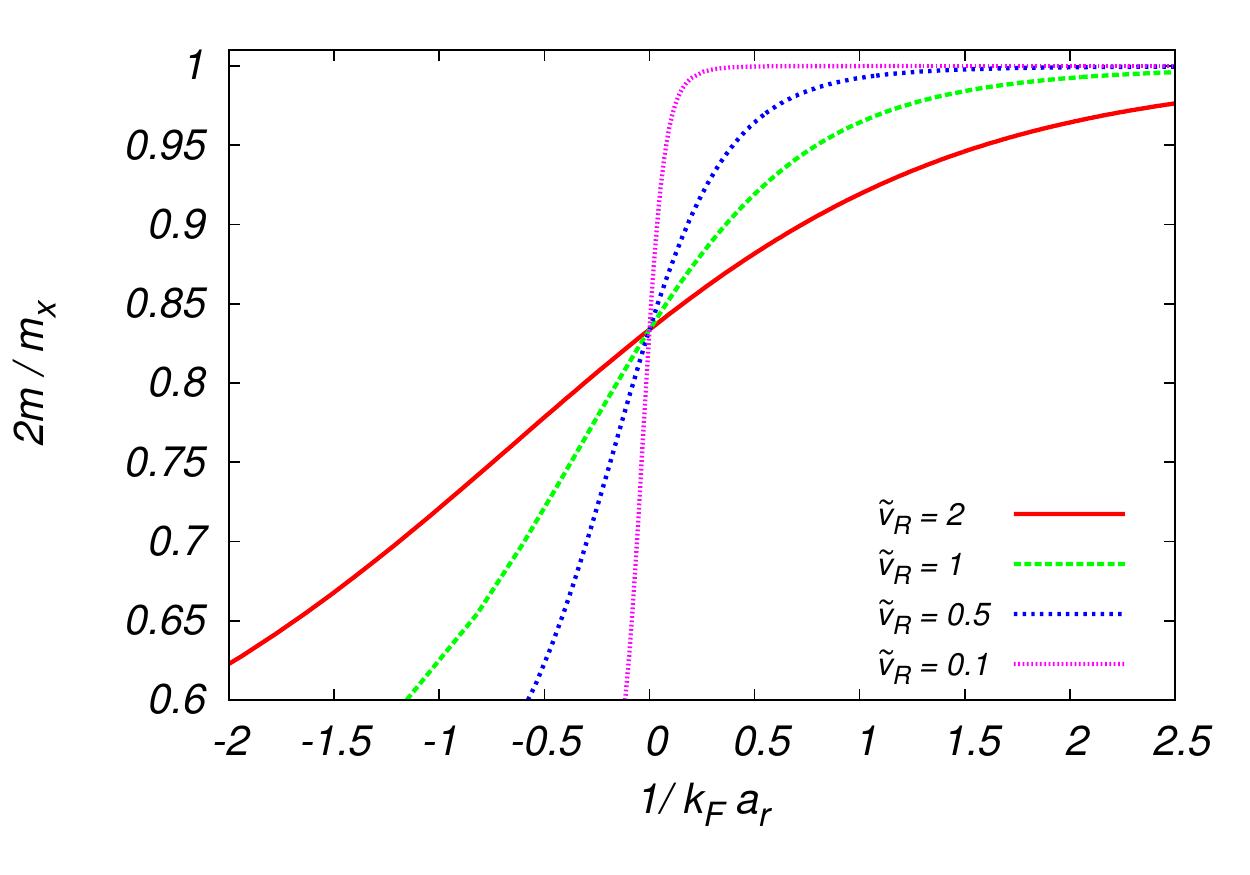}
\caption{Effective mass ratio ${2m}/{m_x}$, as a function of the coupling ${1}/{\kf\ar}$, for different values of purely Rashba coupling $\tvr$, from Eqs.~(\ref{mass_R}) and (\ref{y_R}). For only Rashba coupling $m_x=m_y$.}
\label{mx_vs_y_vr}
\end{figure}

\bea
\tilde{\cal A}=\frac{1}{2\pi^2}\int_0^\infty dp_z\int_0^\infty dp_\perp p_\perp
\frac{(\tilde{\cal E}_b-2p_\perp^2-2p_z^2)^2+4 (2\tvr p_\perp)^2}{\left[(\tilde{\cal E}_b-2p_\perp^2-2p_z^2)^2-4 (2\tvr p_\perp)^2\right]^2}
=\frac{-\sqrt{-{\tilde{\cal E}_b}}}{16\sqrt{2}\,\pi\big(\tilde{\cal E}_b+2\tvr^2\big)}\\
\nonumber \tilde{\cal B}_x=\tilde{\cal B}_y=\frac{1}{2\pi^2}\int_0^\infty dp_z\int_0^\infty dp_\perp 
\frac{8\,p_\perp^3}{\left[(\tilde{\cal E}_b-2 p_\perp^2-2p_z^2)^2-4 (2\tvr p_\perp)^2\right]^2(\tilde{\cal E}_b-2p_\perp^2-2p_z^2)}\\
=\frac{1}{256\sqrt{2}\,\pi \tvr^4}\frac{1}{\sqrt{-\tilde{\cal E}_b}}\left[\log\left(1+\frac{2\tvr^2}{\tilde{\cal E}_b}\right)-\frac{2\tvr^2}{\tilde{\cal E}_b+2\tvr^2}\right]
\eea
where, from Eq.~(\ref{calE}), in the presence of a bound state, we assume that
\be
\tilde{\cal E}_b\le -2 \tvr^2
\ee
The final analytical expressions for the effective masses are, therefore,
\be
\label{mass_R}
\frac{2m}{m_{x}}=\frac{2m}{m_{y}}=1-\frac{1}{2\tilde{\cal E}_b}\left[\big(\tilde{\cal E}_b+2\tvr^2\big) \log\left(1+\frac{2\tvr^2}{\tilde{\cal E}_b}\right)-2\tvr^2\right]
\ee
which, together with Eq.~(\ref{y_Eb_vr}) written in terms of $\tilde{\cal E}_b$ 
\be
\label{y_R}
y=\frac{\sqrt{-\tilde{\cal E}_b}}{\sqrt{2}}
-{\tvr} \arctanh\left(\tvr \frac{\sqrt{2}}{\sqrt{-\tilde{\cal E}_b}}
\right)
\ee
in order to get their behaviors along the crossover, see Fig.~\ref{mx_vs_y_vr}. This result is in agreement with the $T$-matrix approach \cite{a6}. 
At the special point $y=\frac{1}{\kf \ar}=0$, also called \emph{unitary limit}, the ratio $2\tvr^2/{\tilde{\cal E}_b}\equiv -c_o$ is the solution of the following equation
\be
\sqrt{c_o}=\tanh(1/\sqrt{c_o})
\ee
which is $c_o\approx 0.6948$. As a result $\frac{2m}{m_{x}}=\frac{2m}{m_{y}}$ at $y=0$ is the same for any value of $\tvr$, 
\be
\left. \frac{2m}{m_{x}}\right|_{{\ar^{-1}}=0}=1-\frac{1}{2}\left[c_o+\big(1-c_o\big) \log\left(1-c_o\right)
\right]
\approx 0.8387
\ee
which is the nodal point shown in Fig.~\ref{mx_vs_y_vr}. For the general case of mixed Rashba and Dresselhaus terms the effective 
masses are given by Eqs.~\eqref{calmx}, \eqref{calmy}, together with Eqs.~\eqref{calA}-\eqref{calBy}.

\subsubsection{Full crossover}
\begin{figure}
\centering
\includegraphics[width=0.45\textwidth]{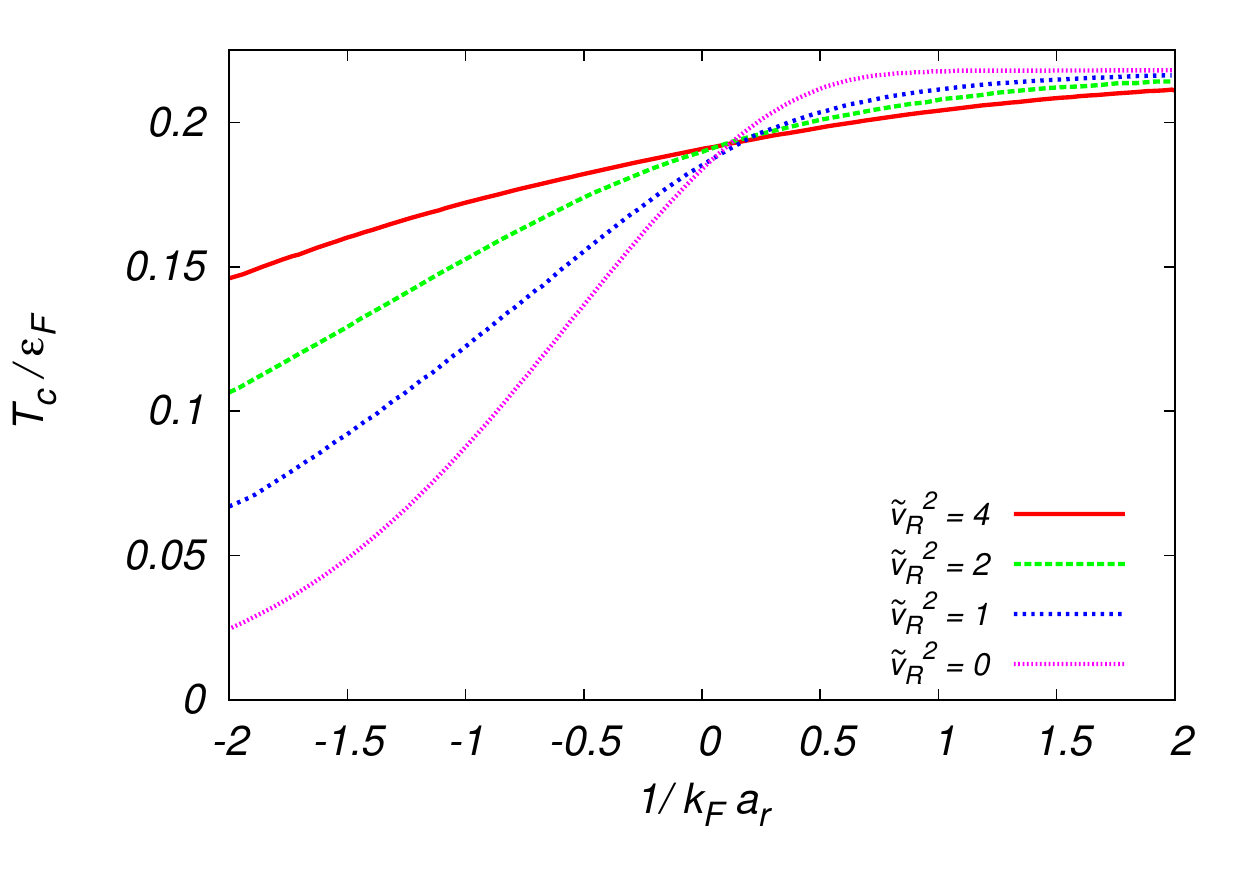}
\includegraphics[width=0.45\textwidth]{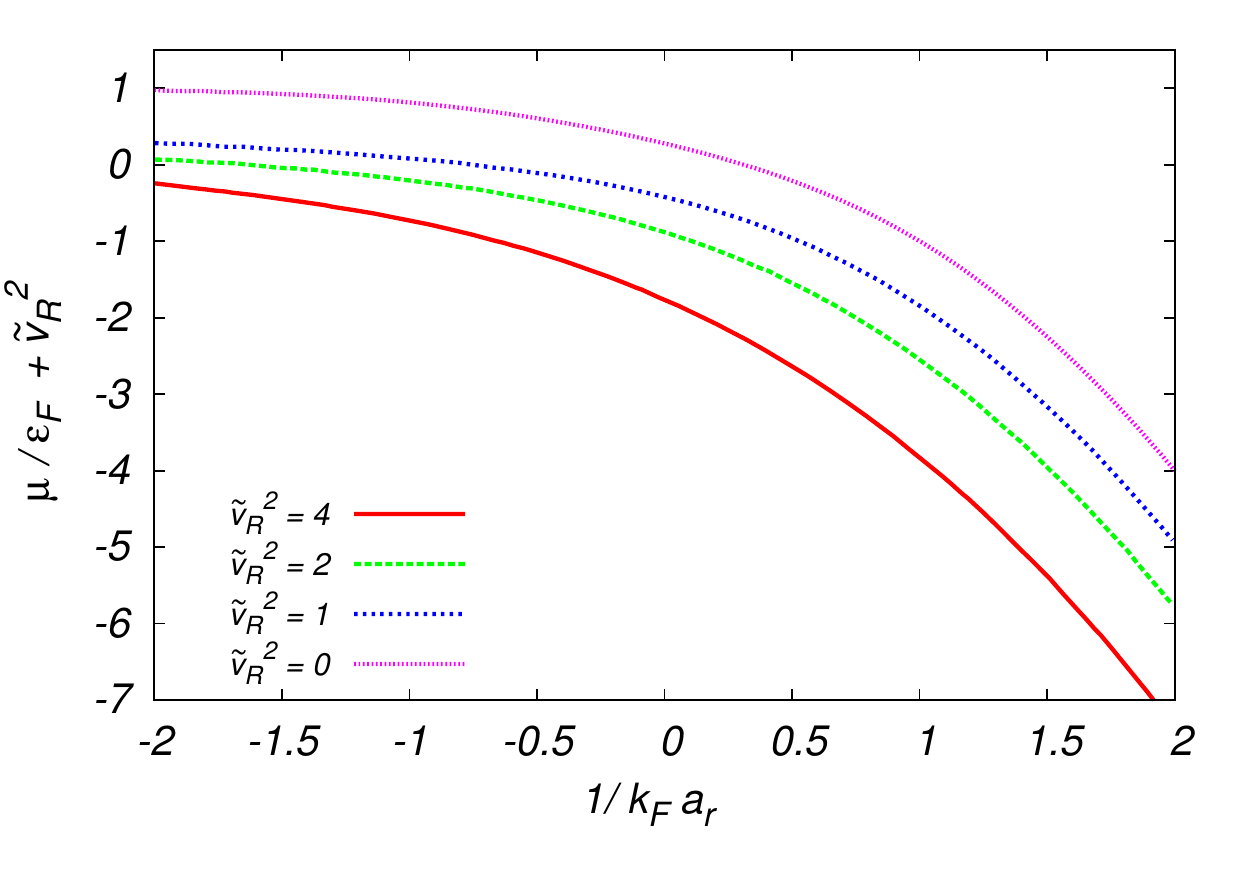}
\caption{Critical temperature $T_c$ and chemical potential $\mu$, at $T_c$, with the inclusion of the Gaussian fluctuations within the bosonic approximation, for different values of $\tvr$ in the purely Rashba case.}
\label{fig:TC_SO}
\end{figure}

\paragraph{Bosonic approximation.} 
We found that the inverse of the scattering matrix, expanding at low momenta around $\mathbf{q}\!=\!0$, has the form 
 \begin{equation}
	T_{2B}^{-1}(\mathbf{q},i\nu)\approx i\nu-\epsilon_{b}(\mathbf{q})=i\nu-{\cal E}_{b}-\frac{q_{x}^{2}}{2m_x}-\frac{q_{y}^{2}}{2m_{y}}-\frac{q_{z}^{2}}{4m}
 \end{equation} 
with ${\cal E}_{b}$, $m_{x}$ and $m_{y}$ that depend on $y=\frac{1}{\kf\ar}$.
This result for the two-body problem can be exploited to determine the properties of the vertex function in the strong coupling limit 
 \be
\left. \Gamma^{-1}(\mathbf{q},i\nu)\right|_{\mu\rightarrow-\infty}=T_{2B}^{-1}(\mathbf{q},i\nu+2\mu)
 \ee
getting, in this regime,
 \be
 	\Gamma^{-1}(\mathbf{q},i\nu)\approx
	i\nu+2\mu-\epsilon_{b}(\mathbf{q})=
	i\nu-\left(\frac{q_{x}^{2}}{2m_x}+\frac{q_{y}^{2}}{2m_{y}}+\frac{q_{z}^{2}}{4m}-\mub\right)
	\label{pole_structure}
 \ee
 where $\mub=2\mu -{\cal E}_b/2\simeq 2\mu -{E}_b/2$. We find, therefore, that 
the partition function approaches that of a set of free bosons, with anisotropic masses. It is straightforward to show that the condensation temperature for this kind of systems is actually rescaled by the effective masses
 \begin{equation}	
 \label{Tcnb}
	T_{c}=\frac{2\pi}{m_{b}}\left(\frac{\bose}{2\zeta(3/2)}\right)^{\frac{2}{3}}
	\left(\frac{\mb}{m_{x}}\right)^{\frac{1}{3}}
	\left(\frac{\mb}{m_{y}}\right)^{\frac{1}{3}}
 \end{equation}
where $\mb=2m$ and, in the deep BEC, $\bose=\frac{n}{2}=\frac{1}{2}\frac{(2m\Ef)^{3/2}}{3\pi^{2}}$, therefore
 \begin{equation}	
	\frac{T_{c}}{\Ef}\approx0.218
	\left(\frac{2m}{m_{x}}\right)^{\frac{1}{3}}
	\left(\frac{2m}{m_{y}}\right)^{\frac{1}{3}}.
 \end{equation}
 We find that in the BEC limit the effective masses approach those of two fermions, 
 \begin{equation}
 \label{->1}
	\frac{2m}{m_{\alpha}}\underset{BEC}{\xrightarrow{\hspace*{0.6cm}}}1
 \end{equation}
 with $\alpha=x,y$, 
for any mixture of Rashba and Dresselhaus couplings. For the case of only Rashba (or Dresselhaus) coupling Eq.~(\ref{->1}) can be read out from Eq.~(\ref{mass_R}) in the limit of $\tilde{\cal E}_b\rightarrow \infty$.\\
For two fermions with same momenta in a spin singlet state, composing a boson, the spin-orbit coupling cancels out and a free particle spectrum with mass $\mb=2m$ emerges.\\
From Eq.~(\ref{Tcnb}), solving in terms of $\bose$, we have
 \begin{equation}
	\bose=\zeta(3/2)\left(\frac{mT_{c}}{\pi}\right)^{3/2}
	\left(\frac{m_{x}}{2m}\right)^{\frac{1}{2}}
	\left(\frac{m_{y}}{2m}\right)^{\frac{1}{2}}
 \end{equation}
As done in Eq.~(\ref{n2bosonic}), let us suppose now that the number of bosonic excitations composed by pairs of fermions is $\bose \approx 2 n^{(2)}$, while the rest of fermions remains unpaired so that we have
 \begin{equation}
	n^{(2)} \approx 2\zeta(3/2)\left(\frac{mT_{c}}{\pi}\right)^{3/2}
	\left(\frac{m_{x}}{2m}\right)^{\frac{1}{2}}
	\left(\frac{m_{y}}{2m}\right)^{\frac{1}{2}} .
 \end{equation} 
With this approximation the set of equations to solve is the following 
 \begin{equation}
	\begin{aligned}
	&\hspace{-0.25cm}\frac{m}{4\pi\ar}=-\frac{1}{4\Omega}\sum_{\mathbf{k}}
	\left\{
	\frac{ \tanh(\frac{\xik-\gammak}{2T_{c}})  }{\xik-\gammak}
	+\frac{ \tanh(\frac{\xik+\gammak}{2T_{c}}) }{\xik+\gammak}
	-\frac{2}{\epsilon_{\mathbf{k}}}
	\right\}\\
	&\hspace{-0.25cm} n= 
	\frac{1}{2\cal V}\sum_{\mathbf{k}}\left\{2
		-\tanh\left(\frac{\xik-\gamma_{\kv}}{2T_{c}} \right)
		-\tanh\left(\frac{\xik+\gamma_{\kv}}{2T_{c}} \right)\right\}
	+ 2 \zeta({3}/{2})\left(\frac{mT_{c}}{\pi}\right)^{3/2}
	\hspace{-0.13cm} \left(\frac{m_{x}}{2m}\,
	\frac{m_{y}}{2m}\right)^{1/2}
	\end{aligned}
 \end{equation}
In the continuum limit, from Eqs.~(\ref{y_eq0}), (\ref{n_eq0}), at fixed density $n=\frac{\kf^{3}}{3\pi^{2}}$, after rescaling ${\bf p}=\frac{1}{\kf}{\bf k}$, $\tmu=\frac{\mu}{\Ef}$, $\tTc=\frac{T_{c}}{\Ef}$, $\tvr=\frac{\vR}{\vf}$,  $\tvd=\frac{\vd}{\vf}$,  
	$\tilde{\gamma}_\mathbf{p}(\tvr,\tvd)=2\gamma_\mathbf{k}(\vR, \vd)$, 
the above equations can be written in the following form
 \begin{eqnarray}
	&&y=I_{y}^{\textsc{so}}(\tmu,\tTc,\tvr,\tvd)\\
&& 1=I_n^{\textsc{so}}(\tmu,\tTc,\tvr,\tvd)+
	3\,\zeta(3/2)\sqrt{\frac{\pi}{2}}\,\tTc^{\,3/2}
	\left(\frac{m_{x}}{2m}\,
	\frac{m_{y}}{2m}\right)^{{1}/{2}}
 \end{eqnarray}
This set of equations can be easily solved numerically. Some results are reported in Fig.~\ref{fig:TC_SO} for 
the most relevant case with pure Rashba coupling.
These results have to be contrasted with those reported in Fig.~\ref{fig.bosonicapprox} in the absence of spin-orbit couplings. 

\paragraph{Further improvements.} The existence of a bound state also in the BCS side of the crossover allowed us to extend the results of the effective masses also in this region, although the pole structure of the vertex function $\Gamma(\q)$, presented in 
Eq.~\eqref{pole_structure}, has been derived in the strong coupling limit. To refine this result it should be sufficient to expand at second order in the momentum $\qv$ the full $\Gamma^{-1}(\q)$ showed in Eqs.~\eqref{gamma_222},  \eqref{chinew}, with $h=0$. 
The only difference with respect to what done so far is to include the expansion of the hyperbolic tangents of some functions, $f_\qv=(\xi_{\kv\pm\qv/2}\pm\gamma_{\kv\pm\qv/2})/2$, appearing in Eq.~\eqref{chinew}, up to second order in \qv
\bea
&&\hspace{-0.5 cm}\tanh\left({f_\qv}/{T_c}\right)=\tanh\left({f_0}/{T_c}\right)+
\textrm{sech}\left({f_0}/{T_c}\right)^2 {\nabla f_\qv\big{|}_0\hspace{-0.1cm}\cdot\qv}/{T_c}+\\
\nonumber&&+\sum_{ij}\left[\textrm{sech}\left({f_0}/{T_c}\right)^2 \partial_{i}\partial_j f_\qv\big{|}_0/2
-\textrm{sech}\left({f_0}/{T_c}\right)^2\tanh\left({f_0}/{T_c}\right)\partial_i f_\qv\partial_j f_\qv\big{|}_0
\right] {q_iq_j}/{T_c^2}+O(q^3)
\eea
This expansion, with respect to the vacuum limit considered so far, can be seen somehow as the addition of thermal corrections to the zero-temperature results. These corrections are relevant in the intermediate regime, since in the strong coupling limit $f_\qv/T_c\rightarrow \infty$, therefore $\tanh({f_\qv}/{T_c})\rightarrow 1$, recovering the results in the bosonic approximation, while in the weak coupling limit the quantum fluctuation themselves are not relevant at all, as discussed previously.\\
Alternatively one can calculate the exact Gaussian fluctuations solving numerically the full equations in Eqs.~\eqref{gapTcnew}, \eqref{mean_field_SOnew} together with Eqs.~\eqref{delta_new}, \eqref{gamma_222},  \eqref{chinew}, as done in the absence of the spin-orbit couplings. 
However, in the presence of SO the rotational symmetry is broken and the numerics is extremely much more involved and computational costly.

\section{Conclusions}

We reviewed the calculations for the critical temperature along the BCS-BEC crossover with and without spin-orbit couplings in the path integral formalism. The final equations, with the inclusion of quantum fluctuations at the Gaussian level have been written in the most general case with Rashba, Dresselhaus and Zeeman terms, even though most of the results presented are done for the most relevant case with only Rashba term. We found that in the weak coupling limit $T_c$ is strongly enhanced, consistently with the increase of the gap \cite{Articolo_tesi2} which implies an increased binding energy promoting the formation of fermionic pairs, therefore, raising the critical temperature.
In the intermediate regime the rapid increase of $T_c$ predicted by the mean-field theory is softened by the evolution of 
the effective masses, originated by the quantum Gaussian fluctuations, of the composite bosons satisfying $m_{x},m_{y}>2m$ which lower the critical temperature. 
Intriguingly, either the binding energy and the effective masses have finite values also in the BCS regime. For the Rashba case this finding can be found analytically, in particular, at the unitary limit we found that the masses do not depend on the spin-orbit parameter.
Finally, in the deep BEC, we found that $T_c$ goes to the same value as that without SO, consistently with the 
strong singlet-pair formation.

\appendix

\section{Scattering amplitude}
\label{app.A}
Here we derive Eq.~(\ref{eq.20}), from Eq.~(\ref{eq.19}), first introducing an identity 
 \begin{eqnarray}
\nonumber \psi_{\mathbf{p}}(\mathbf{r})&=&
\frac{e^{\frac{i}{\hbar}\mathbf{r}\cdot\mathbf{p}}}{(2\pi\hbar)^{3/2}}
+\int d^{3}r'\langle\mathbf{r}|\frac{1}{2\epsilon_{\mathbf{p}}-\hat{H}_{0}+i\epsilon}|\mathbf{r}'\rangle
\langle\mathbf{r}'|\hat{V}|\psi_{\mathbf{p}}\rangle\\
\nonumber &=&\frac{e^{\frac{i}{\hbar}\mathbf{r}\cdot\mathbf{p}}}{(2\pi\hbar)^{3/2}}+
\int d^{3}r'\langle\mathbf{r}'|\hat{V}|\psi_{\mathbf{p}}\rangle
\int d^{3}p'\frac{\langle\mathbf{r}|\mathbf{p}'\rangle\langle\mathbf{p}'|\mathbf{r}'\rangle}
     {2\epsilon_{\mathbf{p}}-2\epsilon_{\mathbf{p}'}+i\epsilon}\\
&=&\frac{e^{\frac{i}{\hbar}\mathbf{r}\cdot\mathbf{p}}}{(2\pi\hbar)^{3/2}}+
\int d^{3}r'\langle\mathbf{r}'|\hat{V}|\psi_{\mathbf{p}}\rangle
\frac{m}{(2\pi\hbar)^{3}}\int d^{3}p'\frac{e^{\frac{i}{\hbar}\mathbf{p}'\cdot(\mathbf{r}-\mathbf{r}')}}
 {p^{2}-{p'}^{2}+i\epsilon}
 \label{eq.A3}
 \end{eqnarray} 
 Let us consider the last integral in spherical coordinates
 \begin{eqnarray}
\nonumber  \frac{m}{(2\pi\hbar)^{3}}\int d^{3}p'\frac{e^{\frac{i}{\hbar}\mathbf{p}'\cdot(\mathbf{r}-\mathbf{r}')}}{p^2-p'^{2}+i\epsilon}&=&
 \frac{2\pi m}{(2\pi\hbar)^{3}}\int_{-1}^{1} d(\cos \theta)\int_{0}^{\infty} dp' \ p'^{2}\frac{e^{\frac{i}{\hbar}p'|\mathbf{r}-\mathbf{r}'|\cos\theta}}{p^2-p'^{2}+i\epsilon}\\
\nonumber  &&\hspace{-1.3cm}=\frac{m}{i|\mathbf{r}-\mathbf{r}'|(2\pi\hbar)^{2}}\int_{0}^{\infty} dp'\frac{p'}{p^2-p'^{2}+i\epsilon}
	\left(e^{\frac{i}{\hbar}p'|\mathbf{r}-\mathbf{r}'|}
	-e^{-\frac{i}{\hbar}p'|\mathbf{r}-\mathbf{r}'|}\right)\\	
\nonumber  &&\hspace{-1.3cm}=\frac{m}{i|\mathbf{r}-\mathbf{r}'|(2\pi\hbar)^{2}}\int_{-\infty}^{\infty} dp'
	\frac{p' \,e^{\frac{i}{\hbar}p' |\mathbf{r}-\mathbf{r}'|}}{p^2-p'^{2}+i\epsilon}\\
	&&\hspace{-1.3cm}=-\frac{m}{4\pi\hbar^{2}}
	\frac{e^{\frac{i}{\hbar}p |\mathbf{r}-\mathbf{r}'|}}{|\mathbf{r}-\mathbf{r}'|}
 \end{eqnarray}
 which, inserted in Eq.~(\ref{eq.A3}), gives Eq.~(\ref{eq.19}). 
In the large $r$ limit, using the expansion  $\frac{1}{|\mathbf{r}-\mathbf{r}'|}=\frac{1}{r}+O(r^{-2})$, we get
 \begin{equation}
\psi_{\mathbf{p}}(\mathbf{r})=
\frac{e^{\frac{i}{\hbar}\mathbf{r}\cdot\mathbf{p}}}{(2\pi\hbar)^{3/2}}
-\frac{m}{4\pi r\hbar^{2}}e^{\frac{i}{\hbar}pr}\int d^{3}r'
e^{-\frac{i}{\hbar}p\frac{\mathbf{r}\cdot\mathbf{r}'}{r}}
\langle\mathbf{r}'|\hat{V}|\psi_{\mathbf{p}}\rangle
+O(r^{-2})
 \end{equation}
Defining $\mathbf{q}=\frac{p}{r}\mathbf{r}$, we finally obtain
 \begin{eqnarray}
\nonumber \psi_{\mathbf{p}}(\mathbf{r})&=&\frac{e^{\frac{i}{\hbar}\mathbf{r}\cdot\mathbf{p}}}{(2\pi\hbar)^{3/2}}
-\frac{m}{4\pi r\hbar^{2}}e^{\frac{i}{\hbar}pr}\int d^{3}r'
e^{-\frac{i}{\hbar}\mathbf{q}\cdot\mathbf{r}'}
\langle\mathbf{r}'|\hat{V}|\psi_{\mathbf{p}}\rangle
+O(r^{-2})\\
\nonumber &=&\frac{e^{\frac{i}{\hbar}\mathbf{r}\cdot\mathbf{p}}}{(2\pi\hbar)^{3/2}}
-\frac{m(2\pi\hbar)^{3/2}}{4\pi r\hbar^{2}}e^{\frac{i}{\hbar}pr}\int d^{3}r'
\langle\mathbf{q}|\mathbf{r}'\rangle
\langle\mathbf{r}'|\hat{V}|\psi_{\mathbf{p}}\rangle
+O(r^{-2})\\
\nonumber &=&\frac{e^{\frac{i}{\hbar}\mathbf{r}\cdot\mathbf{p}}}{(2\pi\hbar)^{3/2}}
-\frac{m(2\pi\hbar)^{3/2}}{4\pi r\hbar^{2}}e^{\frac{i}{\hbar}pr}
\langle\mathbf{q}|\hat{V}|\psi_{\mathbf{p}}\rangle
+O(r^{-2})\\
&=&\frac{1}{(2\pi\hbar)^{3/2}}
\left\{
e^{\frac{i}{\hbar}\mathbf{r}\cdot\mathbf{p}}
+f(\mathbf{q},\mathbf{p})\frac{e^{\frac{i}{\hbar}pr}}{r}
\right\}
+O(r^{-2})
 \end{eqnarray}
where $f(\mathbf{q},\mathbf{p})$ is defined in Eq.~(\ref{eq.f}).

\section{Partial wave expansion}
\label{app.B}

Let us consider the scattering state at large distances 
 \begin{equation}
   \psi_{\mathbf{p}}(\mathbf{r})=e^{i\mathbf{p}\cdot\mathbf{r}}
   +f(\mathbf{q},\mathbf{p})\frac{e^{ip r}}{r}
   +O(r^{-2})
 \end{equation}
 where $\mathbf{q}=p\mathbf{r}/r$ and $p\rightarrow \hbar p$, to get rid of $\hbar$ in the expression, and where we drop, for simplicity, the prefactor 
$1/(2\pi)^3$. 
 For elastic scattering on a short-range spherically symmetric potential, the scattering amplitude can depend now only on the modulus $p$ and the angle $\theta=\arccos(\frac{\mathbf{q}\cdot\mathbf{p}}{p^{2}})$. It is so correct to perform a spherical wave expansion of the function $f(p,\theta)$ as in Eq.~(\ref{f_expand}).  
Let us expand the plane wave also on this basis, 
 \begin{equation}
   e^{i\mathbf{p}\cdot\mathbf{r}}=\sum_{\ell=0}^{\infty}(2\ell+1)i^{\ell}\jmath_{\ell}(pr)P_{\ell}(\cos \theta)
 \end{equation} 
The $\jmath_{\ell}(pr)$ are the usual spherical Bessel functions, whose large distance expansion is
 \begin{equation}
   i^{\ell}\,\jmath_{\ell}(pr)=i^{\ell}\frac{e^{ipr-\frac{\ell\pi}{2}}-e^{-ipr+\frac{\ell\pi}{2}}}{2ipr}+O(r^{-2})=\frac{e^{ipr}-e^{-i(pr-\ell\pi)}}{2ipr}+O(r^{-2})
 \end{equation}
The scattering wavefunction is then a superposition of incoming and outgoing spherical waves
 \begin{equation}
 \label{psi_expand}
   \psi_{\mathbf{p}}(\mathbf{r})=
   \frac{1}{2ipr}\sum_{\ell=0}^{\infty}(2\ell+1)
   \left[
     -e^{-i(pr-\ell\pi)}
     +(1+2ipf_{\ell}(p))e^{ipr}
     \right]
   P_{\ell}(\cos\theta)
   +O(r^{-2})
 \end{equation}
We found, then, that the complex amplitude of the outgoing spherical wave is modified by the presence of the interaction, however, to preserve normalization, its modulus has to be constant. A simple way to implement this condition is to introduce a phase $\delta_\ell (p)$ defined in Eq.~(\ref{f_cond}).
The scattering amplitude at fixed angular momentum is then given by Eq.~(\ref{amplitude_delta}). Inserting this result in Eq.~(\ref{psi_expand}) we get
 \begin{eqnarray}
\nonumber   \psi_{\mathbf{p}}(\mathbf{r})&=&
   \frac{1}{2ipr}
   \sum_{\ell=0}^{\infty}(2\ell+1)i^{\ell} 
   \left[
     +e^{2i\delta_{\ell}(p)}e^{i(pr+\frac{\ell\pi)}{2}}
     -e^{-i(pr-\frac{\ell\pi}{2})}
     \right]P_{\ell}(\cos\theta)
   +O(r^{-2})\\
\nonumber   &=&\sum_{\ell=0}^{\infty}(2\ell+1)i^{\ell}e^{i\delta_{\ell}(p)} 
   \frac{\sin\left(p r-\ell\pi+\delta_{\ell}(p)\right)}{pr}
   P_{\ell}(\cos\theta)
   +O(r^{-2})\\
   &\equiv& \sum_{\ell=0}^{\infty}(2\ell+1)i^{\ell}\,
   \psi_{\ell,\mathbf{p}}(\mathbf{r})
   P_{\ell}(\cos\theta)
   +O(r^{-2})    
    \label{scattering_state}  
 \end{eqnarray}
 from which we can read the coefficients of the wavefunction expansion. 
The scattering properties of particles are defined in relation to the noninteracting case by the phase between the free wave and the scattered one. This is valid at large distances where $\delta_{\ell}$ contains all the informations about the scattering.
The scattering amplitude and the cross section can be written as follows
 \begin{equation}
   f_{\ell}(p)=\frac{e^{i\delta_{\ell}(p)}\sin(\delta_{\ell}(p))}{p}
 \end{equation}
 Using the orthogonality of the Legendre polynomials, $\int^1_{-1}P_\ell(x)P_n(x)dx=\frac{2\,\delta_{n\ell}}{2\ell+1}$, after integrating the differential cross section $d\sigma/d\Omega=|f(p,\theta)|^2$, one gets 
 \begin{equation}
   \sigma(p)=\sum_{\ell=0}^{\infty}
   \frac{4\pi}{p^{2}}(2\ell+1)\sin^{2}(\delta_{\ell}(p))
      \end{equation} 
For identical spinless (polarized) particles it is also necessary to (anti)symmetrize the wavefunction 
 \begin{equation}
   \Psi(\mathbf{r}_{1},\mathbf{r}_{2})=\xi\Psi(\mathbf{r}_{2},\mathbf{r}_{1})
   \longrightarrow
   \psi_{\mathbf{p}}(\mathbf{r})=\xi\psi_{\mathbf{p}}(-\mathbf{r})
 \end{equation}
 where $\xi=+1$ for bosons and $\xi=-1$ for fermions. The unnormalized wavefunction has then two contributions
 \begin{equation}
   \psi_{\mathbf{p}}(\mathbf{r})=
   e^{i\mathbf{p}\cdot\mathbf{r}}+\xi e^{-i\mathbf{p}\cdot\mathbf{r}}
   +f(p,\theta)\frac{e^{ipr}}{r}
   +\xi f(p,\pi-\theta)\frac{e^{-ipr}}{r}
   +o(r^{-2})
 \end{equation}
therefore the differential cross section becomes
 \begin{equation}
   \frac{d\sigma}{d\Omega}=|f(p,\theta)+\xi f(p,\pi-\theta)|^{2}
 \end{equation}
Since the Legendre polynomials have parity $(-1)^{\ell}$, some partial wave contributions to the cross section vanish because of the statistics so that the final results are
 \begin{equation}
 \label{cross_nospin}
   \begin{matrix}
     \sigma(p)=\frac{8\pi}{p^{2}}\sum\limits_{\ell \, even}(2\ell+1)\sin^{2}(\delta_{\ell}(p))\, \;\;
      \; \textrm{for spinless bosons},\\
      \sigma(p)=\frac{8\pi}{p^{2}}\sum\limits_{\ell \, odd}(2\ell+1)\sin^{2}(\delta_{\ell}(p))\, \;\;
       \;\textrm{for spinless fermions}.
	\end{matrix}
  \end{equation}
In the presence of particles with spin, instead, it is necessary to take into account the spin part of the wavefunction. For example, for fermions, if the scattering happens in a singlet (antisymmetric) state $|S\ket$, we should impose that the wavefuction is symmetric in the position space
 \begin{equation}
   |\psi_{\mathbf{p}}(\mathbf{r})\ket=
   \left[e^{i\mathbf{p}\cdot\mathbf{r}}+e^{-i\mathbf{p}\cdot\mathbf{r}}
     +f(p,\theta)\frac{e^{ipr}}{r}
     +f(p,\pi-\theta)\frac{e^{-ipr}}{r}
     \right]|S\ket
 \end{equation}
By these considerations the cross sections for singlet ($\sigma_{S}$) and triplet ($\sigma_{T}$) spin-$1/2$ fermions result
 \begin{equation}
 \label{cross_spin}
   \begin{matrix}
     \sigma_{S}(p)=\frac{8\pi}{p^{2}}\sum\limits_{\ell \, even}(2\ell+1)\sin^{2}(\delta_{\ell}(p))\,,
      \\
      \sigma_{T}(p)=\frac{8\pi}{p^{2}}\sum\limits_{\ell \, odd.}(2\ell+1)\sin^{2}(\delta_{\ell}(p))\,.
      \end{matrix}
 \end{equation}

 \section{Asymptotic solutions of the Schr\"odinger equation}
\label{app.C} 
 
We saw that in the long range limit all the physics is contained in the phase shift 
which can be obtained by imposing the scattering solution 
getting the form for the wavefunction as in Eq.~\eqref{scattering_state}.
One can derive, therefore,  the shift $\delta_{\ell}(p)$ solving the Schr\"odinger equation, 
$ \left(-\frac{\nabla^{2}}{m}+{V}(r)\right)\psi(\mathbf{r}) = E\psi(\mathbf{r})$, with $E=p^2/m$, which can be written in spherical coordinates
 \begin{align}
   \left[
     \frac{1}{r^{2}}\frac{\partial}{\partial r}\left(r^{2}\frac{\partial}{\partial r}\right)
     -\frac{\hat{L}^{2}}{r^{2}}
     +p^{2}
     -{U}(r)
     \right]
   \psi(\mathbf{r}) &=0
 \end{align}
with ${U}(r)=m{V}(r)$. The wavefunction can be expanded in the eigenstates of $\hat{L}^{2}$ and $\hat{L}_{z}$ 
 \begin{equation}
   \psi_{\mathbf{p}}(\mathbf{r})=
   \sum_{\ell=0}^{\infty}\sum_{m=-\ell}^{\ell}
   Y_{\ell}^{m}(\theta,\varphi)\frac{u_{p,\ell}(r)}{r}
 \end{equation}
so that one can write the radial Schr\"odinger equation 
 \begin{equation}
   \left[
     \frac{1}{r^{2}}\frac{d}{dr}\left(r^{2}\frac{d}{dr}\right)
     +p^2+\frac{\ell(\ell+1)}{r^2}-U(r)
     \right]\frac{u_{p,\ell}(r)}{r}=0
     \label{radialSchr}
 \end{equation}
Expanding the first term
\begin{equation}
\label{exp_deriv}
   \frac{1}{r^{2}}\frac{d}{dr}\left(r^{2}\frac{d}{dr}\frac{u_{p,\ell}(r)}{r}\right)=
   \frac{1}{r}\frac{d^{2}u_{p,\ell}(r)}{dr^{2}}+
   u_{p,\ell}(r)\frac{1}{r^{2}}\left[2r\frac{d}{dr}+r^{2}\frac{d^{2}}{dr^{2}}\right]
   \left(\frac{1}{r}\right)
 \end{equation}
 we recognize in the second r.h.s. term of Eq.~(\ref{exp_deriv}) the Poisson equation $\nabla^2_{\bf{r}} \left(\frac{1}{r}\right)=-4\pi\delta(\bf{r})$, so that Eq.~(\ref{radialSchr}) can be rewritten in the form
 \begin{equation}
   \frac{1}{r}\frac{d^{2}u_{p,\ell}(r)}{dr^{2}}
   -\frac{u_{p,\ell}(r)}{r^{2}} \,\delta(r)
   +\left[p^2+\frac{\ell(\ell+1)}{r^2}-U(r)\right]\frac{u_{p,\ell}(r)}{r}=0
   \label{eq_radial}
 \end{equation}
The relevance of the Dirac delta term has been discussed in Ref.~\citep{radial_boundary}. If the potential did not contain a delta function, Eq.~(\ref{eq_radial}) would be the usual radial equation, but then we should impose the boundary condition
 \begin{equation}
   -\delta(r)\frac{u_{p,\ell}(r)}{r^{2}}=0
 \end{equation}
which implies 
 \begin{equation}
   \int d^{3}r\frac{u_{p,\ell}(r)}{r^{2}}\delta(r)=
   -4\pi\int drr^{2}\frac{u_{p,\ell}(r)}{r^{2}}\delta(r)=-4\pi u_{p,\ell}(0)=0
 \end{equation}
 We have, therefore, to solve 
 \begin{equation}
   \frac{d^{2}u_{p,\ell}(r)}{dr^{2}}
   +\left[p^2+\frac{\ell(\ell+1)}{r^2}-U(r)\right]u_{p,\ell}(r)=0 \hspace{1cm} (r>0)
 \end{equation}
 with the boundary condition $u_{p,\ell}(0)=0$. 
Considering a \emph{short range} potential, meaning that it can be neglected beyond some radius
 \begin{equation}
   V(r)\approx0 \hspace*{1cm} \forall \ \ r>r^{*}
 \end{equation}
the problem is split in two domains
 \begin{eqnarray}
&&   \frac{d^{2}u_{p,\ell}(r)}{dr^{2}}
   +\left[p^2+\frac{\ell(\ell+1)}{r^2}\right]u_{p,\ell}(r)=0 \hspace{1cm} (r>r^{*})\\
&&   \frac{d^{2}u_{p,\ell}(r)}{dr^{2}}
   +\left[p^2+\frac{\ell(\ell+1)}{r^2}-U(r)\right]u_{p,\ell}(r)=0 \hspace{1cm} (r<r^{*})
 \end{eqnarray}
 always with the condition $u_{p,\ell}(0)=0$. 
 The radial Schr\"odinger equation, for $r>r^{*}$, is a linear differential equation of the second order, so two independent solutions are expected. \cite{Taylor} These are the Riccati-Bessel functions $\hat{\jmath}_{\ell}(z)$, with $z=pr$, 
 \begin{equation}
   \hat{\jmath}_{\ell}(z)=z\,\jmath_{\ell}(z)=\left(\frac{\pi z}{2}\right)^{1/2}J_{\ell+1/2}(z)=
   z^{\ell+1}\sum_{n=0}^{\infty}\frac{(-z^{2}/2)^{n}}{n!(2\ell+2n+1)!!}
 \end{equation}
where $\jmath_{\ell}(z)$ are the spherical Bessel functions and $J_{\lambda}(z)$ the ordinary Bessel functions. Their relevant asymptotic behaviors are 
 \begin{eqnarray}
 \label{asymptotic_j}
   &&  \hat{\jmath}_{\ell}(z)\underset{z\rightarrow0}{\sim}\frac{z^{\ell+1}}{(2\ell+1)!!}\\
   &&  \hat{\jmath}_{\ell}(z)\underset{z\rightarrow\infty}{\sim}\sin\left(z-\frac{\ell\pi}{2}\right)
 \end{eqnarray}
A second set of solutions are given by the Riccati-Neumann functions $\hat{n}_{\ell}(z)$:
 \begin{equation}
   \hat{n}_{\ell}(z)
   =(-)^{\ell}\left(\frac{\pi z}{2}\right)^{1/2}J_{-\ell-1/2}(z)=
   z^{-\ell}\sum_{n=0}^{\infty}\frac{(-z^{2}/2)^{n}(2\ell+2n-1)!!}{n!}
 \end{equation}
whose asymptotic behaviors are
 \begin{eqnarray}
 \label{asymptotic_n}
&&     \hat{n}_{\ell}(z)\underset{z\rightarrow0}{\sim}z^{-\ell}(2\ell-1)!!\\
&&     \hat{n}_{\ell}(z)\underset{z\rightarrow\infty}{\sim}\cos\left(z-\frac{\ell\pi}{2}\right) 
  \end{eqnarray}
Some examples of Riccati functions for the first angular momenta are reported in this table.
\begin{table}[h!]
 \begin{center}
 \begin{tabular}{|c||c|c|c|}
 \hline
   &$\ell=0$&$\ell=1$&$\ell=2$\\\hline\hline
$\hat{\jmath}_{\ell}(z)$&$\sin{z}$&$\frac{1}{z}\sin{z}-\cos{z}$&$\left(\frac{3}{z^{2}}-1\right)\sin{z}-\frac{3}{z}\cos{z}$\\\hline
$\hat{n}_{\ell}(z)$&$\cos{z}$&$\frac{1}{z}\cos{z}+\sin{z}$&$\left(\frac{3}{z^{2}}-1\right)\cos{z}+\frac{3}{z}\sin{z}$\\
\hline
 \end{tabular}
 \end{center}
\caption{The Riccati-Bessel, $\hat{\jmath}_{\ell}(z)$, and the Riccati-Neumann, $\hat{n}_{\ell}(z)$, functions for the first angular momenta $\ell=0,1,2$.}
\end{table}

\noindent The complete solution, for $r>r^{*}$, is then given by a linear superposition of $\hat{\jmath}_{\ell}(pr)$ and $\hat{n}_{\ell}(pr)$
 \begin{equation}
   u_{p,\ell}(r)=A_{p,\ell} \,\hat{\jmath}_{\ell}(pr)+B_{p,\ell} \,\hat{n}_{\ell}(pr)\,, \hspace{0.5cm} \textrm{for}\; r>r^{*}
   \label{scatt_solution1}
 \end{equation}
 with $A_{p,\ell}$ and $B_{p,\ell}$ some coefficients. 
In the noninteracting problem the solution is fixed by the boundary condition $u_{p,\ell}(0)=0$ that excludes $\hat{n}_{\ell}(pr)$  because of its divergence at the origin, getting
 \begin{equation}
   u^{0}_{p,\ell}(r)\propto\hat{\jmath}_{\ell}(pr)\underset{r\rightarrow\infty}{\sim}
   \sin\left(pr-\frac{\ell\pi}{2}\right)\,.
 \end{equation}
In the interacting case instead, the Schr\"odinger equation has to be solved also for $r<r^{*}$ and then the continuity of the wavefunction and of its first derivative should be imposed at $r\simeq r^{*}$. This is clearly the difficult part of the calculation and an exact solution is known only for a small number of potentials. 
In Appendix \ref{app.B} it has been shown that the scattering solution has the specific form (\ref{scattering_state})  at large distances. Imposing this behavior to the general solution (\ref{scatt_solution1}) 
 \begin{equation}
   u_{p,\ell}(r)\underset{r\rightarrow\infty}{\sim}A_{p,\ell}\,\sin\left(pr-\frac{\ell\pi}{2}\right)+B_{p,\ell}\,\cos\left(pr-\frac{\ell\pi}{2}\right) = C_{p,\ell} \,\sin\left(pr-\frac{\ell\pi}{2}+\delta_\ell(p)\right)
 \end{equation}
allows us to express the whole solution in terms of the phase shift. As a result, up to an overall prefactor the solution of the radial equation, for $r>r^{*}$, is given by
 \begin{equation}
   u_{p,\ell}(r)\propto \, \hat{\jmath}_{\ell}(pr)+\tan\big(\delta_{\ell}(p)\big)\, \hat{n}_{\ell}(pr)\,, \hspace{0.5cm} \textrm{for}\;  r>r^{*}.
   \label{scatt_solution}
 \end{equation}
Interestingly, we notice that the condition for having in Eq.~(\ref{scatt_solution}) both the contributions of the same order, also in the limit $p\rightarrow 0$, using Eqs.~(\ref{asymptotic_j}) and (\ref{asymptotic_n}), is that $\delta_\ell(p) \sim  p^{2\ell+1}$, which is Eq.~(\ref{delta_psmall}). 
The expansion of the radial wavefunction \eqref{scatt_solution} at $\ell=0$, at low energy, gives
 \begin{equation}
   u_{p,0}(r)\underset{p\rightarrow0}{\propto} 1-\frac{r}{\as}\,, \hspace{1cm} r>r^{*}
 \end{equation}
where $a_s$ is defined in Eq.~(\ref{def_as}), 
therefore the corresponding wavefunction is
 \begin{equation}
   \psi_p(r)\underset{p\rightarrow0}{\propto} \frac{1}{r}-\frac{1}{\as}\,, \hspace{1cm} r>r^{*}
 \end{equation}
The same form for the wavefunction can be obtained imposing the so-called Bethe-Peierls boundary conditions
 \begin{equation}
   \left.\frac{1}{r\psi_p(r)}\frac{d\left(r\psi_p(r)\right)}{dr}\right|_{r=0}=\frac{1}{\as}\,.
   \label{BPboundary}
 \end{equation}
One should evaluate the scattering length for a realistic interatomic potential, however in many cases this requires difficult calculations. For this reason it is often convenient to use a simpler pseudo-potential that can reproduce the same scattering amplitude at low energies. 
Fortunately, the relevant experimental parameter in ultracold atomic physics, tunable by the so-called Feshbach resonance technique, is the scattering length $a_s$.

\subsubsection*{Contact potentials}
The simplest short-range potential is the contact interaction described by a Dirac-delta function in a regularized form or in a bare form.
\paragraph*{Regularized delta-funcion.}
 Let us consider the following potential
 \begin{equation}
 \label{regular_delta}
   V(\mathbf{r})=g\,\delta^{3}(\mathbf{r})\frac{\partial}{\partial r}r
 \end{equation}
and write down the $s$-wave radial equation from Eq.~\eqref{eq_radial}
 \begin{equation}
   \frac{1}{r}\frac{d^{2}u_{p,\ell}(r)}{dr^{2}}
   -\frac{u_{p,\ell}(r)}{r^{2}}\,\delta(r)
   +\left[p^2+\frac{\ell(\ell+1)}{r^2}-\frac{mg}{4\pi r^{2}\hbar^{2}}\delta(r)\frac{\partial}{\partial r}r\right]\frac{u_{p,\ell}(r)}{r}=0
 \end{equation}
 Recasting the terms we have
 \begin{equation}
   \frac{1}{r}\frac{d^{2}u_{p,\ell}(r)}{dr^{2}}
   +\left[p^2+\frac{\ell(\ell+1)}{r^2}\right]\frac{u_{p,\ell}(r)}{r}
   -\delta(r)\frac{1}{r^2}\left[
   u_{p,\ell}(r)   +
   \frac{mg}{4\pi \hbar^{2}}\frac{du_{p,\ell}(r)}{dr}\right]=0
 \end{equation}
For $\ell=0$, the $s$-wave term, we have to solve the following set of equations 
 \begin{eqnarray}
   \frac{d^{2}u_{p,0}(r)}{dr^{2}}+p^2u_{p,0}(r)=0\\
   4\pi u_{p,0}(0)+\frac{mg}{\hbar^{2}}\frac{du_{p,0}}{dr}(0)=0
 \end{eqnarray}
The solution of the first equation, as already seen, is  
 \begin{equation}
 \label{up0r}
   u_{p,0}(r)\propto \frac{1}{2ip}\left( -e^{-ipr}+e^{2i\delta_{0}(p)}e^{ipr}\right)=
   \frac{e^{i\delta_{0}(p)}}{p}\sin\left(pr+\delta_{0}(p)\right)
 \end{equation}
so that the boundary condition can be written as it follows
 \begin{equation}
   \frac{u_{p,0}'(0)}{u_{p,0}(0)}=p \,\cot(\delta_{0}(p))=-\frac{4\pi \hbar^{2}}{mg}
 \end{equation}
 Using the definition of the scattering length as in Eq.~(\ref{cot_delta}), with zero effective range, one finds that, in order 
 to model a realistic potential at low energy it is sufficient to use Eq.~(\ref{regular_delta}) with 
 \begin{equation}
   g=\frac{4\pi\hbar^{2}}{m}\as\,.
 \end{equation}

\paragraph*{Bare delta-function.}
Another relevant contact potential is simply
  \begin{equation}
    V(\mathbf{r})=g\,\delta^3(\mathbf{r})
 \end{equation}
This potential is widely employed in the literature, 
even though it leads to ultraviolet divergences. The radial equation, in this case, becomes
 \begin{equation}
   -\frac{\delta(r)}{r^{2}}\left[u_{p,\ell}(r)+\frac{mg}{4\pi\hbar^{2}}\frac{u_{p,\ell}(r)}{r}\right]
   +\frac{1}{r}\frac{d^{2}u_{p,\ell}(r)}{dr^{2}}
   +\left[p^2+\frac{\ell(\ell+1)}{r^2}\right]\frac{u_{p,\ell}(r)}{r}=0
 \end{equation}
For the $s$-wave term we have to solve the following equations
 \begin{eqnarray}
\label{prima}
 &&  \frac{d^{2}u_{p,0}(r)}{dr^{2}}+p^2u_{p,0}(r)=0\\
 &&  \frac{\delta(r)}{r^{2}}\left[u_{p,0}(r)+\frac{mg}{4\pi\hbar^{2}}\frac{u_{p,0}(r)}{r}\right]=0
 \end{eqnarray}
Integrating the second equation we get
 \begin{equation}
   u_{p,0}(0)+\frac{mg}{4\pi\hbar^{2}}\int dr\delta(r)\frac{u_{p,0}(r)}{r}=0
 \end{equation}
Expanding $u_{p,0}(r)$ around $r=0$, only the first two terms survive, and the equation becomes
 \begin{equation}
 \label{boundary_bare}
   u_{p,0}(0)
   +\frac{mg}{4\pi\hbar^{2}}u'_{p,0}(0)
   +\frac{mg}{4\pi\hbar^{2}}u_{p,0}(0)\int dr\frac{\delta(r)}{r}=0
 \end{equation}
Using the solution of Eq.~(\ref{prima}) which is 
Eq.~(\ref{up0r}), we get again $u'_{p,0}(0)=u_{p,0}(0)\,p\cot(\delta_p(0))=-u_{p,0}(r)/a_s$, 
and dividing by $u_{p,0}(r)$ we obtain
 \begin{equation}
   \frac{1}{g}=\frac{m}{4\pi\hbar^{2}\as}
   -\frac{m}{4\pi\hbar^{2}}\int dr\frac{\delta(r)}{r}\,.
 \end{equation}
where the last term diverges. In order to find a clearer understanding of the relation between $g$ and $a_s$ one has to resort to 
the formalism of the two-body scattering matrix  $T_{2B}$, see Sec.~\ref{Matrice di scattering a due corpi}

\section{Scattering problem with spin-orbit coupling}
\label{Accoppiamento Spin-Orbita}

In this Appendix we will report the results of Refs.~\citep{Zhang1,Zhang2}, using also the same notation. 
The system of fermions with spin (or pseudo-spin) in the presence of spin-orbit interaction can be characterized by the following single particle Hamiltonian
 \begin{equation}
   \hat{H}_{1b}=\frac{\hbar^{2}{P}^{2}}{2m}+\lambda\mathbf{M}\cdot\mathbf{{P}}+Z
   \label{H1b}
 \end{equation}
where $\mathbf{M}=\left(\sum_{i}a_{i}\sigma_{i} , \sum_{i}b_{i}\sigma_{i} , \sum_{i}c_{i}\sigma_{i}\right)$ and with $\sigma_{i}$ the usual Pauli matrices for the spin. We will treat later, for simplicity, only the case $\lambda=1$, $\mathbf{M}=\left(\vd\sigma_{y} , \vR\sigma_{x} ,0\right)$ and $Z=0$, so a spin-orbit coupling made of a Rashba and a Dresselhaus terms. However some important results can be obtained from the most general Hamiltonian \eqref{H1b}. 
The Hilbert space of the problem can be expressed as 
  $ \mathcal{H}=\mathcal{H}_{r}\otimes\mathcal{H}_{s_{1}}\otimes\mathcal{H}_{s_{2}}$, 
the product of $\mathcal{H}_{r}$, the relative motion space and the spin spaces. For clarity different brackets are associated to different spaces, 
 $  |\cdot\rangle\!\rangle\in\mathcal{H}$, 
  $ |\cdot)\in\mathcal{H}_{r}$, 
   $|\cdot\rangle\in\mathcal{H}_{s_{1}}\otimes\mathcal{H}_{s_{2}}$, 
so that the position dependent states of the spin is such that $(\mathbf{r}|\psi\rangle\!\rangle=|\psi(\mathbf{r})\rangle$. 
The same study on two-body problem, seen before, should be done also in this case, always for a spherically symmetric and spin-independent potential $\hat{U}({X}^{(1)}\!-\!{X}^{(2)})$, where ${X}^{(i)}$ are the coordinates of the 
particles. The total two-body Hamiltonian is then
 \begin{equation}
   \hat{H}_{2b}=\hat{H}_{1b}^{(1)}+\hat{H}_{1b}^{(2)}+\hat{U}({X}^{(1)}\!-\!{X}^{(2)})
 \end{equation}
This can be easily separated in the Hamiltonian of the center of mass and Hamiltonian of the relative motion. The latter Hamiltonian is 
 \begin{equation}
   \hat{H}=\frac{{p}^{2}}{m}+\lambda\mathbf{c}\cdot\mathbf{{p}}+B(\mathbf{K})+\hat{U}(\mathbf{r})\equiv\hat{H}_{0}+\hat{U}(\mathbf{r})
   \label{spinorbitH}
 \end{equation}
with the definitions of the relative quantities,
  $\mathbf{{p}}=\frac{\mathbf{{P}}^{(1)}-\mathbf{{P}}^{(2)}}{2},\hspace{0.3cm}
  \mathbf{{K}}=\mathbf{{P}}^{(1)}+\mathbf{{P}}^{(2)},\hspace{0.3cm}
  \mathbf{r}=\mathbf{{X}}^{(1)}-\mathbf{{X}}^{(2)},\hspace{0.3cm}
\mathbf{c}=\mathbf{M}^{(1)}-\mathbf{M}^{(2)}, \hspace{0.3cm}
B(\mathbf{{K}})=Z^{(1)}+Z^{(2)}+\lambda\mathbf{K}\cdot\left(\mathbf{M}^{(1)}+\mathbf{M}^{(2)}\right), \hspace{0.3cm}
m=\frac{m_{1}m_{2}}{m_{1}+m_{2}}$.\\
The total momentum $\mathbf{K}$ is conserved during the scattering process and it can be treated as a constant. 
A basis of eigenstates for the free Hamiltonian is given by the tensor product of eigenstates of the momentum operator and of spin eigenstates $|\psi_{\mathbf{k},\alpha}^{(0)}\ketket=|\mathbf{k})\otimes|\mathbf{k},\alpha\rangle$ with $\alpha=0,1,2,3$, labeling the four configurations of two spins, also momentum dependent, which in relative position representation are
$   (\mathbf{r}|\psi_{\mathbf{k},\alpha}^{(0)}\ketket=
   |\psi_{\mathbf{k},\alpha}^{(0)}(\mathbf{r})\ket=
   \frac{1}{(2\pi)^{3/2}}e^{i\mathbf{k}\mathbf{r}}|\mathbf{k},\alpha\rangle$. 
Since the momentum part of the eigenfunction is known (plane waves as in the previous case) the eigenproblem can be restricted exclusively to the spin part. 
In the relative coordinates the Hamiltonian $\hat{H}_{0}$ can be decomposed in the following form
$   \hat{H}_{0}=\hat{H}_{1b}^{(1)}\otimes\mathbb{1}_{2\times2}+\mathbb{1}_{2\times2}\otimes\hat{H}_{1b}^{(2)} \label{H0spin}$
with
 \begin{equation}
   \begin{matrix}
     \hat{H}_{1b}^{(1)}=\frac{p^{2}}{2m}\mathbb{1}_{2\times2}+\lambda\mathbf{{M}}\cdot\mathbf{p}+Z+\frac{\lambda}{2}\mathbf{{M}}\cdot\mathbf{K}\vspace{0.2cm}\\
     \hat{H}_{1b}^{(2)}=\frac{p^{2}}{2m}\mathbb{1}_{2\times2}-\lambda\mathbf{{M}}\cdot\mathbf{p}+Z+\frac{\lambda}{2}\mathbf{{M}}\cdot\mathbf{K}
     \end{matrix}
 \end{equation}
Because of the structure of the operators $\mathbf{{M}}$ and $Z$ the single particle Hamiltonian can be written, with $\xi=(\mathbf{K},\lambda,a_{i},b_{i},c_{i},d_{i})$, as
 \begin{equation}
   \hat{H}_{1b}^{(1)}=
   \frac{p^{2}}{2m}\mathbb{1}_{2\times2}+
   \beta(\mathbf{p},\xi)\sigma_{x}+
   \gamma(\mathbf{p},\xi)\sigma_{y}+
   \eta(\mathbf{p},\xi)\sigma_{z}=
   \begin{pmatrix}
     \frac{p^{2}}{2m}+\eta & \beta-i\gamma\\
     \beta+i\gamma & \frac{p^{2}}{m}-\eta
     \end{pmatrix}
 \end{equation}
The coefficients of the Pauli matrices depend on the specific parameters of the model and linearly by the momentum; for example, $\eta$ should have the form 
$\eta(\mathbf{p},\xi)=\mathbf{v}(\xi)\!\cdot\!\mathbf{p}+q(\xi)$, for some $\mathbf{v}$ and $q$ model dependent set of parameters.
The eigenvalues of $\hat{H}_{1b}^{(1)}$ are given by 
 \begin{equation}
   E^{(1)}_{\pm}=\frac{p^{2}}{2m}\pm\sqrt{\eta^{2}+\beta^{2}+\gamma^{2}} \equiv
   \frac{p^{2}}{2m}\pm\varepsilon(\mathbf{p},\xi)
 \end{equation}
whose asymptotic behavior at large momenta, for the linearity of 
$\eta$, $\beta$ and $\gamma$, is given by 
 \begin{equation}
   \varepsilon(\mathbf{p},\xi)=\sqrt{\alpha^{2}+\beta^{2}+\gamma^{2}}\underset{p\rightarrow\infty}{\sim}\mathbf{m}(\xi)\!\cdot\!\mathbf{p}\,.
\label{asimptoticE}
 \end{equation}
For convention $|\spinup\ket_{1}$ is the eigenvector associated to $E_{+}$ and $|\spindown\ket_{1}$ to $E_{-}$. The same problem is identically solved for $H_{1b}^{(2)}$ with the exception of a sign which has to be considered and so the eigenvalues will have the same form but with a different sign in the spin-orbit part of the spectrum. The eigenvectors $|\spinup\ket_{2}$ and $|\spindown\ket_{2}$ are identically assigned and the energies are
 \begin{equation}
   E^{(2)}_{\pm}=\frac{p^{2}}{2m}\pm\varepsilon(-\mathbf{p},\xi)
 \end{equation}
With a proper unitary transformation it is always possible to put the Hamiltonian $H_0$ 
in a diagonal form
 \begin{equation}
   \hat{H}_{0}=
   \begin{pmatrix}
     \frac{p^{2}}{2m}+\varepsilon(\mathbf{p},\xi)&0\\
     0&\frac{p^{2}}{2m}-\varepsilon(\mathbf{p},\xi)
     \end{pmatrix}\otimes\mathbb{1}_{2}+\mathbb{1}_{2}\otimes
   \begin{pmatrix}
     \frac{p^{2}}{2m}+\varepsilon(-\mathbf{p},\xi)&0\\
     0&\frac{p^{2}}{2m}-\varepsilon(-\mathbf{p},\xi)
     \end{pmatrix}
 \end{equation}
The eigenvalues of the total Hamiltonian can be obtained building a basis of eigenstates; with a simple change in the notation $|m\ket_{i}=|m_{i}\ket$, with $m\!= \ \uparrow,\downarrow$, this can be done building the basis 
 \begin{equation}
   |m_{1},m_{2}\ket=|m_{1}\ket\otimes|m_{2}\ket
 \end{equation}
The associated eigenvalues are therefore
 \begin{equation}
   \begin{matrix}
     \vspace{0.1cm}
     \hat{H}_{0}|\spinup\!,\spinup\ket=
     \left[\frac{p^{2}}{m}+\varepsilon(\mathbf{p},\xi)+\varepsilon(-\mathbf{p},\xi)\right]|\spinup\!,\spinup\ket\\ \vspace{0.1cm}
     \hat{H}_{0}|\spinup\!,\spindown\ket=
     \left[\frac{p^{2}}{m}+\varepsilon(\mathbf{p},\xi)-\varepsilon(-\mathbf{p},\xi)\right]|\spinup\!,\spindown\ket\\ \vspace{0.1cm}
     \hat{H}_{0}|\spindown\!,\spinup\ket=
     \left[\frac{p^{2}}{m}-\varepsilon(\mathbf{p},\xi)+\varepsilon(-\mathbf{p},\xi)\right]|\spindown\!,\spinup\ket\\
     \hat{H}_{0}|\spindown\!,\spindown\ket=
     \left[\frac{p^{2}}{m}-\varepsilon(\mathbf{p},\xi)-\varepsilon(-\mathbf{p},\xi)\right]|\spindown\!,\spindown\ket\\
     \end{matrix}
 \end{equation}
The four states of spin are labeled by $\alpha=0,1,2,3$ and, for simplicity, we will indicate with $t=(\alpha,\mathbf{k})$, which implicitly depends on $\xi$, the specific eigenstate of the free Hamiltonian with eigenvalue $E_t$. To obtain a renormalization relation similar to Eq.~(\ref{gfixing}) it is necessary to write the Lippmann-Schwinger equation for the stationary scattering state
 \begin{equation}
   |\psi_{t}(\mathbf{r})\ket=
   |\psi_{t}^{(0)}(\mathbf{r})\ket+
   (\mathbf{r}|\frac{1}{E_{t}-\hat{H}_{0}+i0^{+}}\hat{V}|\psi_{t}\ketket
 \end{equation}
Inserting a completeness relation in the position space
 \begin{eqnarray}
\nonumber   |\psi_{t}(\mathbf{r})\ket&=&
   |\psi_{t}^{(0)}(\mathbf{r})\ket+
   \int d^{3}r'(\mathbf{r}|\frac{1}{E_{t}-\hat{H}_{0}+i0^{+}}|\mathbf{r}')\hat{V}(\mathbf{r}')|\psi_{t}(\mathbf{r}')\ket\\
   &=&|\psi_{t}^{(0)}(\mathbf{r})\ket+
   \int d^{3}r'G_{0}(E_{t};\mathbf{r},\mathbf{r}')\hat{V}(\mathbf{r}')|\psi_{t}(\mathbf{r}')\ket
 \end{eqnarray}
The main difficulties in the presence of a SO term are in the Green's function
 \begin{equation}
   G_{0}(E;\mathbf{r},\mathbf{r}')=
   \sum_{\alpha}\int d^{3}k(\mathbf{r}|
   \frac{1}{E-\hat{H}_{0}+i0^{+}}
|\psi_t^{(0)}\ketket \brabra \psi_t^{(0)}|\mathbf{r}')
   =\sum_{\alpha}\int d^{3}k
   \frac{e^{i\mathbf{k}\cdot(\mathbf{r}-\mathbf{r}')}}{E-E_{t}+i0^{+}}
   |\mathbf{k},\alpha\ket\bra\mathbf{k},\alpha|.
 \end{equation}
The breaking of the rotational symmetry induced by the SO term makes hard the evaluation of the Green's function and its asymptotic behavior at large distances, therefore, the treatment done before is not easily adaptable to this case. 
However, as proved in Ref.~\citep{Zhang1}, the scattering amplitude for the 
two-body problem with SO coupling, in the low energy limit, also for a realistic short range potential (with $r^{*}$ the range) is
 \begin{equation}
\nonumber   
f(t',t)=-(2\pi)^{3}
   \frac{\bra\psi_{t'}^{(0)}(0)|S\ket\bra S|\psi_{t}^{(0)}(0)\ket}{1/\ar+iE_{t}^{1/2}-4\pi\bra S|F(E_{t})|S\ket}
   \label{scattering_amplitude_SO}
 \end{equation}
where the state $|S\rangle=\left( |\uparrow \downarrow\rangle-|\downarrow 
\uparrow\rangle\right)/\sqrt{2}$ is the singlet spin state and the operator  
$F(E)=\frac{1}{(2\pi)^3}\int dk^3\sum_{\alpha}\left[\frac{|\alpha,\mathbf{k}
\rangle\langle \alpha,\mathbf{k}|}{E+i0^+-E_{t}}-\frac{|\alpha,\mathbf{k}
\rangle\langle \alpha,\mathbf{k}|}{E+i0^+-k^2}\right]$. 
Specifically the authors of Ref.~\citep{Zhang2} were able to prove that, for $rk\ll1$, $r^{*}k\ll1$ and $\lambda r^{*}\ll1$, 
the scattering wavefunction must have the form 
 \begin{equation}
   |\psi_{t}(\mathbf{r})\ket\propto\left(\frac{1}{r}-\frac{1}{\ar}\right)|S\ket
   -i\frac{\lambda}{2}\cdot\left(\frac{\mathbf{r}}{r}\right)|S\ket 
 \end{equation}
This is a natural generalization of the Bethe-Peierls boundary conditions, briefly introduced in \eqref{BPboundary}. This condition can be exploit to deduce the scattering amplitude of the problem in the same approximations. However to obtain the renormalization condition for a contact potential only the zero energy case is needed, which is shown to be
$   f(t', t)\underset{k\rightarrow0}{=}
   -\ar\bra\alpha'|S\ket\bra S|\alpha\ket$, 
quite similar to the previous case, with the only difference that $\as$ should be replaced by the scattering length $\ar$ which now contains also a dependence to the parameters of the SO coupling.
Interestingly, also in this case, a bound state near the scattering threshold exists for energies $E_{b}$ that satisfy 
 \begin{equation}
   1/\ar+iE_{b}^{1/2}-4\pi\bra S|F(E_{b})|S\ket=0\,.
 \end{equation}

\subsubsection*{Renormalization condition for a contact potential}
\label{Rinormalizzazione nel caso spin-orbita}

As done previously, in order to obtain the correct renormalization relation for the contact potential in the presence of a spin-orbit coupling, one can exploit the Lippmann-Schwinger equation and impose that a pseudo-potential reproduces the same scattering amplitude of a realistic potential in the low energy limit. Looking at Eq.\eqref{scattering_amplitude_SO}, the natural choice is to consider a simple contact potential that acts only on the singlet state $|S\ket=\frac{1}{\sqrt{2}}\left(|\spinup\!\spindown\ket-|\spindown\!\spinup\ket\right)$
 \begin{equation}
   \hat{V}_{\text{eff}}(\mathbf{r})=-g\delta(\mathbf{r})\otimes|S\ket\bra S|
 \end{equation}
The Lippmann-Schwinger equation with the effective potential is given by
 \begin{equation}
\label{TSO}
   \hat{T}_{2B}(z)=\hat{V}_{\text{eff}}+\hat{V}_{\text{eff}}\frac{1}{z-\hat{H}_{0}+i0^{+}}\hat{T}_{2B}(z)
 \end{equation}
It is also useful to remember that the scattering amplitude in terms of the T-matrix is 
 \begin{equation}
   f(t', t)=-2\pi^{2}m\brabra\psi^{(0)}_{t'}|T_{2B}(E_t)|\psi^{(0)}_{t}\ketket
 \end{equation}
If two eigenstates of the free Hamiltonian are applied on the right- and on the left-hand sides of Eq.~(\ref{TSO}), denoting for simplicity $|\alpha, \mathbf{k}\rangle$ by $|\alpha\rangle$, the scattering amplitude is recovered
 \begin{equation}
-\frac{1}{2\pi^{2}m}f(t', t)=
-\frac{g}{(2\pi)^{3}}\bra\alpha'|S\ket\bra S|\alpha\ket
+\frac{g}{(2\pi)^{3}}\bra\alpha'|S\ket
\left(
\sum_{\alpha''}\int d^{3}k''\frac{\bra S|\alpha''\ket}{E_t-E_{t''}+i0^{+}}
\frac{1}{2\pi^{2}m}f(t'',t)
\right)
 \end{equation}
Using the zero energy limit for the scattering amplitude 
$f(t',t)\underset{k\rightarrow0}{=}-\ar\bra\alpha'|S\ket\bra S|\alpha\ket$, 
one gets
 \begin{equation}
\frac{\ar}{2\pi^{2}m}=
-\frac{g}{(2\pi)^{3}}
+\frac{g}{(2\pi)^{3}}
\frac{\ar}{2\pi^{2}m}\left(
+\sum_{\alpha''}\int d^{3}k''\frac{\bra\alpha''|S\ket\bra S|\alpha''\ket}{E_{t''}}
\right)
 \end{equation}
Simplifying the expression and making explicit the singlet state the obtained relation between the scattering length $\ar$ and the coupling $g$ can be put in the form
 \begin{equation}
\frac{1}{g}=-\frac{m}{4\pi\ar}
+\frac{1}{(2\pi)^{3}}
\int d^{3}k\frac{1}{2}\left(\frac{1}{E_{\uparrow\downarrow}}+\frac{1}{E_{\downarrow\uparrow}}\right)
 \end{equation}
We can further simplify this relation using $E_{\downarrow\uparrow}(-\mathbf{k})=E_{\uparrow\downarrow}(\mathbf{k})$, getting
 \begin{equation}
\int d^{3}k\frac{1}{2}\left(\frac{1}{E_{\uparrow\downarrow}}+\frac{1}{E_{\downarrow\uparrow}}\right)=
\int d^{3}k\frac{1}{E_{\uparrow\downarrow}}=
\int d^{3}k\frac{1}{k^{2}/m+\varepsilon(\mathbf{k},\xi)-\varepsilon(-\mathbf{k},\xi)}
 \end{equation}
which, in spherical coordinates, can be written as 
 \begin{equation}
   m\int d^{3}k\frac{1}{k^{2}}
  -m \int d\Omega\int_{0}^{\infty}dk
   \left(\frac{\varepsilon(\mathbf{k},\xi)-\varepsilon(-\mathbf{k},\xi)}{k^{2}/m+\varepsilon(\mathbf{k},\xi)-\varepsilon(-\mathbf{k},\xi)}\right)
\label{eqD28}
 \end{equation}
Because of the behavior of the eigenvalues shown in Eq.~\eqref{asimptoticE} the integrand in the last term of Eq.~(\ref{eqD28}), at large momenta, 
will have an asymptotic behavior proportional to $1/k^{2}$, therefore, the integral converges to a constant $C$. We finally get 
 \begin{equation}
\int d^{3}k\frac{1}{k^{2}/m+\varepsilon(\mathbf{k},\xi)-\varepsilon(-\mathbf{k},\xi)}=
m\int d^{3}k\frac{1}{k^{2}}+C
 \end{equation}
The constant term can be easily incorporated in the definition of the coupling $g$, so that the renormalization condition turns to be
 \begin{equation}
\label{rcSO}
\frac{1}{g}=-\frac{m}{4\pi\ar}
+\frac{m}{(2\pi)^{3}}
\int d^{3}k\frac{1}{k^{2}}\,.
 \end{equation}
The presence of a SO coupling does not alter the form of the renormalization condition for a contact pseudo-potential. 
The only caution to be taken is to replace 
$\as$ with $\ar$, a scattering length which depends both on the interatomic potential and on the SO term.

\end{document}